\documentclass[a4paper,11pt]{article}
\pdfoutput=1
\usepackage{jcappub}
\usepackage{subfig}
\usepackage{caption}
\usepackage{hyperref}
\usepackage{amsmath}
\usepackage{mathtools}
\usepackage{tablefootnote}
\usepackage{soul}
\usepackage{color}
\usepackage{verbatim}
\usepackage{float}
\usepackage{bm}
\usepackage{ulem}
\usepackage[dvipsnames]{xcolor}
\usepackage{tikz}
\usetikzlibrary{shapes.geometric, backgrounds, arrows, calc, fit}
\tikzstyle{process} = [rectangle, rounded corners, minimum width=8cm, minimum height=1cm, text centered, draw=black, fill=blue!30]
\tikzstyle{arrow} = [thick,->,>=stealth]

\title{\texttt{CLASS\_GWB}: robust modeling of the astrophysical gravitational wave background anisotropies}

\author[a,b,c]{Nicola Bellomo,}
\emailAdd{nicola.bellomo@austin.utexas.edu}

\author[d,e,f]{Daniele Bertacca,}
\emailAdd{daniele.bertacca@pd.infn.it}

\author[g,1]{Alexander C. Jenkins,%
\note{Current address: Department of Physics and Astronomy, University College London, London WC1E 6BT, United Kingdom}}
\emailAdd{alexander.jenkins@kcl.ac.uk}

\author[d,e,f,h]{Sabino Matarrese,}
\emailAdd{sabino.matarrese@pd.infn.it}

\author[d,e,i]{Alvise Raccanelli,}
\emailAdd{alvise.raccanelli.1@unipd.it}

\author[j]{Tania Regimbau,}
\emailAdd{tania.regimbau@lapp.in2p3.fr}

\author[d,e]{Angelo Ricciardone,}
\emailAdd{angelo.ricciardone@pd.infn.it}

\author[g]{Mairi Sakellariadou}
\emailAdd{mairi.sakellariadou@kcl.ac.uk}

\affiliation[a]{Theory Group, Department of Physics, University of Texas, Austin, TX 78712, USA}
\affiliation[b]{ICC, University of Barcelona, IEEC-UB, Mart\'i i Franqu\`es, 1, E-08028 Barcelona, Spain.}
\affiliation[c]{Dept. de  F\'isica Qu\`antica i Astrof\'isica, Universitat de Barcelona, Mart\'i  i Franqu\`es 1, E-08028 Barcelona, Spain.}
\affiliation[d]{Dipartimento di Fisica e Astronomia ``Galileo Galilei'', Universit\`a degli Studi di Padova, I-35131 Padova, Italy.}
\affiliation[e]{INFN, Sezione di Padova, via F. Marzolo 8, I-35131 Padova, Italy.}
\affiliation[f]{INAF - Osservatorio Astronomico di Padova, Vicolo dell'Osservatorio 5, I-35122 Padova, Italy.}
\affiliation[g]{Theoretical Particle Physics and Cosmology Group, Physics Department, King's College London, University of London, Strand, London WC2R 2LS, UK.}
\affiliation[h]{Gran Sasso Science Institute, Viale F. Crispi 7, I-67100 L'Aquila, Italy.}
\affiliation[i]{Theoretical Physics Department, CERN, 1 Esplanade des Particules, 1211 Geneva 23,
Switzerland.}
\affiliation[j]{LAPP, Universit\'e Grenoble Alpes, USMB, CNRS/IN2P3, F-74000 Annecy, France.}

\abstract{
Gravitational radiation offers a unique possibility to study the large-scale structure of the Universe, gravitational wave sources and propagation in a completely novel way. Given that gravitational wave maps contain a wealth of astrophysical and cosmological information, interpreting this signal requires a non-trivial multidisciplinary approach. In this work we present the complete computation of the signal produced by compact object mergers accounting for a detailed modelling of the astrophysical sources and for cosmological perturbations. We develop the \texttt{CLASS\_GWB} code, which allows for the computation of the anisotropies of the astrophysical gravitational wave background, accounting for source and detector properties, as well as effects of gravitational wave propagation. We apply our numerical tools to robustly compute the angular power spectrum of the anisotropies of the gravitational wave background generated by astrophysical sources in the LIGO-Virgo frequency band. The end-to-end theoretical framework we present can be easily applied to different sources and detectors in other frequency bands. Moreover, the same numerical tools can be used to compute the anisotropies of gravitational wave maps of the sky made using resolved events.
}

\begin{document}
\begin{flushleft}
KCL-PH-TH-2021-70
\end{flushleft}

\maketitle


\section{Introduction}

The way we observe the cosmos completely changed after the first LIGO/Virgo detection of gravitational waves (GWs)~\cite{abbott:firstligodetection, abbott:firstligodetectionproperties}. For the first time we are able to observe physical phenomena beyond the solar system not only via their emitted electromagnetic radiation but also through their gravitational signal. Subsequent GW detections~\cite{abbott:O1O2allevents, venumadhav:O1O2newevents, abbott:O3events} established the birth not only of the GW astronomy as an independent field, but also of multi-messenger astronomy, thanks to those events that present an electromagnetic counterpart~\cite{abbott:multimessenger}.

Needless to say, the detection of GWs opened a new era also in Cosmology. Independently of their intrinsic properties, the detected events represent a true cosmological signal that trace the large-scale structure (LSS) of the Universe~\cite{libanore:gwxlss}. Therefore for the first time we can study the evolution of the Universe using a complementary messenger, i.e., gravitational radiation. In other words we can use current and future GW detectors to probe astrophysical and cosmological properties of GW sources and hosts by using both resolved GW events and unresolved ones, i.e., a gravitational wave background (GWB), which is the main focus of this work.

The total GWB is composed by a cosmological and an astrophysical component. While the former is sourced by GW production mechanisms in the early universe (first-order phase transitions, topological defects, inflation, and second order fluctuations that can produce Primordial Black Holes~\cite{maggiore:earlyuniversegwsources, guzzetti:earlyuniversegwsources, bartolo:earlyuniversegwsources, caprini:earlyuniversegwsources}), the latter is mostly generated by the superposition of unresolved events, mostly inspiraling and merging binaries, see e.g., refs.~\cite{regimbau:astrophysicalsources, romano:astrophysicalsources} and references therein. In the most common scenarios the astrophysical component is expected to dominate over the cosmological one. Accessing the information contained in the cosmological GWB requires subtracting the astrophysical GWB from the total observed signal, therefore characterising the latter becomes of the uttermost importance.

In the recent years there has been a lot of effort in developing tools to detect and disentangle the astrophysical from the cosmological component. The frequency shape represents one powerful discriminator, since it is expected that different sources are characterized by different spectral shapes~\cite{Caprini:2019pxz, Flauger:2020qyi}.  This has been used also to perform a simultaneous reconstruction of both background components both for space-based detectors as LISA~\cite{Karnesis:2019mph, Flauger:2020qyi, Boileau:2020rpg} and for ground-based as LIGO-Virgo-KAGRA and Einstein Telescope-Cosmic Explorer~\cite{Martinovic:2020hru}. However, relying only on the spectral shape may be not enough, considering the large number of sources expected to be present at the same time, especially for third generation detectors. In this paper we explore another powerful observable which is remarkably useful in characterizing GW backgrounds: their anisotropies, and in particular their related angular power spectrum. So far the LIGO-Virgo-KAGRA collaboration succeeded in obtaining upper limits on the isotropic component of the background~\cite{abbott:GWBfirstrun, abbott:GWBsecondrun, abbott:GWBthirdrun}. However, the promising prospects of detecting the (astrophysical) isotropic component of the GWB in the next $\mathcal{O}(5)$ LIGO/Virgo run~\cite{regimbau:GWBdetectionforecast, abbott:GWBdetectionforecast}, combined with the recent claim of the detection of a common red process from the NANOGrav collaboration~\cite{NANOGrav:2020bcs}, makes timely having also a robust and accurate theoretical prediction of the observed anisotropic signal we plan to measure. 

The anisotropies of the GWB have already been computed by several authors, both for its cosmological~\cite{alba:cosmologicalGWB, contaldi:cosmologicalGWB, bartolo:cosmologicalGWB, Bartolo:2019yeu, Geller:2018mwu, ValbusaDallArmi:2020ifo, Adshead:2020bji, Domcke:2020xmn,  Ricciardone:2021kel, Braglia:2021fxn, Dimastrogiovanni:2021mfs} and astrophysical~\cite{cusin:astrophysicalGWBI, cusin:astrophysicalGWBII, cusin:astrophysicalGWBIII, jenkins:astrophysicalGWBI, jenkins:astrophysicalGWBII, bertacca:astrophysicalGWB, canasherrera:astrophysicalGWB, capurri:astrophysicalGWB} component, using analytical and simulation-based methods. The goal of this project is to build upon these works to develop an end-to-end approach that takes the best of those two methods: \textit{(i)} a rigorous analytical treatment of cosmological perturbations and \textit{(ii)} a numerical and simulation-based treatment of those aspects of GW sources and hosts that cannot easily accounted for by analytical prescriptions.

We develop a series of numerical tools able to address the complexity of the problem, which involves different astrophysical (galactic and sub-galactic) and cosmological scales. These tools allow the user to simulate a catalogue of GW events according to different astrophysical models, to establish which events are resolved and which not by a given detector (network) and to compute the expected level of anisotropies generated by the latter events. The former two tasks are performed by \texttt{python} routines which deal with the bulk of the astrophysical sources and hosts information, whereas for the latter we create an extension of the public code \texttt{CLASS}~\cite{blas:class} called \texttt{CLASS\_GWB},\footnote{All the codes described in this work will be released after the article is accepted in the GitHub page \url{https://github.com/nbellomo?tab=repositories}, along with a user manual. In particular~\texttt{CLASS\_GWB} is going to be released as an extension of~\texttt{Multi\_CLASS}~\cite{bellomo:multiclass, bernal:multiclass}.} which allows to compute the angular power spectrum for the GWB taking into account all the projection effects intervening between GW sources and the observer (i.e., \textit{density}, \textit{velocity} and \textit{gravity} terms). We also include, for the first time, a complete modelling of the hosts of the GW sources together with the full time-dependence of the bias and evolution bias effective functions. With all these tools we simulate a catalogue of astrophysical sources expected to be present in the LIGO/Virgo frequency band, i.e., black hole, neutron star and black hole-neutron star binaries (BBH, BNS, BHNS, respectively), according to the latest astrophysical constraints~\cite{abbott:O3events}. We extract the isotropic background energy density contribution and the angular power spectrum expected to be measured at design sensitivity, including all the relativistic corrections and the shot noise contribution. Finally we quantify the impact of different projection effects at large angular scales, which are the only ones at which present and future GW detectors are expected to be sensitive. We confirm that effects, as Kaiser and Doppler, have a non-negligible role and can have an impact on the amplitude of the spectrum up to ten percent.

The framework and the machinery developed in this paper has a far more general scope than the estimation of the astrophysical GWB. First of all the numerical tools developed to simulate GW events and GW detectors can also be applied to the case of resolved events, allowing to build catalog of resolved GW sources. Secondly, they provide the building block to accurately predict the expected measurable signal, also in the case where GW events are used along with other external datasets, (e.g., those provided by LSS surveys). Since the typical astrophysical processes responsible for GW emission are located in galaxies, the GWB of astrophysical origin traces the LSS of our Universe, almost as precisely as galaxies do. By cross-correlating GW maps of the sky with galaxy number count maps we can establish how GW sources are spatially clustered, i.e., how they trace the underlying dark matter field, and how they are temporally distributed when most of the emission is generated. Moreover, by using different galaxy catalogs, we can also measure how GW sources are correlated with different types of galaxies observed in frequency bands ranging from radio to optical and UV, highlighting the importance of a multi-wavelength astronomy. 

Numerous studies already show the promising potential of combining resolved GW events and LSS tracers to constraint different binary formation scenarios~\cite{raccanelli:pbhprogenitors, scelfo:gwxlss}, clustering properties~\cite{calore:gwxlss, libanore:gwxlss, scelfo:gwxlssII} and propagation effects in the context of General Relativity \cite{Laguna:2009re, Hirata:2010ba, Bertacca:2017vod} and Modified Gravity theories~\cite{camera:gwpropagation, namikawa:gwpropagation, mukherjee:gwpropagationI, mukherjee:gwpropagationII, mukherjee:gwpropagationIII, mukherjee:gwpropagationIV, mukherjee:gwpropagationV, Belgacem:2018lbp, Belgacem:2017ihm, LISACosmologyWorkingGroup:2019mwx, Dalang:2019rke, Garoffolo:2019mna, Dalang:2020eaj, Ezquiaga:2020dao, Garoffolo:2020vtd, Tasinato:2021wol, Ezquiaga:2021ler}. Our accurate prediction of the expected signal and the flexibility in modelling GW sources, hosts and detectors, is perfectly suitable both for future GW observatories as Einstein Telescope~\cite{Sathyaprakash:2011bh, Maggiore:2019uih}, Cosmic Explorer~\cite{LIGOScientific:2016wof} and LISA~\cite{LISA:2017pwj}, and for the combination with other LSS survey as EMU~\cite{norris:emu}, Vera Rubin~\cite{abell:lsstsciencebook}, DESI~\cite{aghamousa:desi}, Euclid~\cite{laureijs:euclidsciencebook}, SPHEREx~\cite{dore:spherexwhitepaperI}, Nancy Grace Roman~\cite{spergel:wfirstwhitepaper} and SKA~\cite{bacon:skaredbook}.

The paper is organized as follows: in section~\ref{sec:aGWB_theory} we review the analytical treatment of the astrophysical GWB anisotropies. In section~\ref{sec:sources_and_detectors} we discuss how to build a catalog of unresolved GW events starting from an astrophysical model, while in section~\ref{sec:gw_bias} we present a GW biasing scheme. Then in section~\ref{sec:angular_power_spectrum} we show the expected level of anisotropy, comparing it to the shot noise due to the spatial and temporal clustering of the events. In particular in \S~\ref{subsec:worked_example_Meff}, \ref{subsec:worked_example_bbeff} and~\ref{subsec:worked_example_signalnoise} we provide a worked example of how to apply our framework to a concrete model for the astrophysical GWB. Finally, we draw our conclusions in section~\ref{sec:conclusions}. Appendices~\ref{app:aGWB_poisson_gauge}-\ref{app:gw_waveform}-\ref{app:rotating_coordinate_systems}-\ref{app:halo_properties}-\ref{app:external_modules} are dedicated to technical aspects of this work.


\section{The astrophysical gravitational wave background}
\label{sec:aGWB_theory}

The isotropic astrophysical GWB and its fluctuations are two key observables that can be targeted by a GW observatory. In this section we review the most important results concerning these two quantities following mainly ref.~\cite{bertacca:astrophysicalGWB}, which constitutes the theoretical foundation of the~\texttt{CLASS\_GWB} extension of~\texttt{CLASS}. The authors of ref.~\cite{bertacca:astrophysicalGWB} derive the two observables of interest in a coordinate-independent and gauge invariant way applying the ``Cosmic Rulers'' formalism~\cite{jeong:cosmicrulers, schmidt:cosmicrulers}. They account for all possible local and integrated projection effects intervening in the past GW cone, providing a very accurate prediction of the astrophysical GWB anisotropy at large scales. 
The total present-day GW relative energy density per logarithmic observed frequency~$f_o$ and solid angle~$\Omega_o$ along the line-of-sight~$\hat{\mathbf{n}}$ is~\cite{allen:GWBdefinition, phinney:GWBdefinition}
\begin{equation}
\Omega^\mathrm{tot}_\mathrm{GW}(f_o, \hat{\mathbf{n}}) = \frac{f_o}{\rho_{0c}c^2} \frac{d\rho^\mathrm{tot}_\mathrm{GW}}{df_o d\Omega_o} (f_o, \hat{\mathbf{n}}),
\label{eq:omegatotgw_def}
\end{equation}
where~$\rho^\mathrm{tot}_\mathrm{GW}$ is the total GW energy density, $\rho_{0c}=3H^2_0/(8\pi G)$ is the present-day critical density, $H_0$ is the Hubble expansion rate today, $G$ is Newton's gravitational constant and~$c$ is the speed of light. The total energy density can be decomposed into the sum of the present-day GW relative energy density of all GW sources, i.e., 
\begin{equation}
\Omega^\mathrm{tot}_\mathrm{GW}(f_o, \hat{\mathbf{n}}) = \sum_i \Omega^{[i]}_\mathrm{GW}(f_o, \hat{\mathbf{n}}),
\label{eq:omegatotgw_decomposition}
\end{equation}
where the index~$i$ labels the sources. For each source we characterise the GW relative energy density as~\cite{marassi:ensembleaverage, Regimbau:2008nj, Regimbau:2011rp, Wu:2011ac}
\begin{equation}
\begin{aligned}
\Omega^{[i]}_\mathrm{GW}(f_o, \hat{\mathbf{n}}) &= \frac{f_o}{\rho_{0c}c^2} \int \frac{dz}{H(z)} \left\langle R^{\mathrm{obs},[i]}(z, \hat{\mathbf{n}}) \frac{dE^{[i]}_{\mathrm{GW},e}}{df_{e} d\Omega_{e}}\Big(f_o, z, \hat{\mathbf{n}}\Big) \right\rangle \\
&= \frac{f_o}{\rho_{0c}c^2} \int \frac{dz}{H(z)} \int d\bm{\theta} p^{[i]}(\bm{\theta}|z) \frac{R^{[i]}(z,\bm{\theta}, \hat{\mathbf{n}})}{1+z} \frac{dE^{[i]}_{\mathrm{GW},e}}{df_{e} d\Omega_{e}}\Big(f_o, z, \bm{\theta}, \hat{\mathbf{n}}\Big),
\label{eq:omegagw_ensemble_average}
\end{aligned}
\end{equation}
with $z$ being the redshift, $H(z)$ the Hubble expansion rate, $dE^{[i]}_{\mathrm{GW},e}/df_{e} /d\Omega_{e}$ the GW emitted energy per emitted frequency~$f_{e}=f_{o}(1+z)$ and solid angle~$\Omega_{e}$, $R^{\mathrm{obs},[i]} = R^{[i]}/(1+z)$ and~$R^{[i]}$ the observed comoving merger rate density and the intrinsic comoving merger rate density of $i$-th type of source, respectively. The angle brackets~$\left\langle \cdots \right\rangle$ indicate averages over a set of cosmological and astrophysical parameters~$\bm{\theta}$ characterising both the type of source of interest and its environment, i.e., the ``host''. The conditional probability density functions~$p^{[i]}(\bm{\theta}|z)$ (pdfs) weighs the contribution of single events to the total signal coming from the entire population of GW events of the same kind.

The GW events can be divided into two broad classes: those that are resolved by a detector network, thus called ``resolved events'', and those which are not. In this work we focused on the latter, in fact the astrophysical GWB is made precisely by the incoherent superposition of many of these unresolved events coming from different kind of sources. The definitions we provide in equations~\eqref{eq:omegatotgw_decomposition} and~\eqref{eq:omegagw_ensemble_average} can be characterised to the case of the astrophysical GWB by introducing an ``efficiency'' function~$\varepsilon_\mathrm{res}(z,\bm{\theta})$, which selects resolved events depending on the parameters of the source and on the specific configuration of the detector network. Therefore the isotropic component of the astrophysical GWB reads as 
\begin{equation}
\bar{\Omega}^\mathrm{tot}_\mathrm{GWB}(f_o) = \frac{f_o}{\rho_{0c}c^2} \sum_i \int \frac{dz}{H(z)} d\bm{\theta} p^{[i]}(\bm{\theta}|z) \left[1-\varepsilon_\mathrm{res}(z,\bm{\theta})\right] \frac{\bar{R}^{[i]}(z,\bm{\theta})}{1+z} \overline{\frac{dE^{[i]}_{\mathrm{GW},e}}{df_{e} d\Omega_{e}}}\Big(f_o, z,\bm{\theta}\Big),
\label{eq:omegatot_GWB_background}
\end{equation}
where the symbol~``$\ \overline{\textcolor{white}{a} }\ $'' indicates background quantities and the efficiency function takes the value~$\varepsilon_\mathrm{res}(z,\bm{\theta})=0$ ($\varepsilon_\mathrm{res}(z,\bm{\theta})=1$) if the event described by the set of parameters~$(z,\bm{\theta})$ is unresolved (resolved). In other words, the efficiency function removes resolved events from the GW relative energy density. Along with the efficiency function, we can introduce also an ``average efficiency''
\begin{equation}
\left\langle \varepsilon^{[i]}_\mathrm{res} (z)\right\rangle = \int d\bm{\theta} p^{[i]}(\bm{\theta}|z) \varepsilon_\mathrm{res}(z,\bm{\theta}) = \frac{N^{[i]}_\mathrm{res}(z)}{N^{[i]}_\mathrm{tot}(z)}
\end{equation}
that describes how many GW events we resolve at redshift~$z$ $\left(N^{[i]}_\mathrm{res}\right)$ with respect to the total number of events at the same redshift~$\left(N^{[i]}_\mathrm{tot}\right)$. The average efficiency conveys in a straightforward manner how effective our detector network is at resolving a population of events at given redshift, however in the rest of this work we focus mainly on constructing the event-based efficiency function. Neither the efficiency or the average efficiency depends on the line-of-sight since when we assess whether an event is resolved or unresolved we always assume that the GW propagation happens in an unperturbed Friedmann-Lema\^{i}tre-Robertson-Walker (FLRW) Universe, see also the discussion in section~\ref{subsec:detector}. Note that detector specifications enter only in the efficiency function and not in the conditional pdfs, which are those of the entire population of events.

Besides the isotropic component, we can also investigate the anisotropies in the astrophysical GWB energy density. Starting from equation~\eqref{eq:omegatotgw_decomposition}, we define the total relative fluctuation as~$\Delta^\mathrm{tot}_\mathrm{GWB}(f_o, \hat{\mathbf{n}}) := \left(\Omega^\mathrm{tot}_\mathrm{GWB}-\bar{\Omega}^\mathrm{tot}_\mathrm{GWB}\right)/\bar{\Omega}^\mathrm{tot}_\mathrm{GWB}$ and connect it to the relative fluctuation of the $i$-th type of source~$\Delta^{[i]}_\mathrm{GWB}(f_o, \hat{\mathbf{n}}) := \left(\Omega^{[i]}_\mathrm{GWB}-\bar{\Omega}^{[i]}_\mathrm{GWB}\right)/\bar{\Omega}^{[i]}_\mathrm{GWB} = \Delta\Omega^{[i]}_\mathrm{GWB}/\bar{\Omega}^{[i]}_\mathrm{GWB}$ as
\begin{equation}
\Delta^\mathrm{tot}_\mathrm{GWB}(f_o, \hat{\mathbf{n}}) = \sum_i \frac{\Delta\Omega^{[i]}_\mathrm{GWB}}{\bar{\Omega}^\mathrm{tot}_\mathrm{GWB}} = \sum_i \frac{\bar{\Omega}^{[i]}_\mathrm{GWB}(f_o)}{\bar{\Omega}^\mathrm{tot}_\mathrm{GWB}(f_o)} \Delta^{[i]}_\mathrm{GWB}(f_o, \hat{\mathbf{n}}),
\label{eq:omegatotGWB_fluctuation}
\end{equation}
where the anisotropies of the individual components are weighted by the frequency-dependent coefficient~$\bar{\Omega}^{[i]}_\mathrm{GWB}(f_o)/\bar{\Omega}^\mathrm{tot}_\mathrm{GWB}(f_o) \in [0,1]$.

The explicit form of the fluctuation~$\Delta\Omega^{[i]}_\mathrm{GWB}(f_o, \hat{\mathbf{n}})$ is derived in~\cite{bertacca:astrophysicalGWB} and its computation in the Poisson gauge is reported in appendix~\ref{app:aGWB_poisson_gauge}, hence here we report only the main result and we discuss its interpretation. The notation used hereafter follows the one used in~\texttt{CLASS}, see, e.g., refs.~\cite{ma:perturbationtheory, blas:class, didio:classgal}, to facilitate the comparison with the numerical implementation described also in appendix~\ref{app:aGWB_poisson_gauge}. In the Poisson gauge the spacetime metric of a spatially flat FLRW Universe is given by
\begin{equation}
ds^2 = a^2(\tau)\left[ -(1+2\Psi) d\tau^2 + (1-2\Phi) dx_i dx^i \right],
\end{equation}
where $\tau$ is the conformal time, $x^j$ are comoving coordinates, $a$ is the scale factor, $\Psi$ and~$\Phi$ are the Bardeen potentials. In this gauge we have
\begin{equation}
\begin{aligned}
\Delta\Omega^{[i]}_\mathrm{GWB} (f_o, \hat{\mathbf{n}}) = & \frac{f_o}{\rho_{0c}c^2} \int \frac{dz}{H(z)} \int d\bm{\theta} p^{[i]}(\bm{\theta}|z) \left[1-\varepsilon_\mathrm{res}(z,\bm{\theta})\right] \frac{\bar{R}^{[i]}}{1+z}\overline{\frac{dE^{[i]}_{\mathrm{GW},e}}{df_e d\Omega_e}} \\
&\times \Bigg\{ b^{[i]} D  \\
&\quad + \left(b^{[i]}_\mathrm{evo} - 2 - \frac{\mathcal{H}'}{\mathcal{H}^2}\right) \hat{\mathbf{n}}\cdot\mathbf{V} - \frac{1}{\mathcal{H}}\partial_\parallel(\hat{\mathbf{n}}\cdot\mathbf{V}) - (b^{[i]}_\mathrm{evo}-3) \mathcal{H} V \\
&\quad + \left(3 - b^{[i]}_\mathrm{evo} + \frac{\mathcal{H}'}{\mathcal{H}^2}\right)\Psi + \frac{1}{\mathcal{H}}\Phi' + \left(2 - b^{[i]}_\mathrm{evo} + \frac{\mathcal{H}'}{\mathcal{H}^2}\right) \int_0^{\chi(z)} d\chi \left(\Psi'+\Phi'\right) \\
&\quad + \left(b^{[i]}_\mathrm{evo} - 2 - \frac{\mathcal{H}'}{\mathcal{H}^2}\right)  \left( \Psi_o - \mathcal{H}_0 \int_{0}^{\tau_0} d\tau \left.\frac{\Psi(\tau)}{1+z(\tau)}\right|_o -  \left(\hat{\mathbf{n}}\cdot\mathbf{V} \right)_o \right) \Bigg\},
\end{aligned}
\label{eq:fluctuation_poissongauge}
\end{equation}
where $\chi(z)$ is the comoving distance at redshift~$z$, $\tau_0$ is the conformal time today, $\mathcal{H}=a'/a$ is the Hubble expansion rate in conformal time and $\mathcal{H}_0$ is the present-day Hubble expansion rate. Derivatives with respect to conformal time are indicated by~``$\ '\ $'', while spatial derivatives along the line-of-sight are indicated by~$\partial_\parallel$. Quantities evaluated at observer position are indicated by the subscript~``$\ _{o}\ $''. The quantities $D$ and $\mathbf{V}$ are respectively the gauge-invariant matter density fluctuation and the gauge invariant perturbation velocity, while~$b^{[i]}(z,\bm{\theta})$ and
\begin{equation}
b^{[i]}_\mathrm{evo}(f_o, z, \bm{\theta}):=
-\frac{d}{d\log(1+z)}\log\left[\bar{R}^{[i]}\overline{\frac{dE^{[i]}_{\mathrm{GW},e}}{df_e d\Omega_e}}\right]
\end{equation}
are the bias and the evolution bias of the $i$-th type of GW source, respectively. In analogy with galaxy number counts, the bias function specifies the clustering properties of GW sources, while the evolution bias function characterizes the formation of new sources. In fact GW events does not scale as~$a^{-3}$ because of the formation of new galaxies and of the non-constant star formation rate inside galaxies.

The total astrophysical GWB fluctuation in equation~\eqref{eq:omegatotGWB_fluctuation} can be written using equation~\eqref{eq:fluctuation_poissongauge} in a more compact way as
\begin{equation}
\begin{aligned}
\Delta^\mathrm{tot}_\mathrm{GWB} (\hat{\mathbf{n}}, f_{o}) = & \int  dz \mathcal{M}^\mathrm{eff}(f_o, z) \times \\
&\times \Bigg\{ b^\mathrm{eff} D  \\
&\quad + \left(b^\mathrm{eff}_\mathrm{evo} - 2 - \frac{\mathcal{H}'}{\mathcal{H}^2}\right) \hat{\mathbf{n}}\cdot\mathbf{V} - \frac{1}{\mathcal{H}}\partial_\parallel(\hat{\mathbf{n}}\cdot\mathbf{V}) - (b^\mathrm{eff}_\mathrm{evo}-3) \mathcal{H} V  \\
&\quad + \left(3 - b^\mathrm{eff}_\mathrm{evo} + \frac{\mathcal{H}'}{\mathcal{H}^2}\right)\Psi + \frac{1}{\mathcal{H}}\Phi' + \left(2 - b^\mathrm{eff}_\mathrm{evo} + \frac{\mathcal{H}'}{\mathcal{H}^2}\right) \int_0^{\chi} d\tilde{\chi} \left(\Phi'+\Psi'\right)  \\
&\quad + \left(b^\mathrm{eff}_\mathrm{evo} - 2 - \frac{\mathcal{H}'}{\mathcal{H}^2}\right)  \left( \Psi_{o} - \mathcal{H}_0 \int_{0}^{\tau_0} d\tau \left.\frac{\Psi(\tau)}{1+z(\tau)}\right|_{o} -  \left(\hat{\mathbf{n}}\cdot\mathbf{V} \right)_{o} \right) \Bigg\},
\end{aligned}
\label{eq:relative_fluctuation_poissongauge}
\end{equation}
where we introduced the three effective functions
\begin{equation}
\begin{aligned}
\mathcal{M}^\mathrm{eff}(f_o, z) &= \frac{1}{\bar{\Omega}^\mathrm{tot}_\mathrm{GWB}(f_o)} \sum_i \frac{f_o}{\rho_{0c}c^2} \int d\bm{\theta} p^{[i]}(\bm{\theta}|z)  \frac{\left[1-\varepsilon_\mathrm{res}\right] \bar{R}^{[i]}}{(1+z)H(z)}\overline{\frac{dE^{[i]}_{\mathrm{GW},e}}{df_e d\Omega_e}}, \\
\mathcal{M}^\mathrm{eff}(f_o, z) b^\mathrm{eff}(f_o, z) &= \frac{1}{\bar{\Omega}^\mathrm{tot}_\mathrm{GWB}(f_o)} \sum_i \frac{f_o}{\rho_{0c}c^2} \int d\bm{\theta} p^{[i]}(\bm{\theta}|z) \frac{\left[1-\varepsilon_\mathrm{res}\right] \bar{R}^{[i]}}{(1+z)H(z)}\overline{\frac{dE^{[i]}_{\mathrm{GW},e}}{df_e d\Omega_e}} b^{[i]}(z, \bm{\theta}), \\
\mathcal{M}^\mathrm{eff}(f_o, z) b^\mathrm{eff}_\mathrm{evo}(f_o, z) &= \frac{1}{\bar{\Omega}^\mathrm{tot}_\mathrm{GWB}(f_o)} \sum_i \frac{f_o}{\rho_{0c}c^2} \int d\bm{\theta} p^{[i]}(\bm{\theta}|z) \frac{\left[1-\varepsilon_\mathrm{res}\right] \bar{R}^{[i]}}{(1+z)H(z)}\overline{\frac{dE^{[i]}_{\mathrm{GW},e}}{df_e d\Omega_e}} b^{[i]}_\mathrm{evo}(f_o, z, \bm{\theta}) .
\end{aligned}
\label{eq:effective_functions}
\end{equation}
By definition, $\mathcal{M}^\mathrm{eff}(z)$ is normalized to unity, i.e., $\displaystyle \int dz \mathcal{M}^\mathrm{eff}(f_o, z)\equiv 1$, hence it can be interpreted as a pdf that weighs the cosmological perturbations in curly brackets. The implementation in \texttt{CLASS} of equation~\eqref{eq:relative_fluctuation_poissongauge} is discussed in appendix~\ref{app:aGWB_poisson_gauge}.

The bulk of the cosmological information is contained into the \textit{density}, \textit{velocity}, \textit{gravity} and \textit{observer} terms in the first, second, third and fourth lines of equation~\eqref{eq:relative_fluctuation_poissongauge}, respectively. Each set of terms has a specific ``source'' function, namely the gauge-invariant matter density fluctuation~$D$, the gauge invariant velocity~$\mathbf{V}$ and the Bardeen potentials~$\Psi$ and~$\Phi$. The time- and scale-dependence of these four functions, along with the background expansion history, is fixed by a choice of a cosmological model. None of the terms in equation~\eqref{eq:relative_fluctuation_poissongauge} can be disregarded because \textit{(i)} only by including all of them we are computing a truly gauge-invariant observable, and \textit{(ii)} those terms contribute the most to the total especially at large scale, as shown in ref.~\cite{bertacca:astrophysicalGWB}.

The bulk of astrophysical information is contained into the three effective functions~$\mathcal{M}^\mathrm{eff}$, $b^\mathrm{eff}$ and~$b^\mathrm{eff}_\mathrm{evo}$. These functions contain information on the astrophysical properties of the GW sources, for instance the mass and spin distributions of the binary companions, the emitted GW energy spectrum, the clustering properties of GW events and the properties of the GW observatory under consideration. Even if in this work we focus only on the most standard set of GW sources (BBH, BNS and BHNS), we stress that the formalism introduced in this section is sufficiently flexible to include other types of event. Since the contribution of every class of sources is weighted by its correspondent~$\bar{\Omega}^{[i]}_\mathrm{GWB}(f_o)/\bar{\Omega}^\mathrm{tot}_\mathrm{GWB}(f_o)$ coefficient, not all of them are equally important in a given frequency channel. Therefore, it is quite straightforward to establish whether including or neglecting sources changes significantly the global shape of the~$\mathcal{M}^\mathrm{eff}, b^\mathrm{eff}, b^\mathrm{eff}_\mathrm{evo}$ effective functions, thus the resulting astrophysical GWB total fluctuation. Sections~\ref{sec:sources_and_detectors} and~\ref{sec:gw_bias} contain a complete and detailed recipe of how to compute these functions.

The clustering of GWs can be studied using the same statistical methods already developed for instance for galaxy clustering. In this work we focus on the statistical properties of the two-point function in harmonic space extracted from GW maps of the sky obtained by GW observatories. First of all we expand the astrophysical GWB fluctuation in spherical harmonics as
\begin{equation}
\Delta^\mathrm{tot}_\mathrm{GWB} (f_o, \hat{\mathbf{n}}) = \sum_{\ell m} a^{f_{o}}_{\ell m} Y_{\ell m}(\hat{\mathbf{n}}),
\label{eq:deltatotGWB_spherical_harmonics_decomposition}
\end{equation}
where~$Y_{\ell m}$ and~$a^{f_{o}}_{\ell m}$ are the spherical harmonics and the spherical harmonics coefficients, respectively. The angular power spectrum~$C^{f_{o}f_{o}}_\ell$, i.e., the two-point function in harmonic space, is given by
\begin{equation}
\left\langle a^{f_{o}}_{\ell m} a^{f_{o}*}_{\ell' m'} \right\rangle = \delta^K_{\ell\ell'} \delta^K_{mm'} C^{f_{o}f_{o}}_\ell,
\end{equation}
where~$\delta^K$ is the Kronecker delta, $^*$ indicates the complex conjugate and in this case the angle brackets~$\left\langle\cdots\right\rangle$ represents ensemble average over many realizations. The angular power spectrum is
\begin{equation}
C^{f_of_o}_\ell = 4\pi \int \frac{dk}{k} \mathcal{P}_\zeta(k) \Delta^{f_o}_\ell(k) \Delta^{f_o}_\ell(k),
\end{equation}
where
\begin{equation}
\mathcal{P}_\zeta(k) = A_s \left(\frac{k}{k_\mathrm{pivot}}\right)^{n_s-1}
\end{equation}
is the primordial curvature power spectrum, $A_s$ and~$n_s$ are the scalar perturbations amplitude and tilt, respectively, and~$k_\mathrm{pivot}$ is the pivot scale. The astrophysical GWB transfer function~$\Delta^{f_o}_\ell(k)$ receives contributions from density, velocity, gravity and observer terms (\textit{cf.} equation~\eqref{eq:relative_fluctuation_poissongauge}) and can be decomposed as
\begin{equation}
\Delta^{f_o}_\ell(k) = \Delta^\mathrm{den}_\ell(k) + \sum_{j=1}^3 \Delta^{\mathrm{vel},j}_\ell(k) + \sum_{j=1}^4 \Delta^{\mathrm{gr},j}_\ell(k) + \sum_{j=1}^3 \Delta^{\mathrm{obs},j}_\ell(k),
\label{eq:relative_fluctuation_contributions}
\end{equation}
where the explicit form of the different terms is reported in appendix~\ref{app:aGWB_poisson_gauge}. We created an extension of the public code~\texttt{CLASS}, called~\texttt{CLASS\_GWB},\footnote{The code and the user instructions will be publicly released as an extension of~\href{https://github.com/nbellomo/Multi\_CLASS}{Multi\_CLASS}~\cite{bellomo:multiclass, bernal:multiclass} after the article is accepted.} to compute the astrophysical GWB angular power spectrum for different sets of sources and/or detector networks. \texttt{CLASS\_GWB} accounts for all the projection effects intervening between sources and observer and also for the full time-dependence of the bias and evolution bias effective functions, instead of resorting to bin-averaged values, as in existing versions of~\texttt{CLASS}. Since \texttt{CLASS\_GWB} represents an extension of \texttt{Multi\_CLASS}, it is also possible to compute the cross-correlation of the astrophysical GWB with other cosmological observables, such as the CMB, galaxy clustering or even resolved GWs clustering.


\section{Gravitational wave sources and observatories}
\label{sec:sources_and_detectors}

Computing the effective kernel~$\mathcal{M}^\mathrm{eff}$ requires dealing with several technical aspects of GW sources and detectors. Broadly speaking, we have to simulate a population of GW events, their GW signal and the detector capability to detect them. However, many aspects of GW physics are still unknown despite the recent progress in constraining the properties of GW sources, see, e.g., refs.~\cite{abbott:O12eventsproperties, abbott:O3eventsproperties}. The only possible way to address this underlying uncertainty is to develop a framework that \textit{(i)} is flexible enough to account for a variety of GW sources and detector models and \textit{(ii)} is sufficiently modular so that it is possible to implement any new finding in all GW-adjacent fields, for instance a better characterisation of GW hosts, without disrupting the entire analysis chain.

In this section we present a methodology that meets the requirements described above and we develop a series of numerical tools, implement each and every stage of the analysis. For the sake of providing a quantitative estimate we also show a worked example by computing the effective kernel generated by BBH, BNS and BHNS mergers in the LIGO-Virgo frequency band. Note that other GW sources are expected to contribute to the total astrophysical GWB in this frequency channel, for instance core-collapse supernovae~\cite{buonanno:gwfromsupernovae, crocker:gwfromsupernovaeI, crocker:gwfromsupernovaeII} or rotating neutron stars~\cite{rosado:gwfromrotatingns, lasky:gwfromrotatingns, talukder:gwfromrotatingns}, however in this work we focus only on sources that have already been detected. 


\subsection{Gravitational wave events}
\label{subsec:events_catalog}

The first step of our methodology is generating a catalog of GW events from a fiducial astrophysical model, which should cover both the properties of the sources and of the sources' host. This catalog does not coincide with the catalog of observed events, since our GW observatories are intrinsically limited by noise, selection effects and systematics, however all these effects can be included in the subsequent steps where the detector is modelled. Most importantly, the catalog must provide a fair sample of the underlying GW events population, i.e., it has to contain enough GW events to fairly cover the entire parameter space of the specific astrophysical model of interest. 

At the practical level, selecting an astrophysical model for a GW event population (as BBH, BNS or BHNS mergers) means \textit{(i)} selecting a set of astrophysical parameters~$\bm{\theta}=\{\bm{\theta}_s,\bm{\theta}_\mathrm{host}\}$ that describes the GW source and host, respectively, and \textit{(ii)} selecting the conditional probability density functions (pdfs)~$p^{[i]}(\bm{\theta}|z)=p^{[i]}(\bm{\theta}_s,\bm{\theta}_\mathrm{host}|z)$ for those parameters. These~$\bm{\theta}$ and~$p^{[i]}(\bm{\theta})$ coincide with the set of parameters and pdfs introduced in section~\ref{sec:aGWB_theory} to describe the astrophysical GWB anisotropies.

Depending on the complexity of the astrophysical model we might have time-dependent pdfs and/or parameters statistically non-independent (which is always the case in realistic scenarios). While the former can be easily solved by using different pdfs at different times, the latter requires to establish a hierarchy between the parameters, deciding in which order they have to be sampled and fixing the correct conditional probabilities. Establishing a hierarchy between parameters might involve some degree of arbitrariness, however in this work we propose a method (which might not be unique) that relies on two physical insights:
\begin{enumerate}
\item The astrophysical properties of the sources depend on the properties of the host (and not vice-versa) at the time of the binary formation.
\item The clustering properties of the sources depend the properties of the host at the time of the merger.
\end{enumerate} 
The direct consequence of the first insight is that the parameters of the host should be sampled before those of the source, hence they occupy a higher position in our parameter hierarchy, i.e., $p^{[i]}(\bm{\theta}_s,\bm{\theta}_\mathrm{host}|z)=p^{[i]}(\bm{\theta}_s|\bm{\theta}_\mathrm{host},z) \times p^{[i]}(\bm{\theta}_\mathrm{host}|z)$. The second insight makes evident that there are two typical time-scales: one connected with the source and another one connected to how the sources trace the hosts, thus the large scale structure of the Universe. 

\begin{figure}[t]
\centerline{
\begin{tikzpicture}[node distance=2cm]
\node (pro1) [process, align=center] {Select a category of events between\\ BBH, BNS and BHNS mergers.};
\node (pro2) [process, below of = pro1, align=center] {Sample $\Delta t$ from $p(\Delta t)$.\\ Compute arrival time of the signal.};
\node (pro3) [process, below of = pro2] {Sample $z$ from equation~\eqref{eq:observed_merger_rate_pdf}.};
\node (pro4) [process, below of = pro3, align=center] {Sample $t_d$ from $p(t_d)$.\\ Compute $z_f(z,t_d)$ from equation~\eqref{eq:delay_time}.};
\node (pro5) [process, below of = pro4, align=center] {Sample $M_h(z_f)$ from $p(M_h,z_f)$.\\ Compute $M_h\left(z|M_h(z_f),z_f\right)$.};
\node (pro6) [process, below of = pro5] {Sample $m_1,m_2,\chi_1,\chi_2,\alpha,\delta,\iota,\psi$ from $p(\bm{\theta}_s|\bm{\theta}_\mathrm{host},z_f)$.};
\begin{scope}[on background layer]
	\node[draw, thick, rounded corners, fill=red!30,
          inner ysep=6mm, inner xsep=0mm, minimum width=16cm, 
          fit = (pro2)(pro3)(pro4)(pro5)(pro6)] (gw-event) {};
\end{scope}
\begin{scope}[on background layer]
	\node[draw, thick, rounded corners, fill=green!30,
          inner ysep=4mm, inner xsep=0mm, minimum width=15cm, 
          fit = (pro4)(pro5)] (env-par) {};
\end{scope}
\begin{scope}[on background layer]
	\node[draw, thick, rounded corners, fill=green!30,
          inner ysep=4mm, inner xsep=0mm, minimum width=15cm, 
          fit = (pro6)] (source-par) {};
\end{scope}
\begin{scope}[on background layer]
    \node at ([xshift=-6.3cm, yshift=-0.5cm]gw-event.north) [draw, fill=white, font=\fontsize{15}{0}] {\textbf{GW Event}};
\end{scope}
\begin{scope}[on background layer]
    \node at ([xshift=-6.55cm, yshift=-0.5cm]env-par.north) [draw, fill=white, font=\fontsize{15}{0}] {\textbf{Host}};
\end{scope}
\begin{scope}[on background layer]
    \node at ([xshift=-6.3cm, yshift=-0.5cm]source-par.north) [draw, fill=white, font=\fontsize{15}{0}] {\textbf{Source}};
\end{scope}
\draw [arrow] (pro1) -- (pro2);
\draw [arrow] (pro2) -- (pro3);
\draw [arrow] (pro3) -- (pro4);
\draw [arrow] (pro4) -- (pro5);
\draw [arrow] (pro5) -- (pro6);
\end{tikzpicture}}
\caption{Flowchart representing the parameter hierarchy in the sampling process.}
\label{fig:catalog_flowchart}
\end{figure}
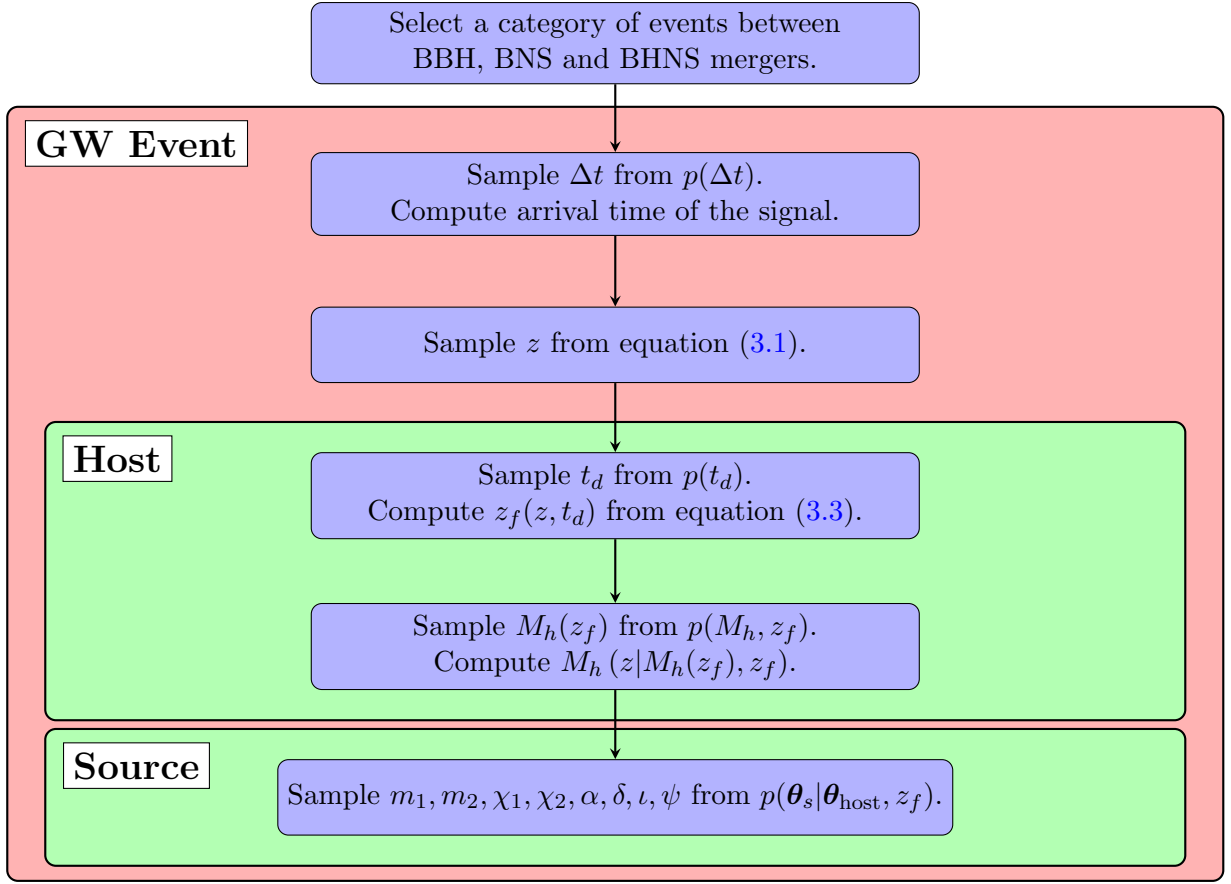

Our sampling strategy is schematically represented by the flowchart in figure~\ref{fig:catalog_flowchart}. It can be summarized as follows: first we decide which event we are interested in between BBH, BNS or BHNS mergers, then we sample the arrival time of the signal to Earth, and the merger redshift, which we choose as the first parameter in our parameter hierarchy. From the observed merger redshift we sample first the host properties both at the binary formation and merger times, and second the source properties at the binary formation time. The sampling of the source parameters follows the procedure outlined in refs.~\cite{regimbau:gwmockcatalogI, regimbau:gwmockcatalogII, regimbau:gwmockcatalogIII} whereas the sampling of the host parameters represents an important novelty of this work.

The first piece of information we need is the redshift $z$ of the merger event, or, equivalently, its conformal distance~$\chi$ from us. This is not a ``pure'' astrophysical parameter, as can be deduced also from the special role the redshift plays in Section~\ref{sec:aGWB_theory}, however it requires a similar treatment for the purpose of generating a catalog of GW events. The redshift of the merger for the $i$-th source is drawn from the observed merger rate pdf
\begin{equation}
\mathcal{P}^{[i]}(z) = \frac{\bar{R}^{[i]}(z)}{1+z} \frac{dV}{dz} \Bigg/ \int dz \frac{\bar{R}^{[i]}(z)}{1+z} \frac{dV}{dz},
\label{eq:observed_merger_rate_pdf}
\end{equation}
where~$dV/dz = 4\pi c \chi^2(z)/H(z)$ is the comoving volume element. The intrinsic merger rate density reads 
\begin{equation}
\bar{R}^{[i]}(z) = \mathcal{A}_1 \int_{t_{d,\mathrm{min}}}^{t(z)} dt_d\ p(t_d) R^{[i]}_f(z_f),
\label{eq:intrinsic_merger_rate_density}
\end{equation}
where~$R_f$ is the binary formation rate density, $z_f$ is the binary formation redshift,\footnote{The concept of ``binary formation'' might be too idealised to describe the real evolution of a binary, which can involve encounters that eject one of the progenitors and acquire a third body. However we do not enter in these details and we assume that once the binary is formed its evolution is not affected by encounter with other bodies or, more in general, by other phenomena.}
\begin{equation}
t_d = - \int_{z_f}^{z} d\tilde{z} \frac{1}{(1+\tilde{z})H(\tilde{z})}
\label{eq:delay_time}
\end{equation}
is the delay time and $p(t_d)$ its pdf, $t_{d,\mathrm{min}}$ is the minimum time delay,
\begin{equation}
t(z)=-\int_{\infty}^{z} d\tilde{z} \frac{1}{(1+\tilde{z})H(\tilde{z})}
\end{equation}
is the age of the Universe at redshift~$z$, and $\mathcal{A}_1 \leq 1$ is a factor that accounts for the fact that not all binaries successfully merge in less than a Hubble time. The binary formation rate density typically tracks the star formation rate density~$R_\star$,\footnote{Binaries of compact objects closely track the star formation rate in the case the compact objects already originated in the binary itself, i.e., if they are field binaries. In the other case, where the binary formed dynamically, this assumption might be less robust. The recent LIGO-Virgo O3 data release showed for the first time that both formation mechanisms contribute to the events we observed~\cite{abbott:O3eventsproperties}. However, for the sake of simplicity, we will assume this approximation holds and that differences in the formation mechanisms can be accounted by changing the normalization factor.} hence we can approximate~$R_f(z) = \mathcal{A}_2 R_\star(z)$ , where~$\mathcal{A}_2 \leq 1$ is a second factor that accounts for the fact that not all stars end up in binaries. Strictly speaking the exact value of~$\mathcal{A}_1$ and~$\mathcal{A}_2$ depends on the underlying stellar evolution model, see e.g., ref.~\cite{marassi:ensembleaverage}. However, given the large uncertainty on these factors, we choose to consider them as constants. Moreover, instead of fixing them individually, we collect these constants into a single one that is fixed by matching the local merger rate density~$R^{[i]}(0)$ to the one measured by the LIGO/Virgo/KAGRA collaboration.

A second quantity which is not a real astrophysical parameter is the arrival time of the signal. In the context of this work, ``arrival time'' refers to the arrival time at the Earth center, as we explain in more detail in
section~\ref{subsec:detector}. Supposing that the arrival of the signals at the Earth is a Poisson process, the pdf of time intervals~$\Delta t$ between two different events follows an exponential probability function~$p^{[i]}(\Delta t) = (\overline{\Delta t^{[i]}})^{-1} e^{-\Delta t/\overline{\Delta t^{[i]}}}$, where~$\overline{\Delta t^{[i]}}$ is the average waiting time between signals and is given by the inverse of the total coalescence rate:
\begin{equation}
\overline{\Delta t^{[i]}} = \left[ \int dz \frac{\bar{R}^{[i]}(z)}{1+z} \frac{dV}{dz} \right]^{-1}.
\label{eq:average_waiting_time}
\end{equation}
Fixing an initial time for the observations~$t_\mathrm{obs}^\mathrm{ini}$, the events will arrive at the Earth at time~$t_1 = t_\mathrm{obs}^\mathrm{ini} + \Delta t_1$, $t_2 = t_1 + \Delta t_2$ and so on, with every time interval~$\Delta t_j$ drawn from the same exponential pdf. The time of arrival defines the relative orientation between the source and the detector, so we need a consistent way to link the Earth orientation to the time. Our choice in this paper is to consider the Greenwich Median Sidereal Time (GMST hereafter). Note from the flowchart in figure~\ref{fig:catalog_flowchart} that this parameter is the first one that is sampled, even though for the sake of consistency we introduced it after the redshift of the merger.

After the time of arrival and redshift of the merger, we sample the properties of the host. In this work we consider as host the dark matter halo and we choose the halo mass~$M_h$ as characterising parameter of the host. This choice describes the minimal set-up needed to consistently assign the events to the hosts, however more complex set-ups are also possible, for instance when we consider galaxies inside the halos as hosts and we characterize them not only by their mass but also by their metallicity~\cite{Dvorkin:2016wac}. The procedure we present is sufficiently flexible to be applied also to these more complex set-ups, however for the sake of simplicity we choose this minimal approach. 

In our framework what governs the merger rate is ultimately the cosmic star formation rate density, which explicitly reads as 
\begin{equation}
R_\star(z) = \int dM_h \left\langle \mathrm{SFR}(M_h,z)\right\rangle \frac{dn_h}{dM_h},
\label{eq:cosmic_sfr_density_per_halo}
\end{equation}
where~$\left\langle\mathrm{SFR}(M_h,z)\right\rangle$ is average halo star formation rate per halo and~$dn_h/dM_h$ is the comoving halo number density. Therefore, first we sample the delay time pdf and we infer the redshift of the binary formation~$z_f$ from equation~\eqref{eq:delay_time}, then we establish the host mass at the time of the binary formation~$M_h(z_f)$ by sampling the pdf $\displaystyle p(M_h,z_f)\propto \left\langle \mathrm{SFR}(M_h,z_f)\right\rangle \frac{dn_h}{dM_h}(M_h,z_f)$. Finally, we compute the mass of the host at the time of the merger~$M_h(z)$ by considering the average growth of dark matter halos.\footnote{Alternatively one can sample the halo mass at merger redshift using as pdf the halo conditional mass function~\cite{lacey:hierarchicalmergersI, lacey:hierarchicalmergersII, zentner:excursionsettheory}, which gives the probability that a halo at redshift~$z$ has mass~$M_h$ knowing that it had a mass~$M_h(z_f)$ at redshift~$z_f$.} In summary, the set of our host properties is given by~$\bm{\theta}_\mathrm{host}=\left\lbrace t_d, z_f, M_h(z_f), M_h(z) \right\rbrace$.

Regarding the astrophysical parameters of the source~$\bm{\theta}_s$, we choose to work with time- and host-independent pdfs, however we note that if we include the metallicity in the set of parameters describing the host, we should consider also its effect on the masses of the progenitors. We leave this aspect of the modelling of the source for future work, however we stress that our methodology can easily take into account this extra layer of complexity. Moreover, despite considering different type of GW source, we work within a reference model where all the sources share the same set of parameters, listed here below.
\begin{itemize}
\item \textbf{Compact object mass}. The masses of the compact objects are indicated by~$m_1$ and~$m_2$, where~$m_2 \leq m_1$ by convention. These compact objects are the result of the evolution of massive stars, therefore the pdfs of their masses depends on assumptions on the initial stellar mass function and on stellar evolution.

\item \textbf{Compact object spin}. The magnitude of the compact object (dimensionless) spins are indicated by~$\chi_1$ and~$\chi_2$, where $\chi_j = c S_j/(G m^2_j)$ and~$S_j$ is the BH spin magnitude. As for the compact object masses, the pdfs of their spins depends on assumptions on the initial stellar mass function and on stellar evolution. The spin alignment with respect to the angular momentum might be a parameter of interest~\cite{abbott:O12eventsproperties, abbott:O3eventsproperties}, however we will not consider it in the rest of this work.

\item \textbf{Sky localization}. The two angles that define the sky localization of the event, along with the merger redshift, describe the 3D localization of the event. We choose as reference frame the geocentric equatorial coordinate system\footnote{This right-handed frame has its origin at the Earth center and its~$\hat{x}$ and~$\hat{z}$ axes pointing towards the vernal point and the North Pole, respectively.} and as sky coordinates the right ascension and declination~$(\alpha,\delta)$.

\item \textbf{Inclination}. Depending on the convention, the inclination angle~$\iota$ quantifies the misalignment between the binary total and/or orbital angular momentum with respect to the line-of-sight.

\item \textbf{Polarization}. The polarization angle~$\psi$ quantifies the alignment between the axes of the GW frame and those of the detector frame.
\end{itemize}


\subsection{Reference frames}
\label{subsec:reference_frames}

\begin{figure}[t]
\centerline{
\includegraphics[width=0.9\columnwidth]{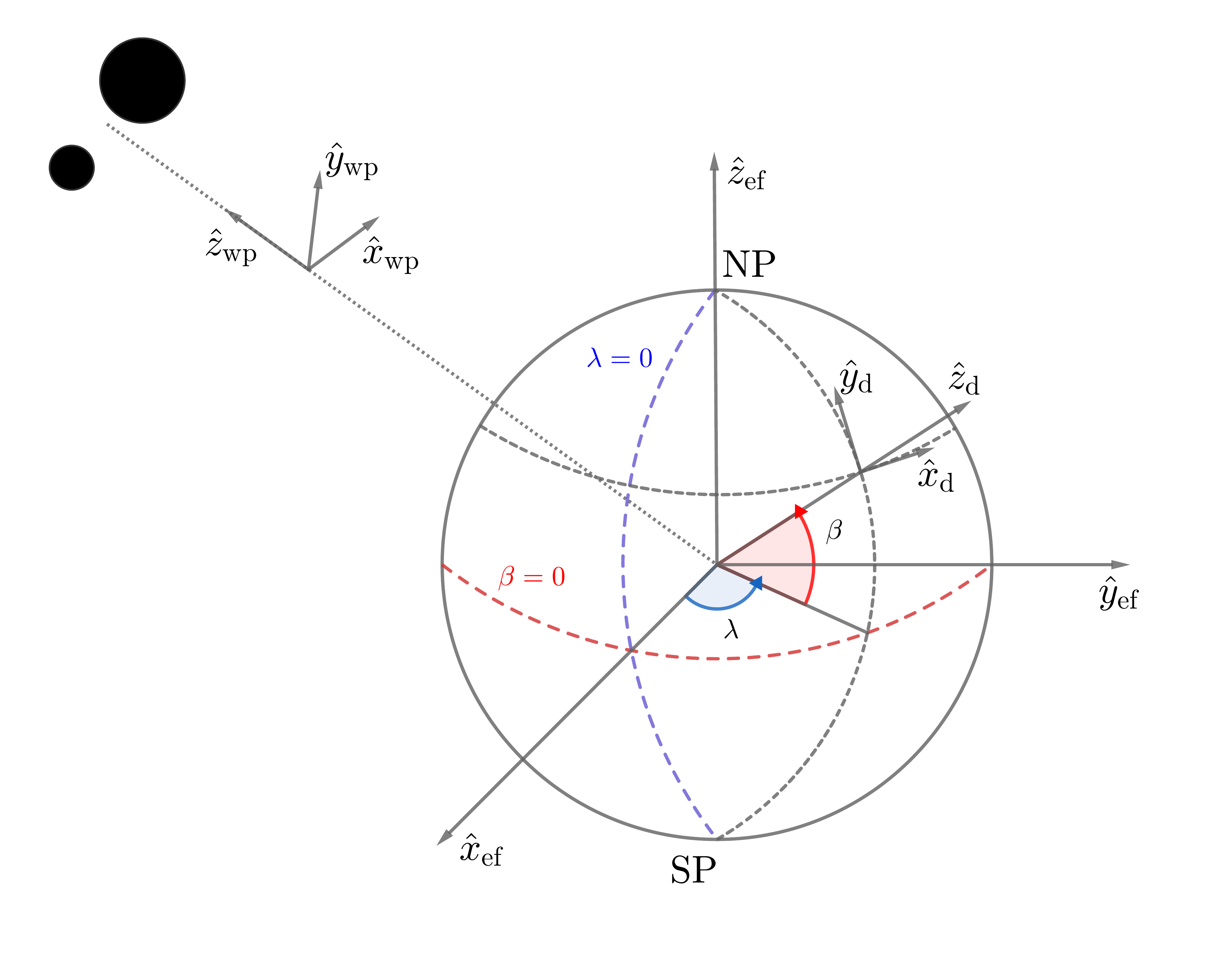}}
\caption{Sketch of the source/GW-Earth-detector system, along with the three right-handed, orthonormal systems of coordinates needed to model the detector response to a passing GW: the wave propagation ($_\mathrm{wp}$), the Earth fixed ($_\mathrm{ef}$) and the detector ($_\mathrm{d}$) frames. The North and South pole are indicated by NP and SP, respectively. The blue and red dashed lines represent the prime meridian and the equator, respectively. The detector is situated at a longitude~$\lambda$ and latitude~$\beta$.}
\label{fig:reference_frame_sketch}
\end{figure}

The second step involves the modelling of the GW signal and of the detectors. One of the key ingredients needed to compute the GW detectability is the knowledge of the relative position of the source with respect to the detector. Therefore, we have to link the position of the source with the position of the detector on the surface of the Earth when the GW arrives. This connection can be achieved by considering three different right-handed, orthonormal coordinate systems: the \textit{wave propagation}, the \textit{Earth fixed} and the \textit{detector} frames, indicated by the subscripts~$_\mathrm{wp},\ _\mathrm{ef},\ _\mathrm{d}$, respectively. We show in figure~\ref{fig:reference_frame_sketch} a sketch of the relative orientation of the three reference frames. The main ideas and results used to move from one coordinate system to the other are summarised in appendix~\ref{app:rotating_coordinate_systems}.

The wave propagation frame has origin on the line connecting the source to the Earth centre, $\hat{z}_\mathrm{wp}$ axis pointing towards the source, and~$\hat{x}_\mathrm{wp}, \hat{y}_\mathrm{wp}$ chosen in order to conveniently write the GW tensor~$H(t)$ in this frame as
\begin{equation}
H_\mathrm{wp}(t) = 
\begin{pmatrix}
h_+(t)      & h_\times(t) & 0 \\
h_\times(t) & -h_+(t)     & 0 \\
0           & 0           & 0 \\
\end{pmatrix}
= h_+(t) e_{+,\mathrm{wp}} + h_\times(t) e_{\times,\mathrm{wp}},
\end{equation}
where $h_+,h_\times$ are the time-dependent amplitudes of the two GW polarization states $(+,\times)$,
\begin{equation}
e_{+,\mathrm{wp}} = \hat{x}_\mathrm{wp}\otimes\hat{x}_\mathrm{wp} - \hat{y}_\mathrm{wp}\otimes\hat{y}_\mathrm{wp} = 
\begin{pmatrix}
1 & 0  & 0 \\
0 & -1 & 0 \\
0 & 0  & 0 \\
\end{pmatrix}, \
e_{\times,\mathrm{wp}} = \hat{x}_\mathrm{wp}\otimes\hat{y}_\mathrm{wp} + \hat{y}_\mathrm{wp}\otimes\hat{x}_\mathrm{wp} = 
\begin{pmatrix}
0 & 1 & 0 \\
1 & 0 & 0 \\
0 & 0 & 0 \\
\end{pmatrix}
\end{equation}
are the polarization tensors and~$\otimes$ represents the outer product.\footnote{Given two column vectors~$\mathbf{a}$ and~$\mathbf{b}$ in~$\mathbb{R}^3$, their outer product is given by~$\mathbf{a} \otimes \mathbf{b} = \mathbf{a}\mathbf{b}^T = 
\begin{pmatrix}
a_1b_1 & a_1b_2 & a_1b_3 \\
a_2b_1 & a_2b_2 & a_2b_3 \\
a_3b_1 & a_3b_2 & a_3b_3 \\
\end{pmatrix}$, where~$^T$ denotes the transpose operator.}

The Earth fixed frame has origin in the Earth centre, but its~$\hat{x}_\mathrm{ef}, \hat{y}_\mathrm{ef}, \hat{z}_\mathrm{ef}$ axes point towards the points~$(\beta,\lambda)=(0^o,0^o), (0^o,90^oE), (90^oN, 0^o)$, respectively, where~$\beta$ is the latitude and~$\lambda$ is the longitude.\footnote{In this work we consider~$\beta\in[-90^oN,90^oN]$ and~$\lambda\in[-180^oE,180^oE]$. The latitude increases going towards the North Pole; the longitude increases going towards East. The equator and the prime meridian are defined as the lines with~$\beta=0^oN$ and~$\lambda=0^oE$, respectively.} This coordinate system rotates with Earth, hence its relative orientation with respect to the wave propagation frame depends on time.

The definition of the detector frame assumes our detectors are L-shaped interferometers and that the Earth is described by the WGS-84 oblate ellipsoidal Earth model, hence hereafter the term latitude will refer to the geodetic latitude.\footnote{Note that the definition of the Earth fixed frame does not change if we use the geocentric latitude instead of the geodetic one.} The origin of this frame is the corner of the interferometer. The~$\hat{x}_\mathrm{d}, \hat{y}_\mathrm{d}$ axes live in a plane that contains the interferometer corner and it is parallel to the plane that \textit{(i)} is tangent to the surface of the ellipsoid and \textit{(ii)} passes by the point on the ellipsoidal that is the closest to the interferometer corner. The~$\hat{x}_\mathrm{d}, \hat{y}_\mathrm{d}$ axes point towards East and North, respectively. The~$\hat{z}_\mathrm{d}$ axis points outwards with respect to the Earth surface.

The total difference in the relative length change of the two arms of a L-shaped interferometric detector is given by the superposition of the (possible) GW signal~$h(t)$ and the detector noise~$n(t)$. The orientation of the detector arms is described by two unit vectors~$\hat{\mathbf{n}}_1$ and~$\hat{\mathbf{n}}_2$. The GW signal in such a detector reads as
\begin{equation}
h(t) = \frac{1}{2} \hat{\mathbf{n}}^T_{1,\mathrm{d}} H_\mathrm{d}(t) \hat{\mathbf{n}}_{1,\mathrm{d}} - \frac{1}{2} \hat{\mathbf{n}}^T_{2,\mathrm{d}} H_\mathrm{d}(t) \hat{\mathbf{n}}_{2,\mathrm{d}},
\label{eq:detected_strain_definition}
\end{equation} 
where~$\hat{\mathbf{n}}_{1,\mathrm{d}}, \hat{\mathbf{n}}_{2,\mathrm{d}}$ are the arms orientation unit vectors (chosen such that~$\hat{\mathbf{n}}_{1,\mathrm{d}} \times \hat{\mathbf{n}}_{2,\mathrm{d}}$ points outwards from the Earth surface) and~$H_\mathrm{d}(t)$ is the GW tensor in the detector frame. The latter can be written as~$H_\mathrm{d} = R_\mathrm{wp-d} H_\mathrm{wp} R^T_\mathrm{wp-d}$, where~$R_\mathrm{wp-d}$ is a time-dependent orthogonal matrix describing a rotation from the wave propagation to the detector frame and~$^T$ represents the transpose operator. Alternatively, equation~\eqref{eq:detected_strain_definition} is often written as~$h(t) = F_+(t) h_+(t) + F_\times(t) h_\times(t)$, where
\begin{equation}
\begin{aligned}
F_+(t) &= \frac{1}{2} \Big[\hat{\mathbf{n}}^T_{1,\mathrm{d}} R_\mathrm{wp-d} e_{+,\mathrm{wp}} R^T_\mathrm{wp-d} \hat{\mathbf{n}}_{1,\mathrm{d}} - \hat{\mathbf{n}}^T_{2,\mathrm{d}} R_\mathrm{wp-d} e_{+,\mathrm{wp}} R^T_\mathrm{wp-d} \hat{\mathbf{n}}_{2,\mathrm{d}} \Big] \\
F_\times(t) &= \frac{1}{2} \Big[\hat{\mathbf{n}}^T_{1,\mathrm{d}} R_\mathrm{wp-d} e_{\times,\mathrm{wp}} R^T_\mathrm{wp-d} \hat{\mathbf{n}}_{1,\mathrm{d}} - \hat{\mathbf{n}}^T_{2,\mathrm{d}} R_\mathrm{wp-d} e_{\times,\mathrm{wp}} R^T_\mathrm{wp-d} \hat{\mathbf{n}}_{2,\mathrm{d}} \Big]
\end{aligned}
\label{eq:f+fx_definition}
\end{equation}
are called antenna response functions. The explicit derivation of the antenna response function for the detector network we are interested in is presented in section~\ref{subsec:detector}.

Even though the characterisation of these reference frames is partially based on the assumption of having L-shaped interferometers, the logic behind is not. Therefore, the same procedure can be applied to other kind of interferometers with different arms configurations, as for the cases of LISA or ET.


\subsection{Gravitational wave waveform}
\label{subsec:waveform}

For our analysis, it is more convenient to work in the frequency domain, hence we need to Fourier transform the GW signal~$h(t)$. As we saw in the previous section, the antenna response is time-dependent because of the rotation of the Earth. However, since we are interested in a class of signals that last at most few seconds, we can consider the antenna response functions as time-independent, since the time-scale of the antenna pattern variation is considerably larger than the duration of the signal.\footnote{Note that similar considerations might not apply for other detectors in other frequency bands, hence this approximation can be used only in certain cases.} The Fourier transform of equation~\eqref{eq:detected_strain_definition} is given by~$h(f_o) = F_+ h_+(f_o) + F_\times h_\times(f_o)$, where the GW polarization state amplitudes in Fourier space read as
\begin{equation}
h_+(f_o) = \frac{1+\cos^2\iota}{2}\ \tilde{h}(f_o), \qquad h_\times(f_o) = \cos\iota\ \tilde{h}(f_o).
\label{eq:inclination_angle_dependence}
\end{equation}
Following ref.~\cite{ajith:gwswaveform}, $\tilde{h}(f_o)$ can be written in terms of an amplitude~$A(f_o)$ and a phase~$\Xi(f_o)$ as
\begin{equation}
\tilde{h}(f_o) = A(f_o) e^{-i\Xi(f_o)},
\label{eq:general_waveform}
\end{equation}
where both the amplitude and the phase depends on the properties of the binary. The authors of ref.~\cite{ajith:gwswaveform} provide fitting formulas for both the amplitude and the phase, however in this work we need only the former. Strictly speaking, the waveform template of~\citep{ajith:gwswaveform} is valid only for BBH coalescence events. However we use the inspiralling fitting formula of this template to model also the inspiralling phase of BHNS and BNS events, whereas the merging and ringdown phases of BHNS and BNS events are neglected since they fall almost outside the LIGO-Virgo frequency band. Moreover, fitting formulas for those stages have to include non-negligible effects due to NS deformability, whereas for the inspiralling part this is not the case since the compact objects are sufficiently far apart.

Given a binary system of compact objects as the one described in section~\ref{subsec:events_catalog}, we can define a set of secondary properties of the system, as the total mass~$M$, reduced mass~$\mu$, symmetric mass ratio~$\eta$, mass difference~$\delta$ and chirp mass~$M_c$ as
\begin{equation}
M=m_1+m_2, \quad \mu=\frac{m_1m_2}{m_1+m_2}, \quad \eta=\frac{\mu}{M}, \quad \delta=\frac{m_1-m_2}{M}, \quad M_c=\mu^{3/5}M^{2/5}.
\end{equation}
In addition, we also define a single spin parameter given by
\begin{equation}
\chi = \frac{1+\delta}{2} \chi_1 + \frac{1-\delta}{2} \chi_2.
\end{equation}

The template for the waveform covers the three stages of the coalescence: inspiralling $(f_o<f_{1,o})$, merging $(f_{1,o}<f_o<f_{2,o})$ and ringdown $(f_{2,o}<f_o<f_{3,o})$. It reads 
\begin{equation}
A(f_o) = C_o f^{-7/6}_{1,o} \left\lbrace
\begin{aligned}
& \left(\frac{f_o}{f_{1,o}}\right)^{-7/6} \left(1 + \alpha_2 v^2 + \alpha_3 v^3 \right) \quad & f_o<f_{1,o}, \\
& \omega_m \left(\frac{f_o}{f_{1,o}}\right)^{-2/3} \left(1 + \epsilon_1 v + \epsilon_2 v^2 \right) \quad & f_{1,o}<f_o<f_{2,o}, \\
& \omega_{r,o} \frac{1}{2\pi} \frac{\sigma_o}{(f_o-f_{2,o})^2+\sigma^2_o/4} \quad & f_{2,o}<f_o<f_{3,o},
\end{aligned} \right.
\label{eq:waveform_template}
\end{equation} 
where~$f_{3,o}$ is just a cut-off frequency and, in the test-mass limit,\footnote{The test-mass limit corresponds to take the limit~$\eta\to 0$.} $f_{1,o}$, $f_{2,o}$ and~$\sigma_o$ reduce to the observed frequency of the last stable orbit~$f_{\mathrm{LSO},o}$, the observed dominant quasi-normal mode~$f_{\mathrm{QNM},o}$~\cite{bardeen:frequencies} and to~$f_{\mathrm{QNM},o}/Q$, where~$Q$ is the ring-down quality factor of a Kerr BH of mass~$M$ and spin~$\chi$~\cite{echeverria:qualityfactor}, respectively. These four frequencies are given by
\begin{equation}
\begin{aligned}
\frac{f_{1,o}}{f_{\mathrm{fund},o}} &= y_{(1),\mathrm{tm}} + \bm{\eta}^T Y_{(1)}\bm{\chi}, \qquad \frac{f_{2,o}}{f_{\mathrm{fund},o}} &= y_{(2),\mathrm{tm}} + \bm{\eta}^T Y_{(2)}\bm{\chi}, \\
\frac{f_{3,o}}{f_{\mathrm{fund},o}} &= y_{(3),\mathrm{tm}} + \bm{\eta}^T Y_{(3)}\bm{\chi}, \qquad \frac{\sigma_{o}}{f_{\mathrm{fund},o}} &= y_{(4),\mathrm{tm}} + \bm{\eta}^T Y_{(4)}\bm{\chi}, \\
\end{aligned}
\label{eq:waveform_frequencies}
\end{equation}
where~$f_{\mathrm{fund},o}=c^{3}/\left[\pi G M(1+z)\right]$ is an observed ``fundamental'' frequency of the system, $\bm{\eta}^T=(1, \eta, \eta^2, \eta^3)$, $\bm{\chi}^T=(1, \chi, \chi^2, \chi^3)$ and the functions~$y_{(k),\mathrm{tm}}$ and the matrices~$Y_{(k)}$ are explicitly reported in appendix~\ref{app:gw_waveform}. The explicit value of the normalization constant~$C_o$, alongside with its derivation, can also be found in appendix~\ref{app:gw_waveform}. The other two normalization constants~$\omega_m,\ \omega_{r,o}$ are chosen such that equation~\eqref{eq:waveform_template} is continuous in~$f_{1,o}$ and~$f_{2,o}$, therefore
\begin{equation}
\omega_m = \frac{1 + \alpha_2 v^2_{(1)} + \alpha_3 v^3_{(1)}}{1 + \epsilon_1 v_{(1)} + \epsilon_2 v^2_{(1)}}, \qquad \omega_r = \frac{\pi\sigma_o}{2} \left(\frac{f_{1,o}}{f_{2,o}}\right)^{2/3} \left[1 + \epsilon_1 v_{(2)} + \epsilon_2 v^2_{(2)}\right] \omega_m,
\end{equation}
where~$v=\left( f_o / f_{\mathrm{fund},o} \right)^{1/3}$ is the Post-Newtonian expansion parameter and~$v_{(j)}=v(f_{j,o})$. Finally, the~$\alpha_j$ and~$\epsilon_j$ parameters read 
\begin{equation}
\begin{aligned}
\alpha_2 = -\frac{323}{224} + \frac{451}{168}\eta, &\qquad \alpha_3 = \left( \frac{27}{8} - \frac{11}{6}\eta \right)\chi, \\
\epsilon_1 = 1.4547\chi - 1.8897, &\qquad \epsilon_2 = -1.8153\chi + 1.6557.
\end{aligned}
\end{equation}


\subsection{Detector sensitivity}
\label{subsec:detector}

The third and final step involves establishing whether the single events are detected by the instrument of interest. Given a detector, the signal-to-noise ratio (SNR) for an event detected by matched filtering with an optimal filter is
\begin{equation}
\rho^2 = 4 \int_0^\infty df_o \frac{|h(f_o)|^2}{S_n(f_o)}. 
\end{equation}
The noise of the detector is typically assumed to be stationary and Gaussian, with zero mean. Under these circumstances, the statistical properties of the noise  are specified by the noise two-point function
\begin{equation}
\left\langle n(f_o) n(\tilde{f}_o) \right\rangle = \frac{1}{2}\delta^D(f_o-\tilde{f}_o) S_n(f_o)
\end{equation}
where~$n(f_o)$ is the Fourier transform of the noise~$n(t)$ and~$S_n(f_o)$ is the one-sided noise spectral density. If the noise in different detectors is uncorrelated, the coherent SNR for a network of detectors is the quadrature sum of the SNR in each individual detector, i.e., $\rho^2_\mathrm{net}=\sum_j \rho^2_{j}$. A GW event is ``detected'' by a GW observatory if its SNR is greater than some SNR threshold~$\rho_\mathrm{det}$ such that the false alarm rate is considered low enough. The typical threshold chosen in the single observatory case is~$\rho_\mathrm{det}=8$, however for a network of detector the threshold value is raised to~$\rho_\mathrm{det}=12$~\cite{kelley:snrnetwork}, which approximately corresponds to a SNR of~$8$ in at least two detectors. For the case of the astrophysical GWB we can distinguish between resolved and detected events. In principle we can discriminate between resolved and unresolved events by introducing a second SNR threshold~$\rho_\mathrm{res}\leq \rho_\mathrm{det}$ that defines at which level we are able to identify a GW signal in the time series\footnote{After identifying the undetected but resolved signal, i.e., a signal with SNR~$\rho_\mathrm{res}<\rho<\rho_\mathrm{det}$, we have to decide what to do with that part of the time series. Subtracting a signal with not well determined parameters introduces systematics in the data analysis~\cite{ferguson:numericalrelativity}, hence probably the best approach is to remove that portion of the data stream. Recent works~\cite{moore:waveformsystematics} also suggest that incorrect waveforms can hide New Physics signals, hence they can impact also the analysis presented in this work.}. Because of the type of sources we included in our analysis and the frequency band of interest we expect very few overlapping events, see, e.g., estimates in ref.~\cite{regimbau:overlappingsignalestimate}, hence the comparison between the SNR of the event and this threshold defines which event is resolved and which is not.\footnote{Computing the observed SNR of an event is non-trivial because of the effects of cosmological perturbations on GW propagation. For simplicity in this work we assume that the GW signal propagates in an unperturbed Friedmann-Lema\^itre-Robertson-Walker Universe, hence the efficiency and average efficiency function do not depend on the line-of-sight. However cosmological perturbations boost/damp the SNR by changing the GW luminosity distance by a small amount~$\delta_L$~\cite{bertacca:gwdlperturbation}. In other words, by going from an unperturbed to a perturbed Universe we observe a shift in the luminosity distance given by~$d_L \to d'_L=d_L(1+\delta_L)$, which induces a shift in the observed SNR given by~$\rho\to\rho'=\rho(1+\delta_\rho)$. The statistical properties of this SNR shift depends on those of the shift in luminosity distance since $\displaystyle \delta_\rho = \frac{d\log\rho}{d\log d_L}\delta_L = -\delta_L $. Therefore~$\delta_\rho$ has zero mean and a non-zero variance~$\left\langle\delta_\rho\delta_\rho\right\rangle = \left\langle\delta_L\delta_L\right\rangle$, which depends on cosmological perturbations and can be computed using a fiducial cosmological model. Moreover certain GW events can be strongly lensed by the intervening large scale structure, even though these events are expected to be very rare and to not contribute significantly to the astrophysical GWB~\cite{capurri:astrophysicalGWB}. For the sake of simplicity in this work we establish the detectability of GW events assuming a unperturbed Universe, however we note that in principle it is possible to obtain a catalog of resolved/unresolved events using the statistical properties of the~$\delta_L$ fluctuations to correct the catalog obtained in the unperturbed Universe.}

\begin{figure}[ht]
\centerline{
\subfloat[][]{\includegraphics[width=0.5\columnwidth]{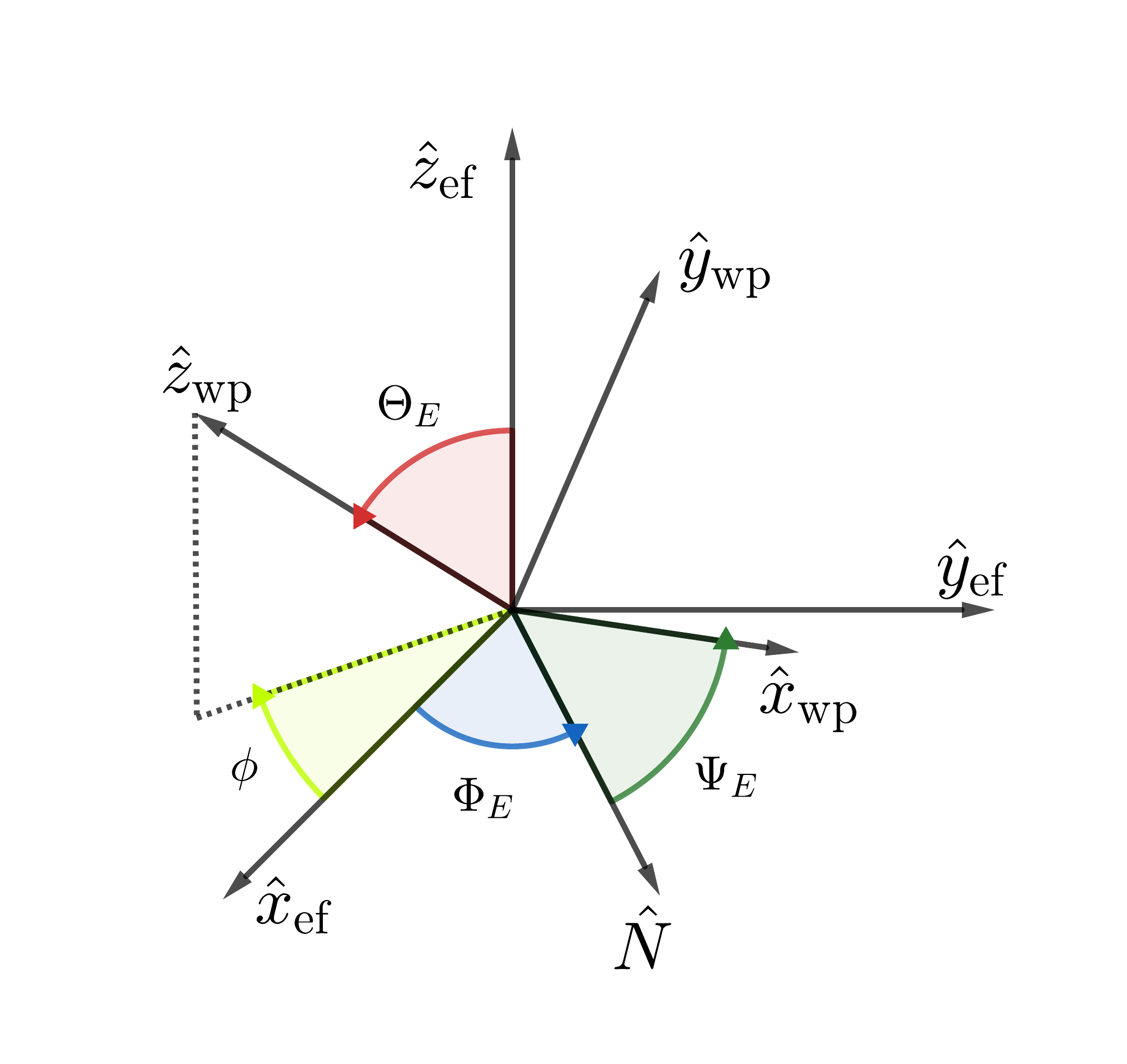}}
\subfloat[][]{\includegraphics[width=0.5\columnwidth]{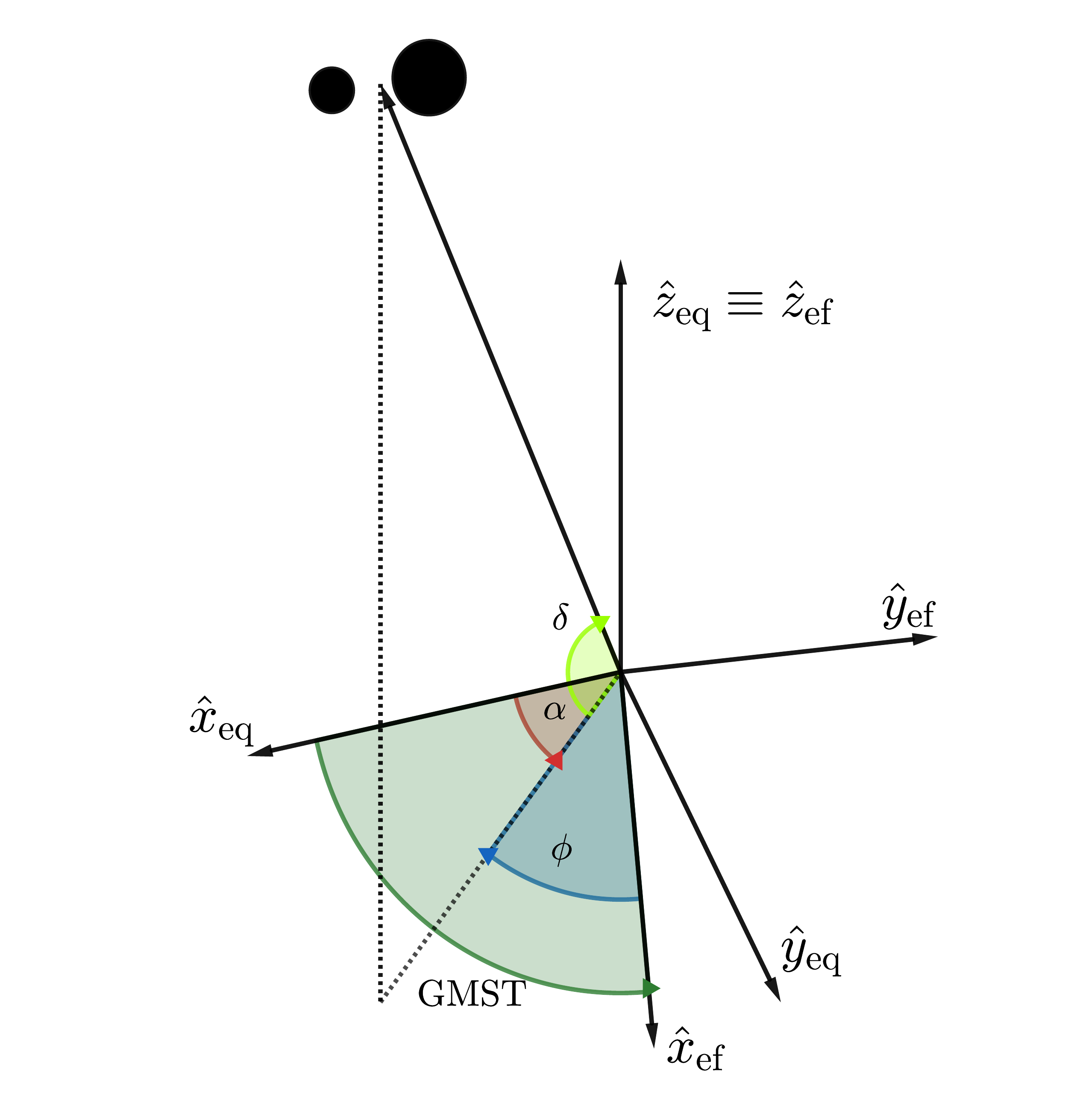}}}
\centerline{
\subfloat[][]{\includegraphics[width=0.5\columnwidth]{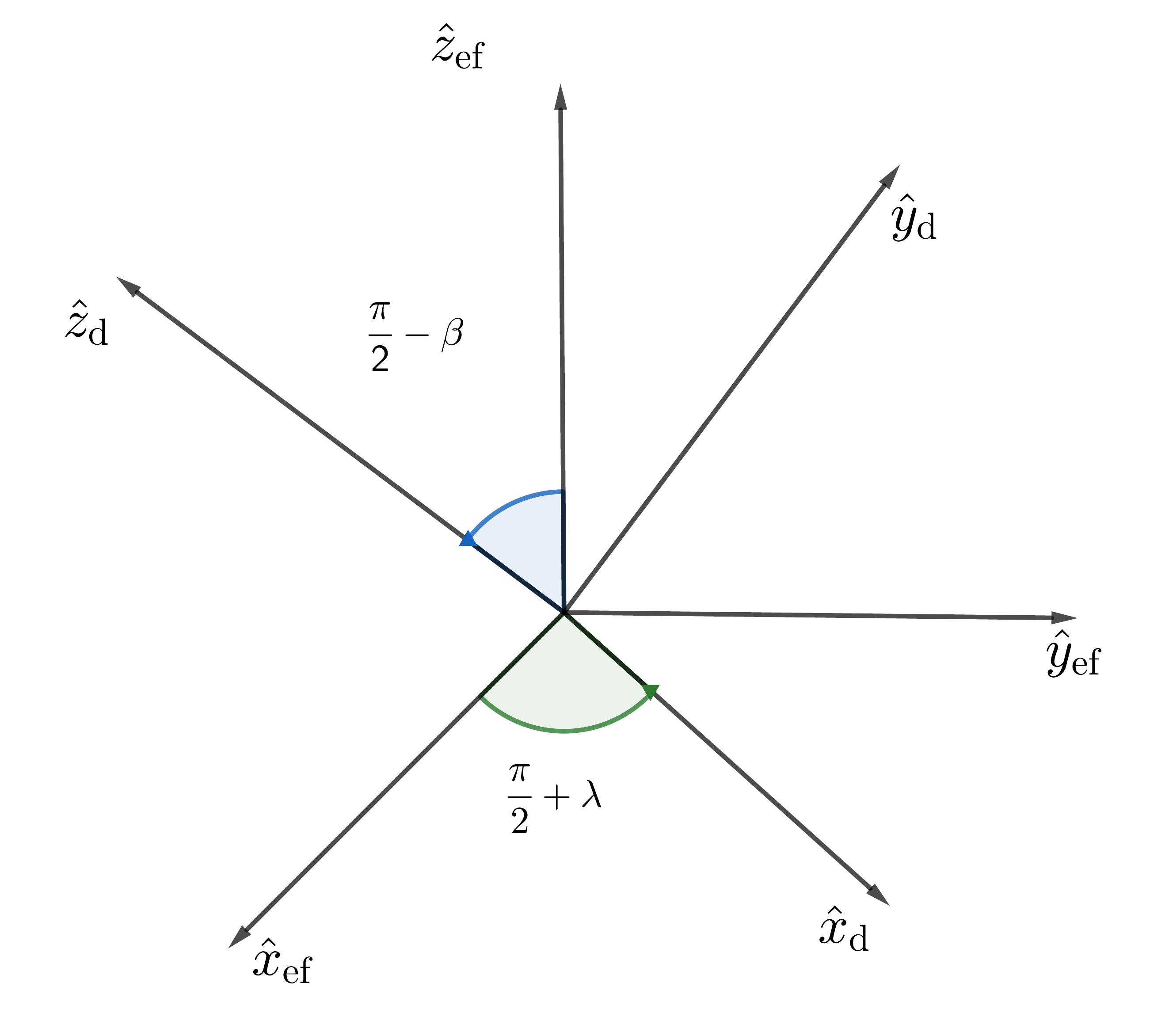}}
\subfloat[][]{\includegraphics[width=0.5\columnwidth]{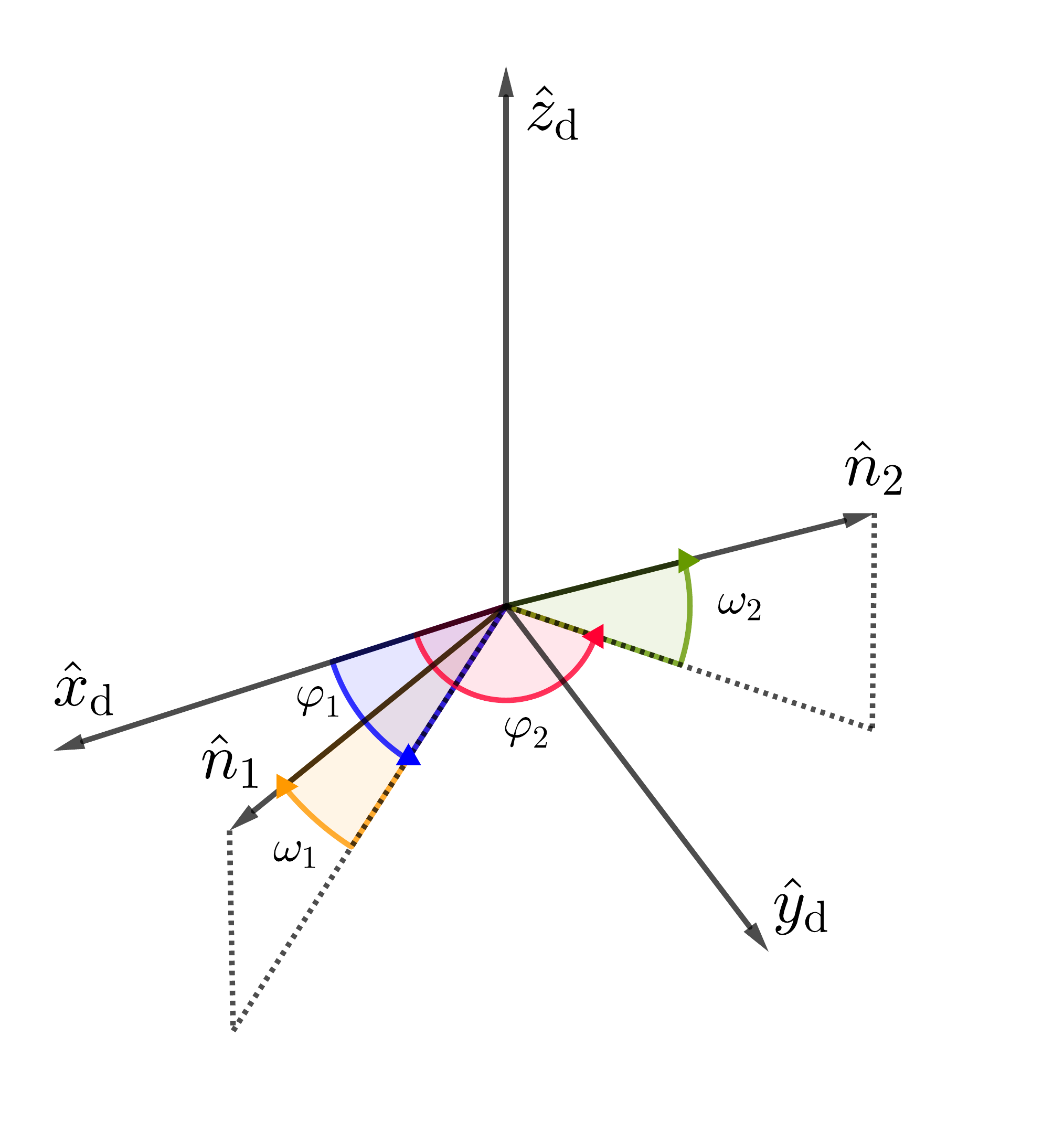}}}
\caption{The relative orientation of different reference frames used to describe the incoming wave and the detector. Panel (a): wave propagation and the Earth fixed frames. Panel (b): equatorial and the Earth fixed frames. Panel (c): Earth fixed and detector frames. Panel (d): arms orientations in the detector frame.}
\label{fig:reference_frames}
\end{figure}

Finally, we explicitly characterize the antenna response functions in equation~\eqref{eq:f+fx_definition} by describing the rotation matrix~$R_\mathrm{wp-d}$. Since rotations are a group, we can separate the problem into two stages, going first from the wave propagation to the Earth fixed frame and then from the Earth fixed to the detector frame, following the procedure described in ref.~\cite{anderson:antennaresponse}.

For practical purposes, it is more convenient to carry out the first rotation in the opposite direction, going form the Earth fixed to the wave propagation frame. The relative orientation of the two coordinate systems can be visualised in panel (a) of figure~\ref{fig:reference_frames}, alongside with the three Euler angles~$\Phi_E$, $\Theta_E$ and~$\Psi_E$ we use to describe the rotation from one system to the other. Without loss of generality, we can introduce the line of nodes vector defined as~$\hat{N} = \hat{z}_\mathrm{ef}\times\hat{z}_\mathrm{wp}$. Therefore the three Euler angles are~$\Phi_E \in [0,2\pi]$ (the angle between~$\hat{x}_\mathrm{ef}$ and~$\hat{N}$, measured counterclockwise), $\Theta_E \in [0,\pi]$ (the angle between $\hat{z}_\mathrm{ef}$ and~$\hat{z}_\mathrm{wp}$) and $\Psi_E \in [0,2\pi]$ (the angle between~$\hat{N}$ and~$\hat{x}_\mathrm{d}$, measured counterclockwise). The rotation matrix connecting these two coordinate systems is built using the fundamental rotational matrices of equation~\eqref{eq:fundamental_rotation_matrices} as
\begin{equation}
R_\mathrm{ef-wp} = R^{\hat{z}_\mathrm{d}}(\Psi_E) R^{\hat{x}_\mathrm{d}}(\Theta_E) R^{\hat{z}_\mathrm{d}}(\Phi_E).
\label{eq:rotation_matrix_wp_ef}
\end{equation}

Then we have to identify the Euler angles with the angles introduced in section~\ref{subsec:events_catalog}, namely right ascension, declination and polarization. First of all we notice that~$\Psi_E$ already encodes the information about the polarization, hence we can directly take it to be the polarization angle, i.e., $\Psi_E=\psi$. The right ascension~$\alpha$ and declination~$\delta$ angles are defined in the equatorial coordinate system (indicated by the~$_\mathrm{eq}$ subscript), rather than in the Earth fixed frame. The equatorial coordinate system has origin in the Earth center, $\hat{z}_\mathrm{eq}$ pointing toward the North Pole, $\hat{x}_\mathrm{eq}$ axis pointing towards the vernal point and~$\hat{y}_\mathrm{eq}$ conveniently chosen to have a right-handed orthonormal basis. This coordinate system does not rotate with Earth, however it shares the same $\hat{z}$-axis and~$\hat{x}-\hat{y}$ plane with the Earth fixed frame, as can be seen in panel~(b) of figure~\ref{fig:reference_frames}. The angle~$\Theta_E$ can be written in terms of the declination angle as~$\Theta_E = \frac{\pi}{2}-\delta$ (recall that~$\Theta_E$ is a positive angle measured from the~$\hat{z}_\mathrm{ef}\equiv\hat{z}_\mathrm{eq}$ axis, while~$\delta\in[-\pi/2,\pi/2]$ is measured from the~$\hat{x}_\mathrm{eq}-\hat{y}_\mathrm{eq}$ plane). The angle~$\Phi_E$ can be connected to the right ascension of the source in two steps. In the first one we note that the $\hat{z}_\mathrm{wp}$ forms a angle~$\phi$ when projected onto the~$\hat{x}_\mathrm{ef}-\hat{y}_\mathrm{ef}$ plane, hence~$\Phi_E = \phi - \frac{\pi}{2}$ since~$\Phi_E$ and~$\phi$ are measured counterclockwise and clockwise from~$\hat{x}_\mathrm{ef}$, respectively (\textit{cf.} panel~(a) of figure~\ref{fig:reference_frames}). In the second one we use the fact that the right ascension of the object is given by~$\alpha = \mathrm{GMST} + \phi$, where~$\mathrm{GMST}$ is the Greenwich Mean Sidereal Time (in radians) and describes how much the Earth fixed frame rotated with respect to when the frame was aligned with the equatorial coordinate system.\footnote{The GMST in radians is given by~$\mathrm{GMST} = \frac{2\pi}{86400}\left[ \mathrm{GMST}(0^\mathrm{h}) + \mathrm{GMST}_\mathrm{day} \right]$, where
\begin{equation*}
\begin{aligned}
\mathrm{GMST}(0^\mathrm{h}) &= 24110.54841 + 8640184.812866T + 0.093104T^2 - 6.2\times 10^{-6}T^3, \\
\mathrm{GMST}_\mathrm{day} &= \left(1.0027379 + 5.9006\times 10^{-11}T - 5.9\times 10^{-15}T^2\right) \times t_\mathrm{day} ,
\end{aligned}
\end{equation*}
and~$T=(\mathrm{JD}-2451545)/36525$, $\mathrm{JD}$ is the Julian date of the day at midnight and~$t_\mathrm{day}$ is the time of the day in UT1 seconds. Obviously the~$\mathrm{GMST}$ can be evaluated modulo~$2\pi$.} Therefore~$\Phi_E = \phi - \frac{\pi}{2} = \alpha - \mathrm{GMST} - \frac{\pi}{2}$. The time-dependence of the antenna response function derives from the time dependence of~$\Phi_E$; however, as previously said, we consider the antenna response function as time-independent for the duration of the GW signal passing through the detector.\footnote{We consider as  ``arrival time'' of our GW signal the arrival time at the center of Earth, neglecting that the signal arrives at slightly different time in each detector location. This assumption introduces a negligible error in the estimated longitude of the detector, which can be  estimated as~$\Delta \lambda = \omega_\oplus \Delta t \sim \omega_\oplus R_\oplus / c \sim 10^{-6}\ \mathrm{rad}$, where~$\Delta t$ is the error in the time of arrival, $\omega_\oplus \simeq 73 \times 10^{-6}\ \mathrm{rad/s}$ and~$R_\oplus \simeq 6400\ \mathrm{km}$ are the Earth angular velocity and radius, respectively. On the other hand, we are aware that the time arrival delay at different detectors in different locations can be used for other applications, as to put constraints on the propagation speed of GWs~\cite{blas:gwsspeed}.}

The relative orientation between the Earth fixed and the detector frames is shown in panel~(c) of figure~\ref{fig:reference_frames}. In particular we appreciate that we can move from the former to the latter by combining two rotations, one counterclockwise around~$\hat{z}_\mathrm{d}$ of angle~$\frac{\pi}{2} + \lambda$ and a second clockwise one around~$\hat{x}_\mathrm{d}$ of angle~$\frac{\pi}{2} - \beta$. Therefore, following the notation of appendix~\ref{app:rotating_coordinate_systems}, the total rotation matrix is given by
\begin{equation}
R_\mathrm{ef-d} = R^{\hat{x}_\mathrm{d}}\left(\frac{\pi}{2}-\beta\right) R^{\hat{z}_\mathrm{d}}\left(\frac{\pi}{2}+\lambda\right).
\end{equation}
By using the rules of rotations presented in appendix~\ref{app:rotating_coordinate_systems} we can then write the GW tensor in the detector frame as
\begin{equation}
H_\mathrm{d} = R_\mathrm{ef-d} H_\mathrm{ef} R^T_\mathrm{ef-d} = R_\mathrm{ef-d} R^T_\mathrm{ef-wp} H_\mathrm{wp} R_\mathrm{ef-wp} R^T_\mathrm{ef-d},
\end{equation}
where we can immediately identify the rotation matrix from the wave propagation to the detector frame to be~$R_\mathrm{wp-d}=R_\mathrm{ef-d} R^T_\mathrm{ef-wp}$.

To complete the characterisation of the antenna response functions given in equation~\eqref{eq:f+fx_definition}, we also need to specify the arm orientation unit vectors~$\hat{\mathbf{n}}_{1,\mathrm{d}}, \hat{\mathbf{n}}_{2,\mathrm{d}}$ in the detector frame. As can be seen in panel~(d) of figure~\ref{fig:reference_frames}, we need four angles: two angles~$(\varphi_1, \varphi_2)$ which describe the angle (measured counterclockwise) between the projection of the arms onto the~$\hat{x}_\mathrm{d}-\hat{y}_\mathrm{d}$ horizontal plane and the~$\hat{x}_\mathrm{d}$ axis; and two angles~$(\omega_1, \omega_2)$ which describe the smallest angle between the arms and the~$\hat{x}_\mathrm{d}-\hat{y}_\mathrm{d}$ horizonal plane. If we introduce a spherical coordinate system centred at the interferometer corner, $(\varphi_1, \varphi_2)$ will be the azimuthal angles of the arms while~$(\omega_1, \omega_2)$ will be the the angles complementary to the polar angle. Therefore the arm orientation vectors read as
\begin{equation}
\hat{\mathbf{n}}_{1,\mathrm{d}} = 
\begin{pmatrix}
\cos\varphi_1 \cos\omega_1 \\
\sin\varphi_1 \cos\omega_1 \\
\sin\omega_1
\end{pmatrix}, \qquad
\hat{\mathbf{n}}_{2,\mathrm{d}} = 
\begin{pmatrix}
\cos\varphi_2 \cos\omega_2 \\
\sin\varphi_2 \cos\omega_2 \\
\sin\omega_2
\end{pmatrix}.
\label{eq:detector_arms_vectors}
\end{equation}
Sometimes the angles~$\chi=(\varphi_1+\varphi_2)/2$ and~$\eta=\varphi_2-\varphi_1$ are used instead of~$(\varphi_1, \varphi_2)$, where~$\chi$ is the angle between the arm bisector projection and the~$\hat{x}_\mathrm{d}$ axis, and~$\eta$ is the arm opening. The two possibilities are completely equivalent, however in the following we use the one shown in equation~\eqref{eq:detector_arms_vectors}.


\subsection{Worked example - Part I}
\label{subsec:worked_example_Meff}

The methodology presented so far applies to a variety of possible sources and detectors, however for the sake of concreteness we present a worked example by choosing a concrete astrophysical model and detector network. In this worked example we treat different aspects of the GW source characterization and of GW propagation and detection, showing how to compute robustly the astrophysical GWB anisotropies despite the many layers of complexity of the problem. 

Our methodology presents several advantages: first of all it works well in synergy with numerical simulations, therefore, for instance, we can use results from those simulations to characterize galaxies and dark matter halos, i.e., the GW host. In other words, because of our modular approach and flexible framework, we are able to continuously improve our GWB estimates every time a new improvement in the characterization of the sources and hosts is available. A second strong point is that this technique is suitable to perform a self-consistent parameter inference, both for astrophysical and for cosmological parameters, which enter in different stages of the analysis. Finally we note that the catalog approach allows to get rid of numerous simplifying assumptions regarding both sources and detectors. In other words, our estimates are less likely to be affected by simple prescription that ultimately might completely bias the final estimate of the astrophysical GWB.

Results presented in this part of the worked example have been obtained using the~\texttt{merger\_rate}, \texttt{catalog}, \texttt{detector} and~\texttt{effective\_functions} routines described in appendix~\ref{app:external_modules}.


\subsubsection{Cosmological model}

In the rest of this work we assume a flat~$\Lambda$CDM model, leaving a self-consistent exploration of Modified Gravity scenario for future work. According to the latest Planck release~\cite{aghanim:planckcosmoparams2018}, our Universe consists of a mixture of radiation, matter and dark energy (here assumed to be a pure cosmological constant) with present-day relative energy densities given by
\begin{equation}
\Omega_{r0}=\frac{\rho_{r0}}{\rho_{0c}}=9.25\times 10^{-5},\quad \Omega_{m0}=\frac{\rho_{m0}}{\rho_{0c}}=0.3153,\quad \Omega_{\Lambda 0}=\frac{\rho_{\Lambda 0}}{\rho_{0c}}=1-\Omega_{m0}-\Omega_{r0},
\end{equation}
where the last equality ensures flatness. The matter sector is made by baryons~$(\Omega_{b0}=0.0495)$ and cold dark matter~$(\Omega_{cdm0}=0.2658)$. The Hubble expansion rate today is assumed to be~$H_0=67.36\ \mathrm{km\ s^{-1} Mpc^{-1}}$. Initial conditions for scalar perturbations have an amplitude and tilt of~$A_s=2.1\times 10^{-9}$ and~$n_s=0.9649$, respectively, at~$k_\mathrm{pivot}=0.05\ \mathrm{Mpc}^{-1}$.


\subsubsection{Merger rate and host model}
\label{subsubsec:host_model}

The model of the host is used in two different steps of our methodology. In the first one we compute the observed merger redshift pdf as outlined in equation~\eqref{eq:observed_merger_rate_pdf}, while in the second one we draw the GW host properties (along with the other GW source properties) from their respective pdfs.

Concerning the first step, we assume the merger rate density~$R^{[i]}(z)$ to be proportional to the cosmic star formation rate density~$R_\star(z)$, as discussed in section~\ref{subsec:events_catalog}. Measurements of this quantity have been provided by numerous experiments employing different techniques and cover a wide redshift range, $0<z<10$, which is the range of interest for the purpose of computing the astrophysical GWB. In this work we resort to the results of the \textsc{UniverseMachine} approach~\cite{behroozi:universemachine}, which is able to capture the connection between star formation history and properties of the host, providing a self-consistent fit to experimental data. The \textsc{UniverseMachine} approach parametrizes the galaxy star formation rate as a function of the halo potential well depth, redshift, and assembly history. This methodology is able to simultaneously match the observed stellar mass function, quenched fractions, cosmic and specific star formation rates, high-redshift UV luminosity functions, high-redshift UV-stellar mass relations, correlations functions, the dependence of the quenched fractions of central galaxies as a function of the environment and the average infrared excess as a function of the UV luminosity. 

In particular, ref.~\cite{behroozi:universemachine} provides the probability of having a given star formation rate~$\mathrm{SFR}$ in a certain halo of mass~$M_h$ at a certain redshift, i.e., $p(\mathrm{SFR}|M_h,z)$. Therefore the cosmic star formation rate density of equation~\eqref{eq:cosmic_sfr_density_per_halo} is given by
\begin{equation}
R_\star(z) := \int dM_h d\mathrm{SFR}\ p(\mathrm{SFR}|M_h,z) \times \mathrm{SFR} \times \frac{dn_h}{dM_h} = \int dM_h \left\langle \mathrm{SFR}(M_h,z)\right\rangle \frac{dn_h}{dM_h},
\label{eq:cosmicsfr_from_halosfr}
\end{equation}
where the comoving halo number density is taken from ref.~\cite{rodriguezpuebla:halomassfunction}\footnote{We use ref.~\cite{rodriguezpuebla:halomassfunction} instead of the classical ref.~\cite{tinker:halomassfunction} since the latter underestimates the number of halos at high redshift~\cite{klypin:multidark}. On the other hand at low redshift the predictions of both references agree.} and reported in appendix~\ref{app:halo_properties}. The star formation rate probability encodes the contributions of quenched and star-forming galaxies and it reads as
\begin{equation}
p(\mathrm{SFR}|M_h,z) = f_Q \mathcal{LN}\left(\log(\mathrm{SFR}_Q), \sigma_Q\right) + (1-f_Q) \mathcal{LN}\left(\log(\mathrm{SFR}_{SF}), \sigma_{SF}\right),
\label{eq:starformationrate_pdf}
\end{equation}
where the subscripts~$_Q,_{SF}$ stands for quenched and star-forming, respectively, $f_Q$ is the fraction of quenched galaxies, 
\begin{equation}
\mathcal{LN}(\mu,\sigma) = \frac{e^{-\left[\log(x)-\mu\right]^2/(2\sigma^2)} }{\sqrt{2\pi} \sigma x}
\end{equation}
indicates a lognormal distribution probability density distribution with median~$e^\mu$ and variance~$\sigma$. The explicit form of these parameters, along with their redshift and halo mass dependence, is reported in appendix~\ref{app:halo_properties}. The lower and upper boundaries of the integral in equation~\eqref{eq:cosmicsfr_from_halosfr} are~$M^\mathrm{min}_{h}=10^8\ M_\odot$ and~$M^\mathrm{max}_{h}(z)=10^{m(z)}\ M_\odot$, with~$m(z)=13.54-0.24z+2.02e^{-z/4.48}$, as used in the~\textsc{UniverseMachine} approach. Since almost exclusively star-forming galaxies contribute to the star formation rate, we have that
\begin{equation}
\left\langle \mathrm{SFR}(M_h,z)\right\rangle \simeq (1-f_Q) \times \mathrm{SFR}_{SF} \times e^{\sigma^2_{SF}/2}.
\label{eq:total_sfr_per_halo}
\end{equation}

We choose $p(t_d) \propto t^{-1}_d$ as delay time probability distribution function, with~$t_{d,\mathrm{min}} = 20\ \mathrm{Myr}$ and $t_{d,\mathrm{min}} = 50\ \mathrm{Myr}$ as minimum time delay for BNS and BBH/BHNS mergers, respectively. Since we are considering only binaries that eventually merge, the normalization of the delay time probability density distribution is redshift-dependent, since the maximum time delay~$t(z)$ depends on the cosmic time when the merger happens. Not all the binaries merge within a Hubble time since many of them have a delay time many times larger than the current age of the Universe~\cite{dominik:maximumdelaytime}. These details about the binary evolution can be encoded into the~$\mathcal{A}_j$ constants. Note also that by choosing as maximum time delay~$t(z)$ some of the events have binary formation redshift~$z_f\gg 10$, however the properties of the hosts and sources in that regime are not well understood. Since removing the events from the catalog after its generation biases the relative arrival time of events, thus the relative orientation between sources and detectors, we change the upper limit of equation~\eqref{eq:intrinsic_merger_rate_density} from~$t(z)$ to~$t(z)-t(z=10)$, selecting in this way only GW merging binaries that have formation redshift~$z_f\leq 10$.

Once we know the cosmic star formation rate density and the time delay pdf we compute the merger rate density as shown in equation~\eqref{eq:intrinsic_merger_rate_density}. The residual constants~$\mathcal{A}_j$ are fixed by matching the local merger rate density to the one measured by the LIGO-Virgo-KAGRA collaboration~\cite{abbott:O3eventsproperties} in the BBH and BHNS case, namely
\begin{equation}
R^\mathrm{BBH}(0) = 19.8\ \mathrm{Gpc^{-3}\ yr^{-1}}, \quad R^\mathrm{BNS}(0) = 320\ \mathrm{Gpc^{-3}\ yr^{-1}},
\end{equation}
whereas for BHNS mergers we choose an arbitrary value of
\begin{equation}
R^\mathrm{BHNS}(0) = 100.0\ \mathrm{Gpc^{-3}\ yr^{-1}},
\end{equation}
which is compatible with the value inferred from the two recently detected BHNS events~\cite{abbott:bhnsevents}. Because of the uncertainty in the BHNS local merger rate, in the following we provide estimates of the astrophysical GWB with and without accounting for these class of events. Note that the inferred value of the local merger rate slightly depends on the mass distribution of the compact objects, however commonly used mass distributions return values that are within the experimental uncertainty~\cite{abbott:O3eventsproperties}.

In the second step, in which we start generating the catalog, first we draw the time interval between consecutive events from the exponential pdf, where the average time interval is computed using equation~\eqref{eq:average_waiting_time} and in our reference model is~$\overline{\Delta t}=765\ \mathrm{s}, 183\ \mathrm{s}, 51\ \mathrm{s}$ for BBH, BHNS and BNS events, respectively. Second we draw the merger redshift from the merger redshift pdf computed in the first step and then we proceed to drawing the other GW host parameters. In particular the halo mass at binary formation is drawn from the pdf~$p\left(M_h(z_f)\right)\propto \left\langle \mathrm{SFR}(M_h,z_f)\right\rangle \frac{dn_h}{dM_h}(z_f)$, whereas the halo mass at merger is obtained using the evolution of the median virial mass~$M_h(M_0,z)$~\cite{behroozi:averagestarformationhistory, rodriguezpuebla:halomassfunction}, where~$M_0$ is the halo mass at~$z=0$ and it is determined by imposing that~$M_h(M_0,z_f)=M_h(z_f)$. We report the analytical formula of the halo mass growth in appendix~\ref{app:halo_properties}.


\subsubsection{Source model}

\begin{table}[t]
\centerline{
\begin{tabular}{|c|c|c|c|}
\hline
Parameter & BBH mergers & BNS mergers & BHNS mergers \\
\hline
\hline
$m_1$ & $\mathcal{BPL}(1.7,6.3,5.6,38.6,76.6)$ & $\mathcal{U}(1\ M_\odot, 2.5\ M_\odot)$ & $\mathcal{DD}(10\ M_\odot)$ \\
$m_2$ & $\mathcal{U}(5.6\ M_\odot, m_1)$ & $\mathcal{U}(1\ M_\odot, 2.5\ M_\odot)$ & $\mathcal{DD}(1.4\ M_\odot)$ \\
$\chi_1$ & $\mathcal{N}(0,0.1)$ & $\mathcal{DD}(0)$ & $\mathcal{N}(0,0.1)$ \\
$\chi_2$ & $\mathcal{N}(0,0.1)$ & $\mathcal{DD}(0)$ & $\mathcal{DD}(0)$ \\
$\alpha$ & $\mathcal{U}(0,2\pi)$ & $\mathcal{U}(0,2\pi)$ & $\mathcal{U}(0,2\pi)$ \\
$\sin\delta$ & $\mathcal{U}(-1,+1)$ & $\mathcal{U}(-1,+1)$ & $\mathcal{U}(-1,+1)$ \\
$\cos\iota$ & $\mathcal{U}(-1,+1)$ & $\mathcal{U}(-1,+1)$ & $\mathcal{U}(-1,+1)$ \\
$\psi$ & $\mathcal{U}(0,2\pi)$ & $\mathcal{U}(0,2\pi)$ & $\mathcal{U}(0,2\pi)$ \\
\hline
\end{tabular}
}
\caption{Probability density distributions for the parameters of the different reference models. The uniform distribution~$p(x)=1/(b-a)$ is labelled by $\mathcal{U}(a,b)$. The Gaussian distribution~$p(x)=e^{-(x-\mu)^2/(2\sigma^2)}/(\sqrt{2\pi}\sigma)$ is labelled by $\mathcal{N}(\mu,\sigma)$. The Dirac delta distribution~$p(x)=\delta^D(x-x_\star)$ is indicated by~$\mathcal{DD}(x_\star)$. The broken power-law distribution~$p(x)\propto x^{-\gamma_1}$ for~$x_\mathrm{min}<x<x_\mathrm{break}$ and~$p(x)\propto x^{-\gamma_2}$ for~$x_\mathrm{break}<x<x_\mathrm{max}$ is indicated by~$\mathcal{BPL}(\gamma_1, \gamma_2, x_\mathrm{min}, x_\mathrm{break}, x_\mathrm{max})$.}
\label{tab:reference_models}
\end{table}

\begin{figure}[t]
\centerline{
\includegraphics[width=0.96\columnwidth]{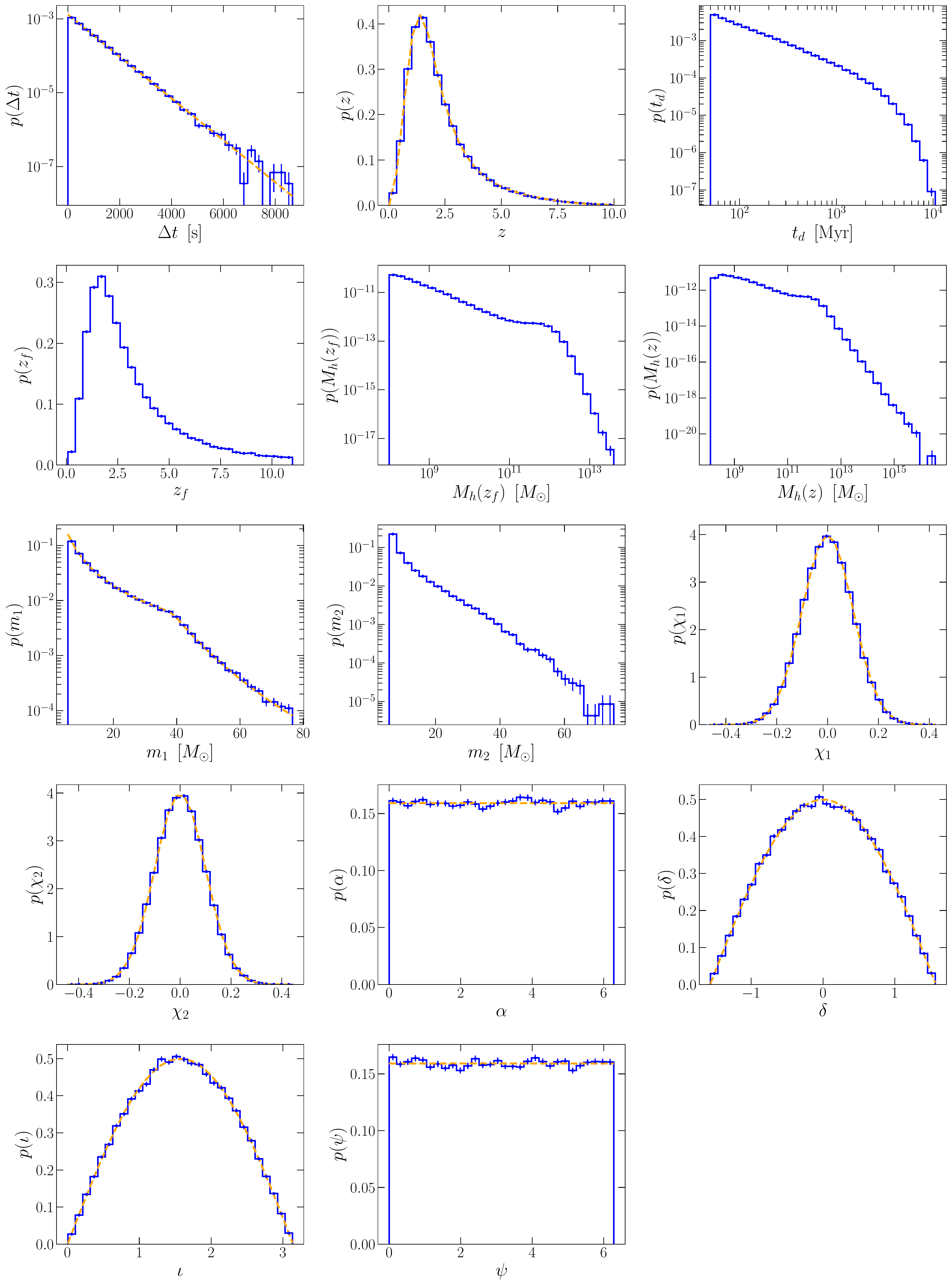}}
\caption{One dimensional pdf (with error bars) for the~$14$ parameters of our reference model. The histograms are drawn using a simulated catalog consisting of~$10^5$ BBH events. The dashed orange lines shows the theoretical pdfs whenever available.}
\label{fig:bhbh_1d_pdf}
\end{figure}

We complete the generation of the catalog by sampling the astrophysical properties of GW events. In this work we consider a reference model for GW sources, i.e., BBH, BNS and BHNS mergers, that consists of eight astrophysical parameters. Their pdfs are summarised in table~\ref{tab:reference_models} and are:

\begin{itemize}
\item \textbf{Compact object mass}. The model for the mass distributions follows the ones implemented by the LIGO-Virgo Collaboration~\cite{abbott:GWBthirdrun}. For BBH events we consider a broken power-law pdf for the primary components
\begin{equation}
p(m_1) \propto 
\left\lbrace \begin{aligned}
& m^{-\alpha_1}_1, &\qquad M_\mathrm{min}<m_1<M_\mathrm{break}, \\
& m^{\alpha_2-\alpha_1}_\mathrm{break} m^{-\alpha_2}_1, &\qquad M_\mathrm{break}<m_1<M_\mathrm{max}, \\
\end{aligned} \right.
\end{equation}
and a uniformly distributed mass~$m_2$ in the mass range~$M_\mathrm{min} \leq m_2 \leq m_1$. We take as reference values~$\alpha_1=1.7$, $\alpha_2=6.3$, $M_\mathrm{min}=5.6\ M_\odot$, $M_\mathrm{break}=38.6\ M_\odot$ and $M_\mathrm{max} = 76.6\ M_\odot$. For BNS events we consider a uniform distribution in the mass range~$1-2.5\ M_\odot$ for both compact objects. For BHNS events we take two Dirac deltas distributions centred at~$10\ M_\odot$ and~$1.4\ M_\odot$ for the BH and the NS, respectively.

\item \textbf{Compact object spin}. The BH spin magnitude is drawn from a Gaussian distribution with $0$ mean and~$0.1$ variance for both compact objects. The NS are assumed to be spinless, hence we take a Dirac delta distribution centered at~$\chi=0$ for them.

\item \textbf{Angles}. For all the events we assume that the right ascension~$\alpha$, the sine of declination~$\sin\delta$, the cosine of the inclination~$\cos \iota$ and the polarization~$\psi$ angles are drawn from a uniform pdf. 
\end{itemize}

We show in figure~\ref{fig:bhbh_1d_pdf} the host and source parameter pdfs of a catalog of~$10^5$ BBH merger events reaching the Earth in a time frame of~$T^\mathrm{BBH}_\mathrm{obs} \simeq 2.4$ years. For several parameters it is possible to directly compare the theoretical pdf with the catalog one and appreciate that they match within the Poissonian errors due to the finite number of events contained in the catalog.\footnote{Consider a set of~$N$ random variables $x_1,x_2,...,x_N$ extracted from a generic probability density distribution~$\mathrm{pdf}(x)$ and suppose to group them into~$B_{i=1,...,M}$ bins. The distribution of the number of events in a given bin follows a Poissonian statistics with constant rate~$\displaystyle N\int_{B_i}dx\ \mathrm{pdf}(x)=N_i$, where~$N_i$ is the number of events in the~$i$-th bin, therefore the error on the count of number of events per bin is expected to be~$\sqrt{N_i}$. The reasoning can be easily extrapolated to multidimensional probability distribution functions.} For other parameters the one dimensional pdf is the sum of many time-dependent one dimensional pdfs, for instance in the case of~$p(t_d)$ or~$p\left(M_h(z)\right)$, hence there is no simple analytical formula that describes it. For the sake of conciseness we do not show the pdfs for the BHNS and BNS catalogs, each one of them containing $10^5$ merger events reaching the Earth in a time frame of~$T^\mathrm{BHNS}_\mathrm{obs} \simeq 0.6$ years and~$T^\mathrm{BNS}_\mathrm{obs} \simeq 0.2$ years, respectively.


\subsubsection{LIGO-Virgo detector network}

\begin{table}[ht]
\centerline{
\begin{tabular}{|c|c|c|c|c|c|c|}
\hline
Detector        & $\beta$ & $\lambda$ & $\varphi_1$ & $\omega_1$ & $\varphi_2$ & $\omega_2$ \\
\hline
\hline
LIGO Hanford    & $0.8108$ &  $-2.0841$ & $2.1991$ & $-6.195\times 10^{-4}$ & $3.7699$ & $1.25\times 10^{-5}$ \\
\hline
LIGO Livingston & $0.5334$ & $-1.5843$ & $3.4508$ & $-3.121\times 10^{-4}$ & $5.0216$ & $-6.107\times 10^{-4}$ \\
\hline
Virgo           & $0.7615$ & $0.1833$ & $1.2316$ & $0.0$ & $2.8024$ & $0.0$ \\
\hline
\end{tabular}}
\caption{Detector latitude~$\beta$, longitude~$\lambda$ and arm orientation angles in radians for the three GW detector composing our detector network.}
\label{tab:detector_constants}
\end{table}

In this work we consider as GW detectors the network of L-shaped interferometer composed by the LIGO Livingston and Hanford, and the Virgo detectors. We report in table~\ref{tab:detector_constants} the coordinates on Earth of the detectors (latitude and longitude) and their arm orientation angles, as described in section~\ref{subsec:reference_frames} and~\ref{subsec:detector}.

\begin{figure}[ht]
\centerline{
\includegraphics[width=\columnwidth]{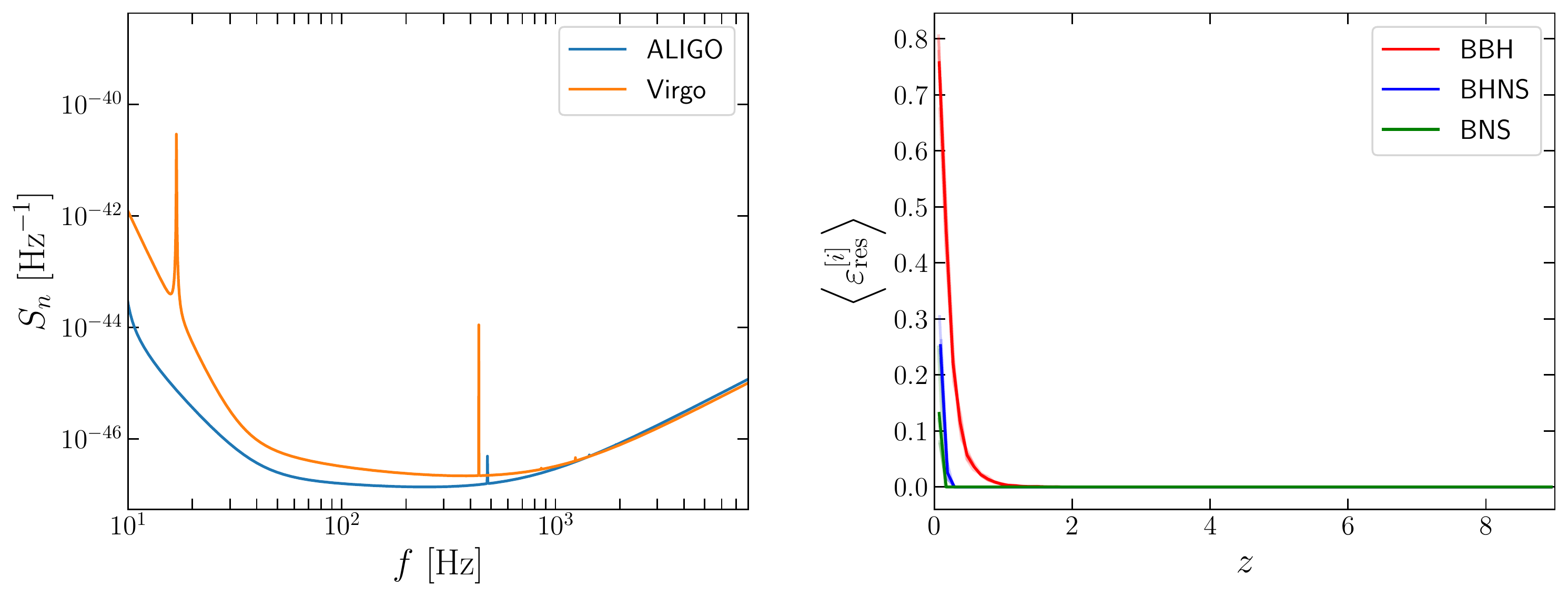}}
\caption{\textit{Left panel:} design sensitivity for the ALIGO and Virgo detector one-sided noise spectral density. \textit{Right panel:} average efficiency in detecting BBH, BHNS and BNS mergers for a SNR threshold of~$\rho_\mathrm{res}=12$. Each solid line is the average of five catalogs of~$10^5$ events. The average efficiency of the individual catalogs is represented as a shaded line.}
\label{fig:sn_epsilon}
\end{figure}

We consider the detector design sensitivity for the one-sided noise spectral density~$S_n$, which we show on the left panel of figure~\ref{fig:sn_epsilon}. The frequency range used for the three detectors is~$\left[10,8000\right]\ \mathrm{Hz}$, and the chosen SNR threshold for the detector network is~$\rho_\mathrm{res}=\rho_\mathrm{det}=12$, following LIGO-Virgo Collaboration approach. The choice of one particular~$\rho_\mathrm{res}$ over another is purely a convention which, in a real comparison with data, should be targeted around the optimal experimental searching strategy. Other choices have also been explored, see, e.g., ref.~\cite{bavera:snrthreshold}. In the right panel of figure~\ref{fig:sn_epsilon} we show the average efficiency of our detector network in detecting BBH, BHNS and BNS mergers. For every type of source we compute the average efficiency~$\left\langle \varepsilon^{[i]}_\mathrm{res} \right\rangle$ for five different catalogs, each consisting of~$10^5$ events. We also compute the average of these five average efficiencies. We observe that differences between individual catalogs are very small, in other words the catalogs provide a representative picture of the constraining power of the detector network.

\subsubsection{Isotropic GWB and effective function~$\mathcal{M}^\mathrm{eff}$}

The relevant quantities of the problem at hand, for instance~$\bar{\Omega}^\mathrm{tot}_\mathrm{GWB}$ or~$\Delta^\mathrm{tot}_\mathrm{GWB}$, require integrating some known function over a multi-dimensional space of parameters. Multi-dimensional integration done by discretising the parameter space always represents a numerical challenge, especially in those cases where we have~$\mathcal{O}(10)$ parameters, since the integrand has to be evaluated multiple times until a given convergence criterion is satisfied. In other words the algorithm has to explore the parameter space, which can have some non-trivial shape and boundaries. On the other hand, a catalog approach based computing the contribution of single events (sampled from the correct pdfs) to the total observable allows to speed up significantly the computation, which can now be performed on a common laptop. The core idea is to evaluated the integrand only in those regions of the parameter space which contribute the most to the integral. Since in our case every point of the parameter space correspond to a physical event, our integral will receive most of its contribution from those events that are more likely to happen over a given time interval. In other words we have an equivalence between
\begin{equation}
\int dz d\bm{\theta} \frac{4\pi c \chi^2(z) \bar{R}^{[i]}(z,\bm{\theta})}{(1+z)H(z)}p^{[i]}(\bm{\theta}|z)[1-\varepsilon_\mathrm{res}(z,\bm{\theta})] \simeq \frac{1}{\overline{\Delta t^{[i]}}} \frac{\sum_{i-\mathrm{th\ catalog}}}{N_\mathrm{events}} = \frac{1}{T^{[i]}_\mathrm{obs}}\sum_{i-\mathrm{th\ catalog}}
\end{equation}
which becomes exact in the limit of a catalog with an infinite number of events~$N_\mathrm{events}\to\infty$.

\begin{figure}[t]
\centerline{
\includegraphics[width=\columnwidth]{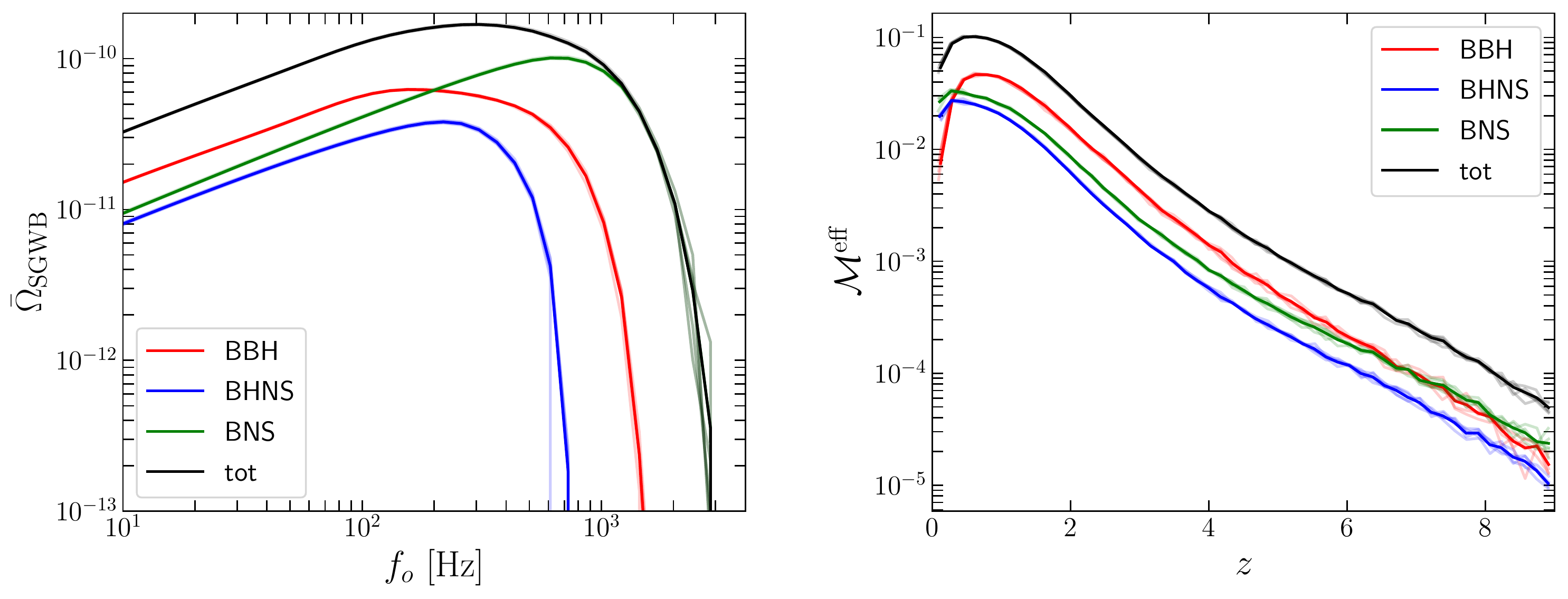}}
\caption{\textit{Left panel:} isotropic astrophysical GWB relative energy density obtained from catalogs of~$10^5$ BBH, BHNS and BNHS events. \textit{Right panel:} total effective kernel for the astrophysical GWB and contributions of individual type of sources at~$f_o=55\ \mathrm{Hz}$. Solid lines are the average of five catalogs, each one consisting of~$10^5$ events. Effective kernels of individual catalogs are represented by a shaded line.}
\label{fig:omegaGWB_Meff}
\end{figure}

Therefore the isotropic astrophysical GWB background relative energy density of equation~\eqref{eq:omegatot_GWB_background} can be rewritten in terms of sum over unresolved events of the catalogs as
\begin{equation}
\begin{aligned}
\bar{\Omega}^\mathrm{tot}_\mathrm{GWB}(f_o) &= \sum_i \int dz d\bm{\theta} \frac{4\pi c \chi^2(z) R^{[i]}(z,\bm{\theta})}{(1+z)H(z)}p^{[i]}(\bm{\theta}|z)[1-\varepsilon_\mathrm{det}(z,\bm{\theta})] \left[ \frac{f_o}{\rho_{0c}c^2} \frac{(1+z)^2}{4\pi c d^2_L(z)} \frac{dE^{[i]}_{\mathrm{GW},e}}{df_{e} d\Omega_{e}} \right] \\
&= \sum_i \frac{1}{T^{[i]}_\mathrm{obs}} \sum_{i-\mathrm{th\ catalog}} \left[ \frac{f_o}{\rho_{0c}c^2} \frac{(1+z)^2}{4\pi c d^2_L} \frac{dE^{[i]}_{\mathrm{GW},e}}{df_{e} d\Omega_{e}} \right]_\mathrm{event} \\
&= \sum_i \frac{1}{T^{[i]}_\mathrm{obs}} \sum_{i-\mathrm{th\ catalog}} \left[ \frac{f^3_o}{8G\rho_{0c}} \left(|h_+(f_o)|^2 + |h_\times(f_o)|^2 \right) \right]_\mathrm{event},
\end{aligned}
\label{eq:omegatot_GWB_catalog}
\end{equation}
where~$T^{[i]}_\mathrm{obs}$ is the observation time interval spanned by the catalog of sources of the~$i$-th type. Since the parameter we choose to generate a catalog is the number of events per catalog, each catalog spans a different time interval because of the different average waiting time~$\overline{\Delta t}$ of each source. To go from the second to the third line of equation~\eqref{eq:omegatot_GWB_catalog} we use equation~\eqref{eq:singleGW_energyspectrum}. We show in the left panel of figure~\ref{fig:omegaGWB_Meff} the isotropic astrophysical GWB and the relative contribution of each type of source. In our reference model the most important contribution in the frequency band LIGO-Virgo is more sensitive is represented by BBH merger, followed by BNS and BHNS mergers. 

Following the same logic we compute the first effective function~$\mathcal{M}^\mathrm{eff}$. As before, its definition in equation~\eqref{eq:effective_functions} can be rewritten in terms of sums over events of the catalog by dividing the events into redshift bins~$\Delta_j = [z_j-\Delta z, z_j+\Delta z]$, where~$z_j$ and~$\Delta z$ are the mean redshift and half-width of the bin, respectively. Therefore we have that
\begin{equation}
\bar{\Omega}^\mathrm{tot}_\mathrm{GWB}(f_o)\mathcal{M}^\mathrm{eff}(f_o, z_j) = \sum_i \frac{1}{T^{[i]}_\mathrm{obs}} \sum_{\substack{i-\mathrm{th\ catalog}, \\ \mathrm{events \ in\ } \Delta_j}} \left[ \frac{f^3_o}{8G\rho_{0c}} \left(|h_+(f_o)|^2 + |h_\times(f_o)|^2 \right) \right]_\mathrm{event}.
\end{equation}
We show in the right panel of figure~\ref{fig:omegaGWB_Meff} the effective kernel for BBH, BHNS and BNS events. As for the background relative energy density, the dominant contribution is given by BBH mergers, followed by BNS and BHNS ones. In other words, anisotropies of the astrophysical GWB probe mostly the physics of BBH mergers. As for the average efficiency, also in this case we report the~$\mathcal{M}^\mathrm{eff}$ effective function for five catalogs of~$10^5$ events for each kind of source to show that differences between catalogs are indeed very small and they occur either at very low or at very high-redshift. These fluctuations are a consequence of the low number of events in these redshift bins.

In our approach we replace integrals over pdfs with sums over events of a catalog obtained using those same pdfs, which is formally correct only in the limit of a catalog with infinite number of events. Even if this would be unfeasible, we can create catalogs with a number of events such that errors due to the finiteness of the catalog are negligible. Obviously the number of events required to span the entire parameter space are implicitly determined by the number of astrophysical parameters of our model. However, as we see both in figure~\ref{fig:sn_epsilon} and~\ref{fig:omegaGWB_Meff}, different catalogs of~$10^5$ events return results consistent with each other. As we show in section~\ref{sec:angular_power_spectrum}, those small differences in the high-redshift tail of the effective kernel produce a completely negligible effect on the total astrophysical GWB anisotropy. We conclude by noting that this entire concept is not designed around any specific reference model, and that it can be adapted to include different parameters and pdfs.


\section{Gravitational wave bias functions}
\label{sec:gw_bias}

Both resolved GW events and the astrophysical GWB trace the large scale structure of the Universe. Therefore gravitational radiation can be used as a new cosmological probe which is at the same time independent from and complementary to the traditional analysis based on electromagnetic radiation, for instance the one done using galaxy surveys.

Astrophysical sources of GWs are located in galaxies, where stellar activity is more intense, and galaxies, in turn, form in dark matter halos. Both galaxies and halos are biased tracers of the underlying dark matter field, which follows a well understood cosmological dynamics. The effective clustering and evolution bias in equation~\eqref{eq:effective_functions} contain the main insight about how GWs effectively trace the large scale structure of the Universe: the total effective clustering/evolution bias is simply the clustering/evolution bias of single events weighted by the contribution of the event to the total astrophysical GWB.

This result is not surprising, in fact we could have found a similar result by using the HOD (halo occupation distribution) formalism, see e.g., ref.~\cite{zheng:hod} and refs. therein, similarly to the approach presented for instance in refs.~\cite{libanore:gwxlss, capurri:astrophysicalGWB}. In the HOD formalism there is some freedom to choose the key parameter describing the properties of the GW host. In many works this has been identified with the stellar mass~$M_\star$, which is reasonable since stellar mass and star formation rate are highly correlated given that~$M_\star(t) \sim \int_0^{t} dt'\ \mathrm{SFR}(t')$. In this case the GW bias for sources of the~$i$-th type contributing to the astrophysical GWB is 
\begin{equation}
b^{[i]}_\mathrm{GWB}(z) = \frac{\displaystyle \int dM_\star \frac{d\bar{\Omega}^{[i]}_\mathrm{GWB}}{d M_\star}(M_\star,z) b_\mathrm{gal}(M_\star,z)}{\displaystyle \int dM_\star \frac{d\bar{\Omega}^{[i]}_\mathrm{GWB}}{d M_\star}(M_\star,z)},
\label{eq:GWB_bias_stellar_mass}
\end{equation}
where
\begin{equation}
\frac{d\bar{\Omega}^{[i]}_\mathrm{GWB}}{d M_\star} = \left\langle \frac{dn^{[i]}_\mathrm{GWB}}{dM_\star}\frac{dE_{\mathrm{GW},e}}{df_e d\Omega_e}\Bigg|M_\star\right\rangle = \frac{dn_\mathrm{gal}}{d M_\star} \times \left\langle N^{[i]}_\mathrm{GWB/gal}\frac{dE_{\mathrm{GW},e}}{df_e d\Omega_e}\Bigg|M_\star\right\rangle,
\end{equation}
$dn^{[i]}_\mathrm{GWB}/dM_\star$ is the unresolved GW number density per stellar mass, $dn_\mathrm{gal}/d M_\star$ is the galaxy number density per stellar mass, $N^{[i]}_\mathrm{GWB/gal}$ is the number of unresolved events per galaxy of mass~$M_\star$, $\left\langle N^{[i]}_\mathrm{GWB} \frac{dE_{\mathrm{GW},e}}{df_e d\Omega_e}\Big|M_\star\right\rangle$ is the unresolved GW average emitted energy in a galaxy with mass~$M_\star$, or, in other words, the astrophysical GWB halo occupation distribution, and~$b_\mathrm{gal}$ is the galaxy bias. 

Other host parameter choices are possible, for instance we can use the halo mass as main parameter characterizing GW hosts, as we propose in this work. Halo mass and stellar mass are almost in a 1-to-1 relation, see e.g., ref.~\cite{behroozi:averagestarformationhistory, behroozi:universemachine}. Since this relation presents a very small dispersion, using~$M_\star$ or~$M_h$ is practically equivalent for our purposes. In this case we have that
\begin{equation}
\begin{aligned}
\frac{dn_\mathrm{gal}}{d M_\star}(M_\star,z) &= \int dM_h \frac{dn_h}{d M_h} \left\langle N_\mathrm{gal/halo}(M_\star)|M_h\right\rangle, \\
b_\mathrm{gal}(M_\star,z) &= \frac{\displaystyle \int d M_h \frac{dn_h}{d M_h} \left\langle N_\mathrm{gal/halo}(M_\star)|M_h\right\rangle b_h(M_h,z)}{\displaystyle \int d M_h \frac{dn_h}{d M_h} \left\langle N_\mathrm{gal/halo}(M_\star)|M_h\right\rangle},
\end{aligned}
\end{equation}
where~$\left\langle N_\mathrm{gal/halo}(M_\star)|M_h\right\rangle$ is the average number of galaxies with mass~$M_\star$ per halo of mass~$M_h$, and~$b_h$ is the halo bias. Therefore, equation~\eqref{eq:GWB_bias_stellar_mass} can be recast as
\begin{equation}
b^{[i]}_\mathrm{GWB}(z) = \frac{\displaystyle \int dM_h \frac{d\bar{\Omega}^{[i]}_\mathrm{GWB}}{dM_h} (M_h,z) b_h(M_h,z)}{\displaystyle \int dM_h \frac{d\bar{\Omega}^{[i]}_\mathrm{GWB}}{dM_h} (M_h,z)},
\label{eq:GWB_bias_halo_mass}
\end{equation}
where
\begin{equation}
\begin{aligned}
\frac{d\bar{\Omega}^{[i]}_\mathrm{GWB}}{dM_h} &= \frac{dn_h}{d M_h} \int dM_\star \left\langle N^{[i]}_\mathrm{GWB/gal} \frac{dE_{\mathrm{GW},e}}{df_e d\Omega_e} \Bigg| M_\star\right\rangle \left\langle N_\mathrm{gal/halo}(M_\star)|M_h\right\rangle \\
&= \frac{dn_h}{d M_h} \left\langle N^{[i]}_\mathrm{GWB/halo} \frac{dE_{\mathrm{GW},e}}{df_e d\Omega_e} \Bigg| M_h\right\rangle \\
&= \left\langle \frac{dn^{[i]}_\mathrm{GWB}}{dM_h} \frac{dE_{\mathrm{GW},e}}{df_e d\Omega_e} \Bigg| M_h\right\rangle.
\end{aligned}
\end{equation}
In other words we explicitly showed that the logic behind equation~\eqref{eq:effective_functions} can be easily traced to the more common framework of the HOD formalism, and that the formalism already includes some flexibility in the way the GW host is parametrised, as we see in equations~\eqref{eq:GWB_bias_stellar_mass} and~\eqref{eq:GWB_bias_halo_mass}. 

Even if for the sake of simplicity we present this procedure by using only one parameter, the argument holds also when multiple parameters are used to describe the host. The same kind of reasoning can be applied to the case of the evolution bias.


\subsection{Worked example - part II}
\label{subsec:worked_example_bbeff}

\begin{figure}[t]
\centerline{
\includegraphics[width=\columnwidth]{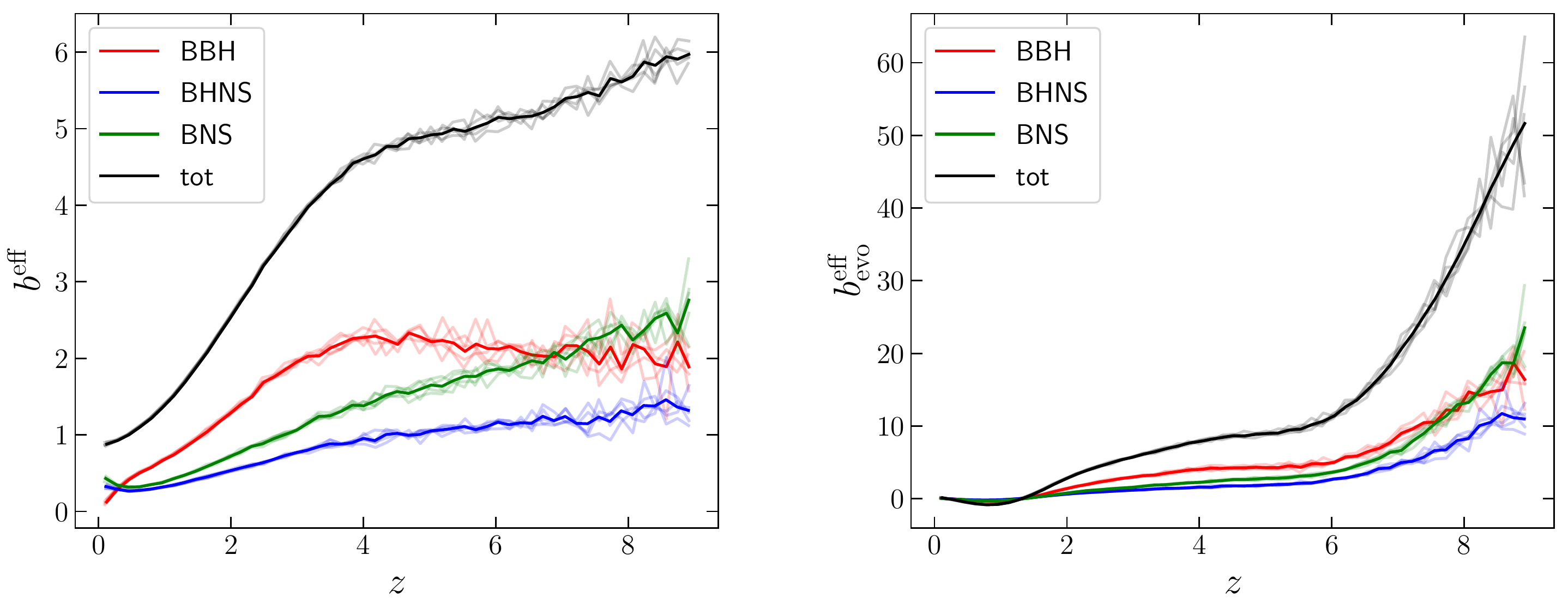}}
\caption{Total effective clustering bias (\textit{left panel}) and evolution bias (\textit{right panel}) for the astrophysical GWB and contributions of individual type of sources at~$f_o=55\ \mathrm{Hz}$. Solid lines are the average of five catalogs, each one consisting of~$10^5$ events. Effective bias functions of individual catalogs are represented by a shaded line.}
\label{fig:effective_bias}
\end{figure}

Following the same approach of section~\ref{subsec:worked_example_Meff}, the effective clustering and evolution bias functions of equation~\eqref{eq:effective_functions} can be written in terms of the sum over catalog events in different redshift bins~$\Delta_j$ as
\begin{equation}
\begin{aligned}
\bar{\Omega}^\mathrm{tot}_\mathrm{GWB} \mathcal{M}^\mathrm{eff}(f_o, z_j) b^\mathrm{eff}(f_o, z_j) &= \sum_i \frac{1}{T^{[i]}_\mathrm{obs}} \sum_{\substack{i-\mathrm{th\ catalog}, \\ \mathrm{events \ in\ } \Delta_j}} \left[ \frac{f^3_o}{8G\rho_{0c}} \left(|h_+(f_o)|^2 + |h_\times(f_o)|^2 \right) b^{[i]} \right]_\mathrm{event}, \\
\bar{\Omega}^\mathrm{tot}_\mathrm{GWB} \mathcal{M}^\mathrm{eff}(f_o, z_j) b^\mathrm{eff}_\mathrm{evo}(f_o, z_j) &= \sum_i \frac{1}{T^{[i]}_\mathrm{obs}} \sum_{\substack{i-\mathrm{th\ catalog}, \\ \mathrm{events \ in\ } \Delta_j}} \left[ \frac{f^3_o}{8G\rho_{0c}} \left(|h_+(f_o)|^2 + |h_\times(f_o)|^2 \right) b^{[i]}_\mathrm{evo} \right]_\mathrm{event}.
\end{aligned}
\label{eq:effective_bias_catalog}
\end{equation}
As in previous sections, we show in figure~\ref{fig:effective_bias} the effective clustering and evolution bias functions for five catalogs of~$10^5$ events for each kind of source (BBH, BHNS and BNS mergers) to demonstrate that differences between catalogs keep being small and occuring in those redshift bins with low number statistics.

\begin{figure}[t]
\centerline{
\includegraphics[width=\columnwidth]{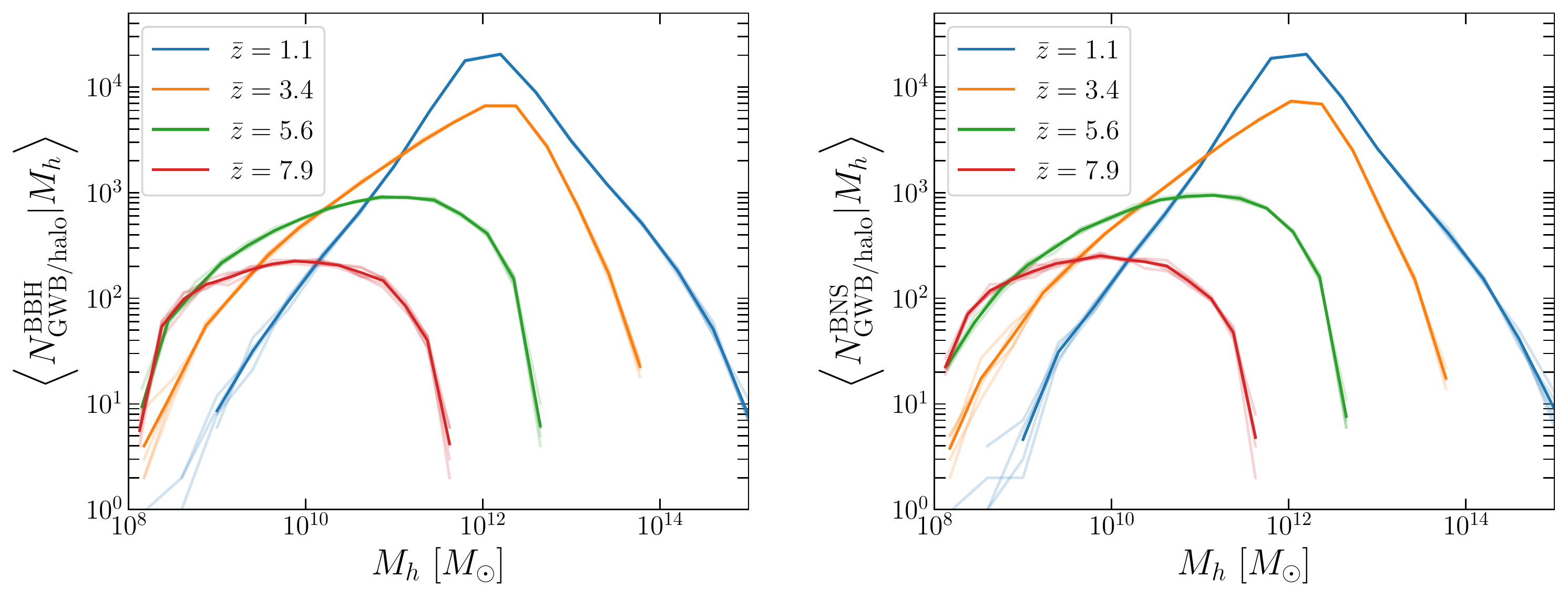}}
\caption{Number of BBH (\textit{left panel}) and BHNS (\textit{right panel}) events per halo in four redshift bins for five catalogs of~$10^5$ events. Numbers refer to an observation time of~$T^\mathrm{BBH}_\mathrm{obs}=2.4\ \mathrm{yr}$, $T^\mathrm{BHNS}_\mathrm{obs}=0.6\ \mathrm{yr}$ and~$T^\mathrm{BNS}_\mathrm{obs}=0.2\ \mathrm{yr}$. Solid lines are the average of five catalogs, shaded lines for individual catalogs.}
\label{fig:dNGWB_dMh}
\end{figure}

One feature of the catalog approach is that it provides insight on key quantities of the HOD formalism. In particular, we have that
\begin{equation}
\frac{d\bar{\Omega}^{[i]}_\mathrm{GWB}}{dM_h} \propto \left\langle N^{[i]}_\mathrm{GWB/halo} \frac{dE_{\mathrm{GW},e}}{df_e d\Omega_e} \Bigg| M_h\right\rangle = \left\langle N^{[i]}_\mathrm{GWB/halo} \Big| M_h\right\rangle \times \left\langle \frac{dE_{\mathrm{GW},e}}{df_e d\Omega_e} \right\rangle_\mathrm{halo},
\end{equation}
where~$\left\langle \frac{dE_{\mathrm{GW},e}}{df_e d\Omega_e} \right\rangle_\mathrm{halo}$ is the average GW emission per halo and in the last equality we use the fact that in our reference model every halo has the same average emission because the source parameter pdfs do not depend on the host. In figure~\ref{fig:dNGWB_dMh} we show the number of BBH and BNS GW events per halo mass in four different redshift bins. The case of BHNS GW events is intermediate between the two presented here: the small differences between the two type of events are due to different delay time pdfs and different detector sensitivities to each type of GW event.

Results presented in this part of the worked example have been obtained using the~\texttt{effective\_functions} routine described in appendix~\ref{app:external_modules}.


\section{Angular power spectrum and shot noise}
\label{sec:angular_power_spectrum}

In section~\ref{sec:aGWB_theory} we focused primarily on the cosmological signal generated by the total astrophysical GWB, however the observed signal extracted from a GW map of the sky is the sum of the true cosmological signal and of the (unavoidable) noise, i.e.,
\begin{equation}
s^{f_o}_{\ell m} = a^{f_o}_{\ell m} + n^{f_o}_{\ell m},
\end{equation}
where~$a^{f_o}_{\ell m}$ is defined in equation~\eqref{eq:deltatotGWB_spherical_harmonics_decomposition} and~$n^{f_o}_{\ell m}$ is the noise harmonic coefficient. Therefore, assuming that signal and noise are uncorrelated, i.e., $\left\langle a^{f_o}_{\ell m} n^{f_o *}_{\ell' m'} \right\rangle = \left\langle n^{f_o}_{\ell m} a^{f_o *}_{\ell' m'} \right\rangle = 0$, the total two-point statistics of the sky map is given by
\begin{equation}
\left\langle s^{f_o}_{\ell m} s^{f_o *}_{\ell' m'} \right\rangle = \delta^K_{\ell\ell'} \delta^K_{m m'} \tilde{C}^{f_o f_o}_\ell = \delta^K_{\ell\ell'} \delta^K_{m m'} \left(C^{f_o f_o}_\ell + N^{f_o f_o}_\ell \right),
\end{equation}
where~$N_\ell$ is the noise angular power spectrum~\cite{Alonso:2020rar, Contaldi:2020rht}. 

Any attempt to forecast the capability of future GW observatories to constraint astrophysical and/or cosmological parameters or to actually try to measure them from existing observations involves some well-understood ``likelihood-based'' method, for instance Fisher matrices or MCMCs. For the case of the astrophysical GWB, which is accurately described by linear theory at large scales, it is reasonable to assume a Gaussian likelihood. In this framework the typical uncertainty is estimated by computing the observed signal covariance matrix, which read either as~\cite{bellomo:multiclass}
\begin{equation}
\begin{aligned}
\mathrm{Cov}_\ell &= \left(C^{f_o f_o}_\ell + N^{f_o f_o}_\ell \right), \\
\mathrm{Cov}_\ell &= \frac{2}{2\ell+1} \left(C^{f_o f_o}_\ell + N^{f_o f_o}_\ell \right)^2,
\end{aligned}
\end{equation}
depending on whether we use a Gaussian likelihood for the spherical harmonic coefficients~$s_{\ell m}$ or for the angular power spectrum~$C_\ell$, respectively. In both cases we observe the same issue: the ``error`` depends on the signal especially at large scales, where the~$\ell^{-1}$ suppression is not effective and the measurement is effectively dominated by cosmic variance. For this reason it is crucial to obtain a robust prediction of the expected signal.

In this work we consider only the simplest contribution to the noise: the shot noise due to the discreteness in space and time of merger events. The derivation of this shot noise has been done in ref.~\cite{jenkins:aGWBshotnoise}, however here we present it using the formalism of our framework. By using equation~\eqref{eq:omegatot_GWB_background} we compute the covariance of the total astrophysical GWB over a given observation time period~$T_\mathrm{obs}$ as
\begin{equation}
\begin{aligned}
& \int d\hat{\mathbf{n}} d\hat{\mathbf{n}}' \mathrm{Cov}\!\left[ \Omega^\mathrm{tot}_\mathrm{GWB}(\hat{\mathbf{n}}), \Omega^\mathrm{tot}_\mathrm{GWB}(\hat{\mathbf{n}}') \right] T^2_\mathrm{obs} \\
&= \left(\frac{f_o}{\rho_{0c} c^2}\right)^2 \sum_{ij} \int \frac{dz d\hat{\mathbf{n}} d\bm{\theta} p^{[i]}(\bm{\theta}|z) \left[1-\varepsilon_\mathrm{det}(z,\bm{\theta})\right]}{(1+z)H(z)} \frac{dz' d\hat{\mathbf{n}}' d\bm{\theta}' p^{[j]}(\bm{\theta}'|z') \left[1-\varepsilon_\mathrm{det}(z',\bm{\theta}')\right]}{(1+z')H(z')} \\
&\qquad\qquad\qquad \times \mathrm{Cov}\! \left[ R^{[i]}(z, \hat{\mathbf{n}}, \bm{\theta}), R^{[j]}(z', \hat{\mathbf{n}}', \bm{\theta}') \right] \frac{dE^{[i]}_{\mathrm{GW},e}}{df_e d\Omega_e}(z,\bm{\theta}) \frac{dE^{[j]}_{\mathrm{GW},e}}{df_e d\Omega_e}(z',\bm{\theta}') T^2_\mathrm{obs} \\
&= \sum_{ij} \int dz d\hat{\mathbf{n}} d\bm{\theta} dz' d\hat{\mathbf{n}}' d\bm{\theta}' \mathrm{Cov}\! \left[ \frac{dN^{[i]}_\mathrm{GWB}}{dzd\Omega d\bm{\theta}}(z, \hat{\mathbf{n}}, \bm{\theta}), \frac{dN^{[j]}_\mathrm{GWB}}{dz'd\Omega' d\bm{\theta}'}(z', \hat{\mathbf{n}}', \bm{\theta}') \right]\\
&\qquad\qquad\qquad \times \left[\frac{f_o}{\rho_{0c} c^2}\frac{(1+z)^2}{cd_L^2(z)}\frac{dE^{[i]}_{\mathrm{GW},e}}{df_e d\Omega_e}(z,\bm{\theta})\right] \left[\frac{f_o}{\rho_{0c} c^2}\frac{(1+z')^2}{cd_L^2(z')}\frac{dE^{[j]}_{\mathrm{GW},e}}{df_e d\Omega_e}(z',\bm{\theta}')\right]\\
&= \sum_{ij} \int dz d\hat{\mathbf{n}} d\bm{\theta} dz' d\hat{\mathbf{n}}' d\bm{\theta}' \overline{\frac{dN^{[i]}_\mathrm{GWB}}{dzd\Omega d\bm{\theta}}}\  \overline{\frac{dN^{[j]}_\mathrm{GWB}}{dz'd\Omega' d\bm{\theta}'}} \left\langle \left(1+\delta^{[i]}_N (z, \hat{\mathbf{n}}, \bm{\theta})\right) \left(1+\delta^{[j]}_N(z', \hat{\mathbf{n}}', \bm{\theta}') \right) \right\rangle\\
&\qquad\qquad\qquad \times \left[\frac{f_o}{\rho_{0c} c^2}\frac{(1+z)^2}{cd_L^2(z)}\frac{dE^{[i]}_{\mathrm{GW},e}}{df_e d\Omega_e}(z,\bm{\theta})\right] \left[\frac{f_o}{\rho_{0c} c^2}\frac{(1+z')^2}{cd_L^2(z')}\frac{dE^{[j]}_{\mathrm{GW},e}}{df_e d\Omega_e}(z',\bm{\theta}')\right]\\
&= \sum_{ij} \int dz d\hat{\mathbf{n}} d\bm{\theta} dz' d\hat{\mathbf{n}}' d\bm{\theta}' \overline{\frac{dN^{[i]}_\mathrm{GWB}}{dzd\Omega d\bm{\theta}}}\  \overline{\frac{dN^{[j]}_\mathrm{GWB}}{dz'd\Omega' d\bm{\theta}'}}  \left[1+\left\langle\delta^{[i]}_N (z, \hat{\mathbf{n}}, \bm{\theta}) \delta^{[j]}_N(z', \hat{\mathbf{n}}', \bm{\theta}')\right\rangle \right] \\
&\qquad\qquad\qquad \times \left[\frac{f_o}{\rho_{0c} c^2}\frac{(1+z)^2}{cd_L^2(z)}\frac{dE^{[i]}_{\mathrm{GW},e}}{df_e d\Omega_e}(z,\bm{\theta})\right] \left[\frac{f_o}{\rho_{0c} c^2}\frac{(1+z')^2}{cd_L^2(z')}\frac{dE^{[j]}_{\mathrm{GW},e}}{df_e d\Omega_e}(z',\bm{\theta}')\right]\\
\end{aligned}
\label{eq:GWB_covariance}
\end{equation}
where 
\begin{equation}
\frac{dN^{[i]}_\mathrm{GWB}}{dzd\Omega d\bm{\theta}} = \frac{p^{[i]}(\bm{\theta}|z) \left[1-\varepsilon_\mathrm{res}\right] R^{[i]}T_\mathrm{obs}}{1+z} \frac{dV}{dzd\Omega} = \overline{\frac{dN^{[i]}_\mathrm{GWB}}{dzd\Omega d\bm{\theta}}} \left[1+\delta^{[i]}_N (z, \hat{\mathbf{n}}, \bm{\theta})\right]
\end{equation}
is the number of GW events contributing to the astrophysical GWB and~$\delta_N$ is a small fluctuation in the number of GW events with zero mean. As pioneered in ref.~\cite{feldman:shotnoise}, where a similar procedure was applied to galaxy clustering, we compute the discretised version of the expectation value in equation~\eqref{eq:GWB_covariance} as
\begin{equation}
\left\langle \delta^{[i]}_N \delta^{[j]}_N \right\rangle \simeq \left\langle \delta^{[i]}_{N,mn} \delta^{[j]}_{N,m'n'} \right\rangle = \left[\xi^{i}_{mm',n}(1-\delta^K_{mm'}) + \left\langle \left(\delta^{[i]}_{N,mn}\right)^2 \right\rangle \delta^K_{mm'}\right] \delta^K_{ij} \delta^K_{nn'},
\label{eq:numberofevents_twopointfunction}
\end{equation}
where the three set of indices $ij$, $mm'$ and~$nn'$ label the sources, volume cells and parameter space cells, respectively. Following ref.~\cite{jenkins:aGWBshotnoise}, we assume that different sources and different parameters are uncorrelated, even if some degree of correlation is expected because of the common stellar processes and of the common star formation history of the host. In equation~\eqref{eq:numberofevents_twopointfunction} $\xi^{[i]}_{mm',n}$ represents the two-point correlation function in real space, i.e., the analogous in real space of the~$C_\ell$s. The second contribution on the RHS of equation~\eqref{eq:numberofevents_twopointfunction} is the shot noise contribution and it reads as
\begin{equation}
\left\langle \left(\delta^{[i]}_{mn}\right)^2 \right\rangle  = \left\langle \left(\frac{ N^{[i]}_{mn}-\bar{N}^{[i]}_{mn}}{\bar{N}^{[i]}_{mn}}\right)^2 \right\rangle =  \frac{ \left\langle \left(N^{[i]}_{mn}\right)^2 \right\rangle - \left(\bar{N}^{[i]}_{mn}\right)^2}{\left(\bar{N}^{[i]}_{mn}\right)^2},
\end{equation}
where the number of events in a given volume cell is a stochastic variable. For the sake of clarity we drop all the indices and we will introduce them again at the end of the calculation. In our framework GW events live in halos with~$M_h \geq 10^8\ M_\odot$, which are spatially distributed according to a Poisson distribution with average~$\left\langle N_h \right\rangle$; moreover the events are also temporally distributed according to a second Poisson distribution with average~$\left\langle N_\mathrm{GWB/halo} \right\rangle$. Therefore the number of events in a given cell is given by
\begin{equation}
N = \sum_1^{N_h} N_\mathrm{GWB/halo}\,,
\end{equation}
where~$N_h$ and~$N_\mathrm{GWB/halo}$ are the number of halos per cell and the number of events per halo. The variable~$N$ follows a compound Poissonian distribution, therefore
\begin{equation}
\begin{aligned}
\left\langle N \right\rangle &= \left\langle N_h \right\rangle \left\langle N_\mathrm{GWB/halo} \right\rangle \equiv \bar{N}, \\
\left\langle N^2 \right\rangle &= \mathrm{Var}(N) + \left\langle N \right\rangle^2 = \left\langle N_h \right\rangle \mathrm{Var}(N_\mathrm{GWB/halo}) + \left\langle N_\mathrm{GWB/halo} \right\rangle^2 \mathrm{Var}(N_h) + \bar{N}^2 \\
&= \bar{N}\left[1+\left\langle N_\mathrm{GWB/halo} \right\rangle \right] + \bar{N}^2 .
\end{aligned}
\end{equation} 
Therefore in any given cell and for every set of sources and parameters the shot noise contribution reads as
\begin{equation}
\left\langle \delta^2 \right\rangle  = \frac{1 + \left\langle N_\mathrm{GWB/halo} \right\rangle}{\bar{N}} = \frac{1}{\bar{N}} + \frac{1}{\left\langle N_{h} \right\rangle}.
\end{equation}

In the continuum limit, after we reintroduce all the indices and we specify the calculation to the case of the relative fluctuation instead of the total one we find that the shot noise contribution is given by
\begin{equation}
\begin{aligned}
& \frac{1}{\left(\bar{\Omega}^\mathrm{tot}_\mathrm{GWB}\right)^2}\sum_{ij} \int dz d\hat{\mathbf{n}} d\bm{\theta} dz' d\hat{\mathbf{n}}' d\bm{\theta}' \overline{\frac{dN^{[i]}_\mathrm{GWB}}{dzd\Omega d\bm{\theta}}}\  \overline{\frac{dN^{[j]}_\mathrm{GWB}}{dz'd\Omega' d\bm{\theta}'}} \\
&\qquad\qquad \times \left[\frac{f_o}{\rho_{0c} c^2}\frac{(1+z)^2}{cd_L^2(z)}\overline{\frac{dE^{[i]}_{\mathrm{GW},e}}{df_e d\Omega_e}}(z,\bm{\theta})\right] \left[\frac{f_o}{\rho_{0c} c^2}\frac{(1+z')^2}{cd_L^2(z')}\overline{\frac{dE^{[j]}_{\mathrm{GW},e}}{df_e d\Omega_e}}(z',\bm{\theta}')\right] \\
& \qquad\qquad \times \left[ \left(\overline{\frac{dN^{[i]}_\mathrm{GWB}}{dzd\Omega d\bm{\theta}}}\right)^{-1} + \left(\frac{dN_h}{dzd\Omega}\right)^{-1}\right] \delta^D(z-z') \delta^D(\hat{\mathbf{n}}-\hat{\mathbf{n}}') \delta^D(\bm{\theta}-\bm{\theta}') \delta^K_{ij}.
\end{aligned}
\label{eq:noise_realspace}
\end{equation}


\subsection{Worked example - part III}
\label{subsec:worked_example_signalnoise}

As done in previous sections, we transform the theoretical expression of the shot noise into a catalog-derived quantity. In this case, by transforming the noise from real space into harmonic space, we find that the shot noise reads as
\begin{equation}
\begin{aligned}
& \left(\bar{\Omega}^\mathrm{tot}_\mathrm{GWB}\right)^2 N_\ell = \sum_i \sum_j \frac{1}{T^{[i]}_\mathrm{obs}} \frac{1}{T^{[j]}_\mathrm{obs}} \sum_{i-\mathrm{th\ catalog}} \sum_{j-\mathrm{th\ catalog}} \\
&\qquad  \left[ \frac{f^3_o}{8G\rho_{0c}} \left(|h_+(f_o)|^2 + |h_\times(f_o)|^2 \right) \right]_{\mathrm{event\ in\ }i} \left[ \frac{f^3_o}{8G\rho_{0c}} \left(|h_+(f_o)|^2 + |h_\times(f_o)|^2 \right) \right]_{\mathrm{event\ in\ }j} \\
& \qquad \left[ \left(\overline{\frac{dN^{[i]}_\mathrm{GWB}}{dzd\Omega d\bm{\theta}}}\right)^{-1} + \left(\frac{dN_h}{dzd\Omega}\right)^{-1}\right] \delta^K_{\mathrm{event\ in\ }i,\mathrm{event\ in\ }j} \\
&= \sum_i \frac{1}{(T^{[i]}_\mathrm{obs})^2} \sum_{i-\mathrm{th\ catalog}} \left\lbrace \left[ \frac{f^3_o}{8G\rho_{0c}} \left(|h_+(f_o)|^2 + |h_\times(f_o)|^2 \right) \right]^2 \left[ \left(\overline{\frac{dN^{[i]}_\mathrm{GWB}}{dzd\Omega d\bm{\theta}}}\right)^{-1} + \left(\frac{dN_h}{dzd\Omega}\right)^{-1}\right]\right\rbrace_{\mathrm{event}},
\end{aligned}
\end{equation}
where the combination of Dirac deltas in equation~\eqref{eq:noise_realspace} in the continuum limit becomes in a catalog approach a Dirac delta of the events, i.e.,
\begin{equation}
\delta^D(z-z') \delta^D(\hat{\mathbf{n}}-\hat{\mathbf{n}}') \delta^D(\bm{\theta}-\bm{\theta}') \delta^K_{ij} = \delta^K_{\mathrm{event\ in\ }i,\mathrm{event\ in\ }j},
\end{equation}
and
\begin{equation}
\begin{aligned}
\overline{\frac{dN^{[i]}_\mathrm{GWB}}{dzd\Omega d\bm{\theta}}} &= \frac{p^{[i]}(\bm{\theta}|z) \left[1-\left\langle\varepsilon_\mathrm{res}\right\rangle\right] \bar{R}^{[i]}T^{[i]}_\mathrm{obs}}{1+z} \frac{dV}{dzd\Omega}, \\
\frac{dN_h}{dzd\Omega} &= \int_{M^\mathrm{min}_h}^{M^\mathrm{max}_h(z)} dM_h \frac{dn_h}{dM_h} \times \frac{dV}{dzd\Omega},
\end{aligned}
\end{equation}
with~$M^\mathrm{min}_h, M^\mathrm{max}_h(z)$ defined in section~\ref{subsubsec:host_model}, and~$dV/dzd\Omega = (1/4\pi) \times dV/dz$.

\begin{figure}[t]
\centerline{
\includegraphics[width=\columnwidth]{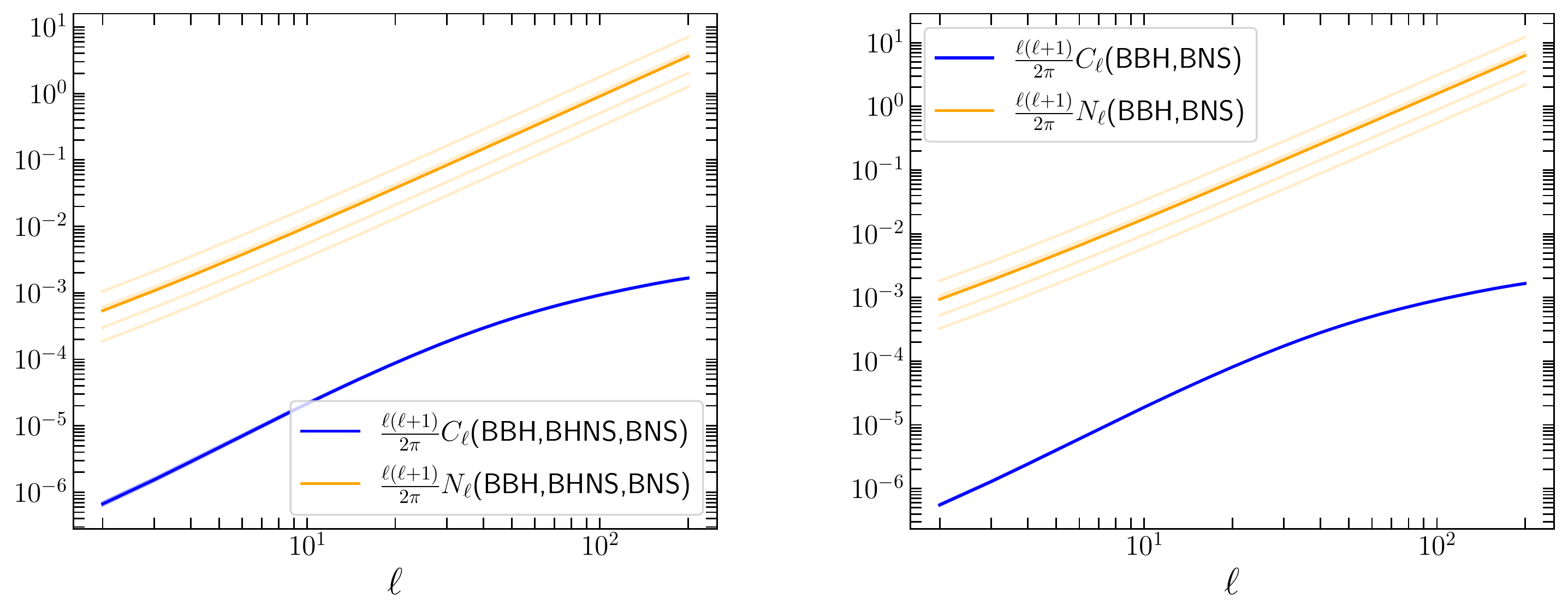}}
\caption{Angular power spectrum (computed including all the relativistic corrections) and shot noise from spatial and temporal clustering with (\textit{left panel}) and without (\textit{right panel}) including BHNS mergers at~$f_o=55\ \mathrm{Hz}$. The shot noise has been normalized to the expected noise in one year.}
\label{fig:cls}
\end{figure}

The analysis of statistical properties of GW maps of the sky is limited to large scales because of the low angular resolution of GW detectors. Even in the case of a network of next generation detectors (as Einstein Telescope and Cosmic Explorer) it will be challenging to localize high SNR events with a solid angle precision better than~$\Delta\Omega\simeq 1\ \mathrm{deg}^2$~\cite{calore:gwxlss, klimenko:gwlocalization, sidery:gwlocalization}, which corresponds (in this extremely optimistic case) to a maximum multipole~$\ell_\mathrm{max}\simeq 180\ \mathrm{deg} / \sqrt{\Delta\Omega}\ \simeq 200$. Hence we present our results on the astrophysical GWB angular power spectrum only for multipoles~$\ell \leq 200$.

We show in figure~\ref{fig:cls} the magnitude of the astrophysical GWB angular power spectrum, computed including all the relativistic corrections, and of the noise with and without the contribution coming from BHNS events. The reader can easily appreciate that the high redshift fluctuations we found in the effective function are not relevant: the curves obtained by the five catalog fundamentally overlap. While the difference in including BHNS events in the signal is at the level of~$10-20\%$ at low multipoles, the difference in the noise is a factor of few. For the reader interested in the angular power spectrum of the fluctuations, this observable can be straightforwardly computed by multiplying the angular power spectrum of the relative fluctuation in figure~\ref{fig:cls} by~$\left(\bar{\Omega}^\mathrm{tot}_\mathrm{GWB}\right)^2$.

\begin{figure}[t]
\centerline{
\includegraphics[width=\columnwidth]{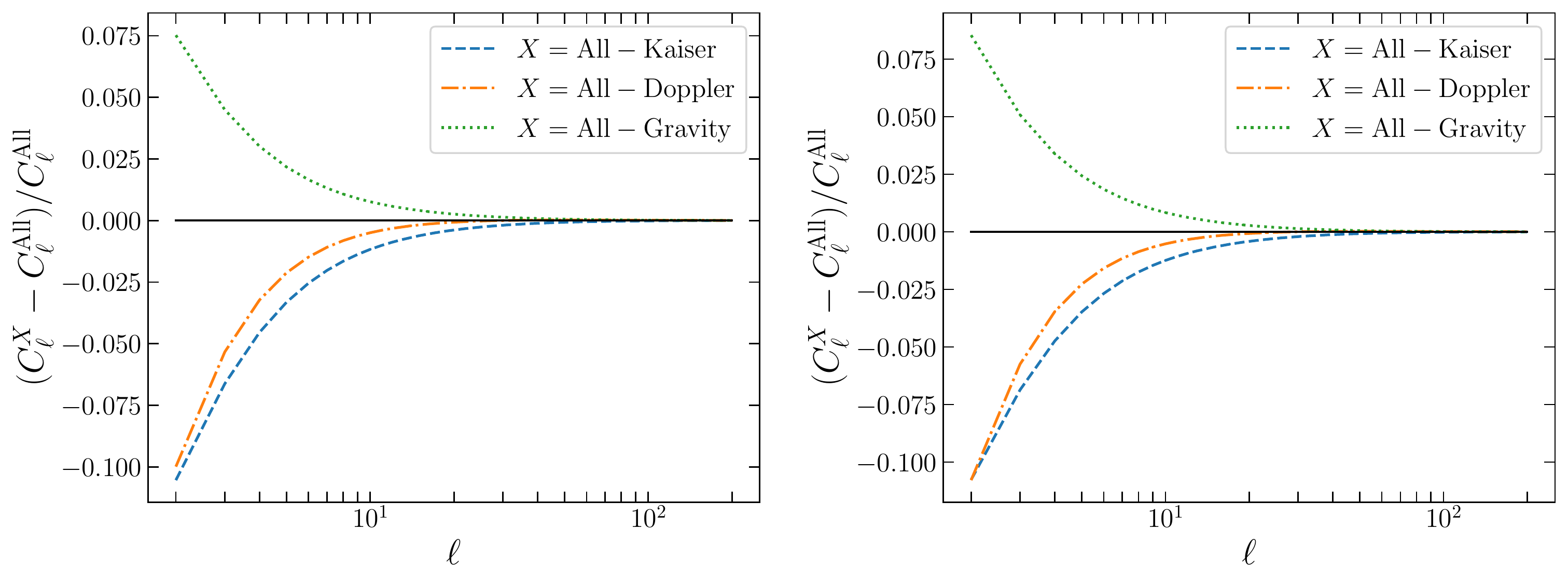}}
\caption{Relative contribution of different projection effects on the angular power spectra with (\textit{left panel}) and without (\textit{right panel}) including BHNS mergers at~$f_o=55\ \mathrm{Hz}$. The solid thin line at zero indicates all projection effects included; the blue dashed line shows the relative difference due to the non-inclusion of the Kaiser term, the dot-dashed orange is due to the non-inclusion of the Doppler effect and the dotted green shows the non-inclusion of gravitational potential terms.}
\label{fig:projection_effects}
\end{figure}

\begin{figure}[t]
\centerline{
\includegraphics[width=\columnwidth]{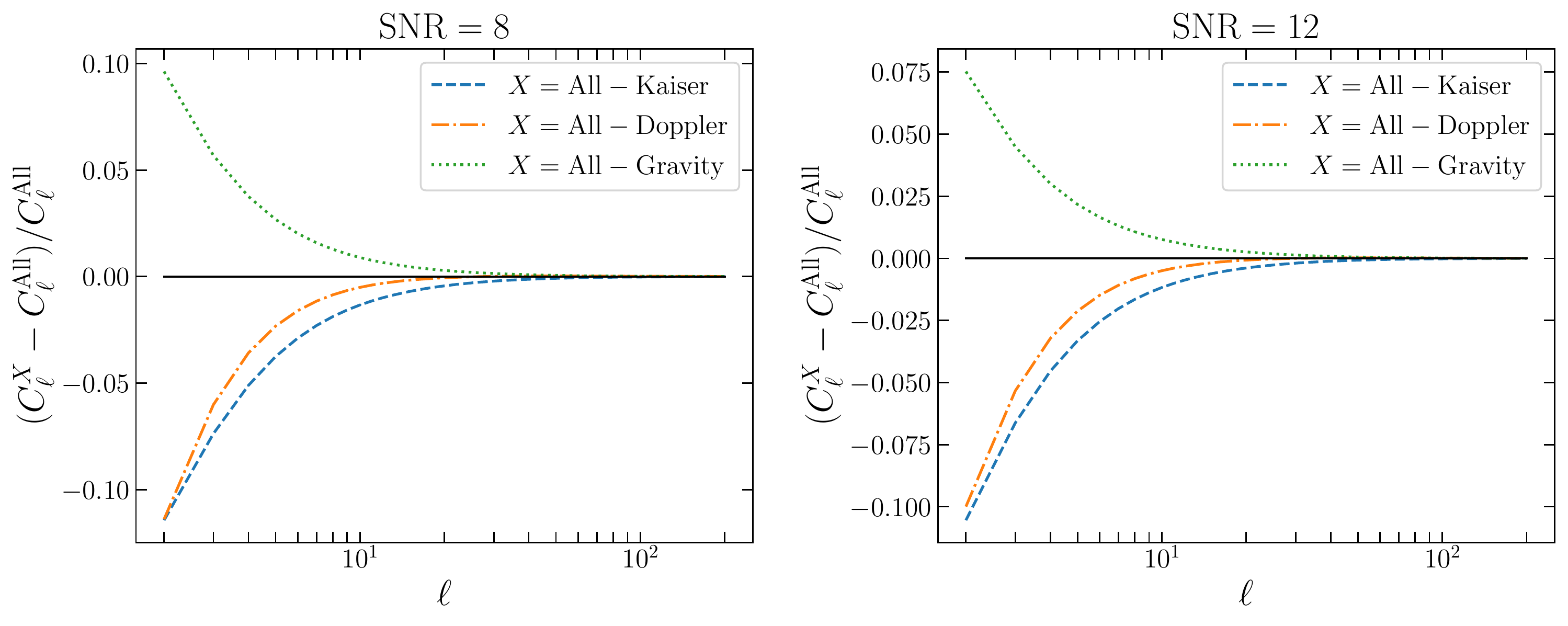}}
\caption{Relative contribution of different projection effects on the angular power spectra with $\mathrm{SNR=8}$ (\textit{left panel}) and $\mathrm{SNR=12}$ (\textit{right panel}). The solid thin lines at zero indicate all projection effects included; the blue dashed lines show the relative difference due to the non-inclusion of the Kaiser term, the dot-dashed orange are due to the non-inclusion of the Doppler effect and the dotted green show the non-inclusion of gravitational potential terms.}
\label{fig:projection_effects_snr}
\end{figure}

In figure~\ref{fig:projection_effects} we show the relative importance of all the different projection effects, namely the Kaiser ($^{\mathrm{vel},1}$ term in equation~\eqref{eq:relative_fluctuation_contributions}), Doppler ($^{\mathrm{vel},2},^{\mathrm{vel},3}$) and Gravity ($^{\mathrm{gr},1},^{\mathrm{gr},2},^{\mathrm{gr},3},^{\mathrm{gr},4}$) contributions, with and without including BHNS events. We confirm that projection effects play a critical role at ultra-large scale/low multipoles, contributing up to few tens of percent of the total angular power spectrum amplitude. In particular, the Kaiser term contribution is the most important one independently of the inclusion or not of BHNS merger events. Moreover we show in figure~\ref{fig:projection_effects_snr} the relative importance of the relativistic terms for different choices of SNR threshold. We compare our benchmark case obtained by choosing~$\rho_\mathrm{res}=\rho_\mathrm{det}=12$ with an optimistic scenario in which we have~$\rho_\mathrm{res}=8<\rho_\mathrm{det}$. In other words we are assuming that we are able to confidently remove parts of the data stream where we identify resolved signals with a~$\mathrm{SNR}$ not high enough to pass the threshold. Note that the relative importance of perturbation effects depends on the choice of astrophysical model and detector specifications, hence these terms can never be removed a priori.

Results presented in this part of the worked example have been obtained using the~\texttt{noise} routine described in appendix~\ref{app:external_modules}, as well as using the proper extension of~\texttt{CLASS}.


\section{Conclusions}
\label{sec:conclusions}

The birth of GW astronomy opened a new era also in Cosmology. Since gravitational radiation traces the underlying large-scale structure of the Universe in a very similar fashion to what electromagnetic radiation does, it allows to study the evolution of cosmic structures using an independent and complementary tracer. In fact existing and future GW detectors, as LISA or Einstein Telescope, can provide both catalogs of resolved GW events and sky maps of the GWB.

Interpreting these new observables represents a true scientific challenge because of the multidisciplinary expertise required to analyse these datasets. Measuring how GWs trace the large-scale structure of the universe requires an extensive knowledge of the properties of GW sources, of their signal generation and propagation over cosmological distances, and finally of how the signal is detected. The robustness of the measurement (and that of the derived constraints on astrophysical and/or cosmological models) ultimately depends on the robustness of the modelling of every aspect previously mentioned.

In this work we systematically analyse all the different key factors, providing both a general recipe that can be applied to a variety of models, and a worked example for the case of the astrophysical GWB expected in the LIGO/Virgo frequency band from BBH, BNS, BHNS mergers. We simulate a catalog of GW events expected to be present in the LIGO/Virgo band according to the latest observational constraints. From this we extract the expected background taking into account the {\it efficiency} of the detector (network) in resolving sources. Then we compute the angular power spectrum and we quantify the impact of the (large scale) projection effects, showing that, for some of them, the impact on the spectrum can reach the $10 \%$ level.  

The results presented in this work have been obtained using the \texttt{CLASS\_GWB} code complemented by other python routines. These numerical tools take into account all the relevant astrophysical dependencies and detector properties, as well as the effects of GW propagation in a perturbed Universe. The methodology we propose is flexible enough to incorporate other kind of sources parametrised by different astrophysical parameters, other detectors covering different frequency ranges and also different cosmological models, which can be implemented in every step of the analysis (and not only in the last one, where we compute the angular power spectra).

Another important key aspect of this work is the modularity of the methodology. Given the amount of different inputs required from different fields, it is tantamount to be able to improve existing estimate any time a new theoretical/observational result is obtained. For instance in our worked example we show how the results of the \textsc{UniverseMachine} model can be effectively used to provide an insight on the properties of GW hosts (and ultimately on their clustering properties). In the same way, simulation of stellar populations can provide important input on GW sources population, i.e., about their relevant astrophysical parameter and their pdfs. We leave the implementation of this aspect for future work.

The GWB represents an important observable for many different fields, from Astronomy to Cosmology up to High Energy Physics. Hence its detection represents one of the priorities of many GW observatories. Thanks to our methodology (and related numerical tools) we are able to produce a very robust prediction for the anisotropies of the GWB. However, how we show in this work, two obstacles stand in our way: GW maps of the sky are very noisy and have low resolution. The first limitation can be circumvent by cross-correlating GW maps of the sky with large-scale structure ones: even if it is hard to extract tight constraints from a ``GWB$\times$GWB'' correlation, the ``Galaxy$\times$GWB'' statistics have a lot of constraining power. Note that the idea of cross-correlating different datasets has already been extensively applied in Cosmology, for instance by using different LSS surveys~\cite{beutler:bossmultitracerI, marin:bossmultitracerII} or LSS surveys in combination with the cosmic microwave background~\cite{raccanelli:crosscorrelation, ho:cmbxlss, hirata:cmbxlss}, neutrinos~\cite{fang:neutrinosxlss}, ultra-high energy cosmic rays~\cite{urban:uhecrxlss, motloch:uhecrxlss} and so on; however accurate predictions of the expected signal are still required to be able to use this technique. 

On the other hand the low resolution of GW sky maps should be viewed as an opportunity instead of a limitation: even if new GW observatories will improve the overall angular resolution, these maps can be used only to study the clustering properties of GW sources at large scales. However this regime has not been fully and thoroughly probed yet and can potentially contain exciting new physics features. The cross-correlation technique can also be used to separate the astrophysical component of the GWB from the cosmological one since the astrophysical GWB probes the evolution of structures at ultra-large scales whereas the cosmological one does not. Moreover, any information coming from ``Galaxies$\times$GWB'' cross-correlation provides a complementary and independent check of the same physics explored by the ``Galaxies$\times$Resolved GW'' probe, which has been showed to have remarkable constraining power.

The methodology presented in this work and the \texttt{CLASS\_GWB} code can have a large application in the context of parameter estimation for both Astrophysics and Cosmology. We plan to explore these aspects in a forthcoming paper.


\section*{Acknowledgments}
NB is partially supported by the Spanish MINECO under grant BES-2015-073372. DB and SM acknowledge partial financial support by ASI Grant No. 2016-24-H.0. ACJ is supported by King's College London through a Graduate Teaching Scholarship. AR~acknowledges funding from MIUR through the ``Dipartimenti di eccellenza'' project Science of the Universe. MS is supported in part by the Science and Technology Facility Council (STFC), United Kingdom, under the research grant ST/P000258/1.


\appendix
\section{Astrophysical GWB fluctuation in Poisson gauge}
\label{app:aGWB_poisson_gauge}

In this appendix we derive the anisotropy of the astrophysical GWB found in ref.~\cite{bertacca:astrophysicalGWB} in the Poisson (or longitudinal or conformal Newtonian) gauge. The metric of a spatially flat Friedmann-Lema\^{i}tre-Robertson-Walker background, perturbed in a general gauge at the linear order, reads as
\begin{equation}
ds^2 = a^2(\tau)\left[ -(1+2A) d\tau^2 - 2B_i d\tau dx^i + (\delta_{ij}+h_{ij}) dx^i dx^j \right],
\end{equation}
where $A$ is a scalar, $B_i = \partial_i B + \hat{B}_i$, with $B$ a scalar and $\hat{B}_i$ a solenoidal vector ($\partial^i \hat{B}_i=0$), and $h_{ij}=2D\delta_{ij} + (\partial_i\partial_j - \delta_{ij}\nabla^2/3)F + \partial_i \hat{F}_j+ \partial_j \hat{F}_i + \hat{h}_{ij}$, with~$D$ and~$F$ scalars, $\hat{F}_i$ a solenoidal vector field and $\hat{h}_{ij}$ a trasverse traceless tensor ($\partial^i \hat{h}_{ij} = \hat{h}^i_{\ i} = 0$).

The astrophysical GWB fluctuation in a general gauge at linear order is given by equation~$(82)$ of ref.~\cite{bertacca:astrophysicalGWB}. According the formalism developed in section~\ref{sec:aGWB_theory}, the fluctuation for a single class of sources reads as
\begin{equation}
\begin{aligned}
\Delta\Omega^{[i]}_\mathrm{GW} (f_o, \hat{\mathbf{n}}) = & \frac{f_o}{\rho_{0c}c^2} \int \frac{dz}{H(z)} \int d\bm{\theta} p^{[i]}(\bm{\theta}|z) \left[1-\varepsilon_\mathrm{res}\right] \frac{\bar{R}^{[i]}}{1+z}\overline{\frac{dE^{[i]}_{\mathrm{GW},e}}{df_e d\Omega_e}} \\
&\times \Bigg\{ \delta^{[i]} + \left(3 - b^{[i]}_\mathrm{evo} + \frac{\mathcal{H}'}{\mathcal{H}^2}\right)A - B_\parallel + \left(b^{[i]}_\mathrm{evo} - 1 - \frac{\mathcal{H}'}{\mathcal{H}^2}\right)v_\parallel \\
&\quad + \frac{1}{\mathcal{H}}\left[ \frac{\mathrm{d}}{\mathrm{d}\bar{\chi}}\left(A-v_\parallel\right) + A' - B'_\parallel -\frac{1}{2}h'_\parallel \right] + 2\left(b^{[i]}_\mathrm{evo} - 2 - \frac{\mathcal{H}'}{\mathcal{H}^2}\right)I \\
&\quad + \left(b^{[i]}_\mathrm{evo} - 2 - \frac{\mathcal{H}'}{\mathcal{H}^2}\right) \left(\delta a_o + A_o - v_{\parallel, o}\right) \Bigg\},
\end{aligned}
\label{eq:omegatotgw_fluctuation_generalgauge}
\end{equation}
where~$'\equiv\partial_\tau $ and~$\partial_\parallel = \hat{\mathbf{n}}\cdot\nabla = n^i\partial_i$ indicate derivatives with respect to conformal time and along the line-of-sight, respectively, $\frac{\mathrm{d}}{\mathrm{d}\bar{\chi}}=-\partial_\tau+\partial_\parallel$, $v_\parallel = n^iv_i = \hat{\mathbf{n}}\cdot\mathbf{v}$, the subscript ``$_o$'' indicates quantities evaluated at observer position and
\begin{equation}
\begin{aligned}
I &= - \frac{1}{2} \int_0^{\chi(z)} d\tilde{\chi} \left( A' - B'_\parallel -\frac{1}{2}h'_\parallel \right), \\
\delta a_o &= - \mathcal{H}_0 \int_{0}^{\tau_0} d\tau \left.\frac{A(\tau)}{1+z(\tau)}\right|_o. \\
\end{aligned}
\end{equation}

The Poisson gauge is defined by $A=\Psi$, $B_i=0$, $D=-\Phi$, $F=\hat{F}_i=\hat{h}_{ij}=0$. In this gauge $v_\parallel=\hat{\mathbf{n}}\cdot\mathbf{v}=\hat{\mathbf{n}}\cdot\mathbf{V}$, where $\mathbf{V}$ is the gauge invariant velocity perturbation, $-h_\parallel/2=-n^in^jh_{ij}/2=\Phi$, $\delta a_o=-\mathcal{H}_0\int_0^{\tau_0}d\tau \left.\Psi(\tau)/\left[1+z(\tau)\right]\right|_o$ and
\begin{equation}
2I = - \int_0^{\chi(z)} d\tilde{\chi} \left(\Psi'+\Phi'\right).
\end{equation}
In General Relativity we have~\cite{challinor:galaxynumberdensity} $\mathbf{v}' + \mathcal{H}\mathbf{v} + \nabla\Psi = 0$ for cold particles, hence $\hat{\mathbf{n}}\cdot\mathbf{V}' + \mathcal{H}\hat{\mathbf{n}}\cdot\mathbf{V} + \partial_\parallel\Psi = 0$. Following ref.~\cite{jeong:galaxynumberdensity}, the overdensity~$\delta^{[i]}$ can be written in the Poisson gauge as
\begin{equation}
\delta^{[i]} = b^{[i]}(\delta^\mathrm{Poisson}_m - 3\mathcal{H}V) - \left(b^{[i]}_\mathrm{evo}-3\right)\mathcal{H} V = b^{[i]}D - \left(b^{[i]}_\mathrm{evo}-3\right)\mathcal{H}V,
\end{equation}
where $D=\delta^\mathrm{Poisson}_m - 3\mathcal{H}V = \delta^\mathrm{Sync.-Com.}_m$ is the gauge invariant matter density fluctuation, $\delta^{\cdots}_m$ is the matter overdensity in different gauges (Newtonian and synchronous comoving), $V$ is the velocity potential, i.e., $\mathbf{V}=\nabla V$. Therefore equation~\eqref{eq:omegatotgw_fluctuation_generalgauge} in Poisson gauge reads as
\begin{equation}
\begin{aligned}
\Delta\Omega^{[i]}_\mathrm{GW} (f_o, \hat{\mathbf{n}}) = & \frac{f_o}{\rho_{0c}c^2} \int \frac{dz}{H(z)} \int d\bm{\theta} p^{[i]}(\bm{\theta}|z) \left[1-\varepsilon_\mathrm{res}\right] \frac{\bar{R}^{[i]}}{1+z}\overline{\frac{dE^{[i]}_{\mathrm{GW},e}}{df_e d\Omega_e}} \\
&\times \Bigg\{ b^{[i]} D  \\
&\quad + \left(b^{[i]}_\mathrm{evo} - 2 - \frac{\mathcal{H}'}{\mathcal{H}^2}\right) \hat{\mathbf{n}}\cdot\mathbf{V} - \frac{1}{\mathcal{H}}\partial_\parallel(\hat{\mathbf{n}}\cdot\mathbf{V}) - (b^{[i]}_\mathrm{evo}-3) \mathcal{H} V \\
&\quad + \left(3 - b^{[i]}_\mathrm{evo} + \frac{\mathcal{H}'}{\mathcal{H}^2}\right)\Psi + \frac{1}{\mathcal{H}}\Phi' + \left(2 - b^{[i]}_\mathrm{evo} + \frac{\mathcal{H}'}{\mathcal{H}^2}\right) \int_0^{\chi(z)} d\tilde{\chi} \left(\Psi'+\Phi'\right) \\
&\quad + \left(b^{[i]}_\mathrm{evo} - 2 - \frac{\mathcal{H}'}{\mathcal{H}^2}\right)  \left( \Psi_o - \mathcal{H}_0 \int_{0}^{\tau_0} d\tau \left.\frac{\Psi(\tau)}{1+z(\tau)}\right|_o -  \left(\hat{\mathbf{n}}\cdot\mathbf{V} \right)_o \right) \Bigg\},
\end{aligned}
\label{eq:omegatotgw_fluctuation_poissongauge}
\end{equation}
where we isolated the \textit{density}, \textit{velocity}, \textit{gravity} and \textit{observer} contributions in the first, second, third and fourth line, respectively.

By Fourier transforming\footnote{For simplicity we keep the same symbol also for Fourier transformed quantities.} equation~\eqref{eq:omegatotgw_fluctuation_poissongauge} we obtain the harmonic transfer functions contributing to the total astrophysical GWB fluctuation of equation~\eqref{eq:relative_fluctuation_contributions}. They are implemented in \texttt{CLASS\_GWB} as\footnote{The harmonic transfer functions can be easily compared with the code keeping in mind that
\begin{equation*}
\frac{aH'}{(aH)^2} = \frac{a}{(aH)^2}\left(\frac{\mathcal{H}}{a}\right)' = \frac{\mathcal{H}'}{\mathcal{H}^2}-1.
\end{equation*}}
\begin{equation}
\begin{aligned}
\Delta^\mathrm{den}_\ell(k) &= \int \frac{dz}{H(z)} W(z) \mathcal{M}^\mathrm{eff}(z) b^\mathrm{eff}(k,z) j_\ell(k\chi) S_D(k,z), \\
\Delta^\mathrm{vel,1}_\ell(k) &= \int \frac{dz}{H(z)} W(z) \mathcal{M}^\mathrm{eff}(z) \frac{1}{\mathcal{H}} \partial^2_{k\chi} j_\ell(k\chi) S_\Theta(k,z), \\
\Delta^\mathrm{vel,2}_\ell(k) &= \int \frac{dz}{H(z)} W(z) \mathcal{M}^\mathrm{eff}(z) \frac{\left(2 - b^\mathrm{eff}_\mathrm{evo} + \frac{\mathcal{H}'}{\mathcal{H}^2}\right)}{k} \partial_{k\chi} j_\ell(k\chi) S_\Theta(k,z), \\
\Delta^\mathrm{vel,3}_\ell(k) &= \int \frac{dz}{H(z)} W(z) \mathcal{M}^\mathrm{eff}(z) \left(b^\mathrm{eff}_\mathrm{evo}-3\right) \frac{\mathcal{H}}{k^2} j_\ell(k\chi) S_\Theta(k,z), \\
\Delta^\mathrm{gr,1}_\ell(k) &= \int \frac{dz}{H(z)} W(z) \mathcal{M}^\mathrm{eff}(z) j_\ell(k\chi) S_\Psi(k,z), \\
\Delta^\mathrm{gr,2}_\ell(k) &= \int \frac{dz}{H(z)} W(z) \mathcal{M}^\mathrm{eff}(z) \left(b^\mathrm{eff}_\mathrm{evo} - 2 - \frac{\mathcal{H}'}{\mathcal{H}^2}\right) j_\ell(k\chi) S_\Phi(k,z), \\
\Delta^\mathrm{gr,3}_\ell(k) &= \int \frac{dz}{H(z)} W(z) \mathcal{M}^\mathrm{eff}(z) \frac{1}{\mathcal{H}} j_\ell(k\chi) S_{\Phi'}(k,z) , \\
\Delta^\mathrm{gr,4}_\ell(k) &= \int \frac{dz}{H(z)} \widehat{W}(z) k \partial_{k\chi}j_\ell(k\chi) S_{(\Psi+\Phi)}(k,z), \\
\end{aligned}
\label{eq:relative_fluctuation_transfers}
\end{equation}
where $X(k,z)=S_X(k,z)\zeta_\mathrm{ini}(k)$ defines the transfer functions~$S_X$ in terms of the initial primordial curvature perturbation and
\begin{equation}
S_\Theta = -k^2S_V,\qquad S_D = S_{\delta_m} + 3\frac{\mathcal{H}}{k^2}S_\Theta.
\end{equation}
In equation~\eqref{eq:relative_fluctuation_transfers} we also included a window function~$W(z)\simeq\Theta_H(z)\Theta_H(z_\mathrm{max}-z)$, where~$\Theta_H$ is the Heaviside theta function, to limit the domain of integration. As long as~$z_\mathrm{max}\simeq \mathcal{O}(10)$, this window function does not introduce any bias in the numerical estimate of the angular power spectrum since the effective kernel~$\mathcal{M}^\mathrm{eff}$ is peaked in a narrower redshift range. We also defined a second window entering in the~$_\mathrm{gr,4}$ contribution that reads as
\begin{equation}
\widehat{W}(\chi) = \int_\chi^\infty d\tilde{\chi} W(\tilde{\chi}) \mathcal{M}^\mathrm{eff}(\tilde{\chi}) \left(2 - b^\mathrm{eff}_\mathrm{e}(\tilde{\chi}) + \frac{\mathcal{H}'(\tilde{\chi})}{\mathcal{H}^2(\tilde{\chi})}\right).
\end{equation}

Notice that the gravity terms of equations~\eqref{eq:relative_fluctuation_poissongauge} and~\eqref{eq:relative_fluctuation_transfers} do in fact match. This rewriting is necessary since \texttt{CLASS} stores the~$\Psi+\Phi$ source function but not the~$(\Psi+\Phi)'$ one. To go from equation~\eqref{eq:relative_fluctuation_poissongauge} to equation~\eqref{eq:relative_fluctuation_transfers} first we rewrite the Integrated Sachs-Wolfe term as
\begin{equation}
\begin{aligned}
\mathrm{ISW} &= \int_0^\infty d\chi W(\chi) \mathcal{M}^\mathrm{eff}(\chi) \left(2 - b^\mathrm{eff}_\mathrm{e}(\chi) + \frac{\mathcal{H}'(\chi)}{\mathcal{H}^2(\chi)}\right) \int_0^\infty d\tilde{\chi} \frac{d(\Psi+\Phi)}{d\tilde{\tau}} j_\ell(k\tilde{\chi}) \Theta_H(\chi-\tilde{\chi}) \\
&= \int_0^\infty d\tilde{\chi} \left[ \int_{\tilde{\chi}}^\infty d\chi W(\chi) \mathcal{M}^\mathrm{eff}(\chi) \left(2 - b^\mathrm{eff}_\mathrm{e}(\chi) + \frac{\mathcal{H}'(\chi)}{\mathcal{H}^2(\chi)}\right) \right] \frac{d(\Psi+\Phi)}{d\tilde{\tau}} j_\ell(k\tilde{\chi}) \\
&= -\int_0^\infty d\tilde{\chi} \widehat{W}(\tilde{\chi}) \frac{d(\Psi+\Phi)}{d\tilde{\chi}} j_\ell(k\tilde{\chi})
\end{aligned}
\end{equation}
where~$\tilde{\chi}=\tau_0-\tilde{\tau}$. Then we integrate by parts this term to find
\begin{equation}
\begin{aligned}
\mathrm{ISW} &= - \left(\widehat{W}(\tilde{\chi}) (\Psi+\Phi) j_\ell(k\tilde{\chi}) \right|_{0}^\infty + \int_0^\infty d\tilde{\chi} (\Psi+\Phi) \left[\frac{d\widehat{W}}{d\tilde{\chi}}j_\ell + \widehat{W} k \partial_{k\tilde{\chi}} j_\ell\right],
\end{aligned}
\end{equation}
where the first term in the RHS vanishes at infinity because of the definition of~$\widehat{W}$ and contributes to the observer terms at~$\tilde{\chi}=0$, the first term in the square brackets is proportional to $\Psi$ and~$\Phi$, hence it has to be consistently summed to the other terms with the same source, generating the ``$_\mathrm{gr,1}$'' and ``$_\mathrm{gr,2}$'' terms, whereas the second term in the square brackets corresponds to the ``$_\mathrm{gr,4}$'' contribution. 

The observer terms
\begin{equation}
\begin{aligned}
\Delta^\mathrm{obs,1}_\ell(k) &= \int \frac{dz}{H(z)} W(z) \mathcal{M}^\mathrm{eff}(z) \left(2 - b^\mathrm{eff}_e + \frac{\mathcal{H}'}{\mathcal{H}^2}\right) \mathcal{H}_0 \int_0^{\tau_0}d\tau \frac{S_\Psi(k,\tau)}{1+z(\tau)} \delta^K_{\ell 0}, \\
\Delta^\mathrm{obs,2}_\ell(k) &= \int \frac{dz}{H(z)} W(z) \mathcal{M}^\mathrm{eff}(z) \left(b^\mathrm{eff}_e - 2 - \frac{\mathcal{H}'}{\mathcal{H}^2}\right) S_\Psi(k,0) \delta^K_{\ell 0}, \\
\Delta^\mathrm{obs,3}_\ell(k) &= \int \frac{dz}{H(z)} W(z) \mathcal{M}^\mathrm{eff}(z) \left(b^\mathrm{eff}_e - 2 - \frac{\mathcal{H}'}{\mathcal{H}^2}\right) \frac{\left.\partial_{y} j_\ell(y)\right|_{y=0}}{k} S_\Theta(k,0), \\
\end{aligned}
\end{equation}
have not been included in \texttt{CLASS\_GWB} for two reasons. The first one is that they contribute only to the monopole~$C_0$ and to the dipole~$C_1$, in fact we have that~$j_\ell(0)=\delta^K_{\ell 0}$ and $\left.\partial_{y} j_\ell(y)\right|_{y=0}\propto \delta^K_{\ell 1}$, whereas in general \texttt{CLASS\_GWB} deals with computing the angular power spectra for multipoles~$\ell \geq 2$. The second one is connected with the fact that it is hard to consistently compute ensemble averages in a single point of spacetime, in this case the observer position, since we cannot estimate the ensemble average as a volume average. Monopole terms are not measurable since we do not know what is the \textit{true} value of the background while dipole terms should be removed as for instance is done with CMB velocity dipole. Moreover, since we live in a galaxy inside a dark matter halo, these effects are cannot be described accurately by cosmological linear theory.


\section{Gravitational wave waveform details}
\label{app:gw_waveform}

We complete the description of the GW waveform by providing the four~$y_{(k),\mathrm{tm}}$ factors and~$Y_{(k)}$ matrices appearing in equation~\eqref{eq:waveform_frequencies}. They read as~\cite{ajith:gwswaveform}
\begin{equation}
\begin{aligned}
y_{(1),\mathrm{tm}} &= 1-4.455(1-\chi)^{0.217} + 3.521(1-\chi)^{0.26}, \\
y_{(2),\mathrm{tm}} &= \frac{1-0.63(1-\chi)^{0.3}}{2}, \\
y_{(3),\mathrm{tm}} &= 0.3236 + 0.04894\chi + 0.01346\chi^2, \\
y_{(4),\mathrm{tm}} &= \frac{\left[1-0.63(1-\chi)^{0.3}\right](1-\chi)^{0.45}}{4},
\end{aligned}
\end{equation}
and
\begin{equation}
\begin{aligned}
Y_{(1)} &= \begin{pmatrix}
0 & 0 & 0 & 0 \\ 
0.6437 & 0.827 & -0.2706 & 0 \\
-0.05822 & -3.935 & 0 & 0 \\
-7.092 & 0 & 0 & 0 \\
\end{pmatrix} , \
Y_{(2)} = \begin{pmatrix}
0 & 0 & 0 & 0 \\ 
0.1469 & -0.1228 & -0.02609 & 0 \\
-0.0249 & 0.1701 & 0 & 0 \\
2.325 & 0 & 0 & 0 \\
\end{pmatrix} , \\
Y_{(3)} &= \begin{pmatrix}
0 & 0 & 0 & 0 \\ 
-0.1331 & -0.08172 & 0.1451 & 0 \\
-0.2714 & 0.1279 & 0 & 0 \\
4.922 & 0 & 0 & 0 \\
\end{pmatrix} , \
Y_{(4)} = \begin{pmatrix}
0 & 0 & 0 & 0 \\ 
-0.4098 & -0.03523 & 0.1008 & 0 \\
1.829 & -0.02017 & 0 & 0 \\
-2.87 & 0 & 0 & 0 \\
\end{pmatrix}.
\end{aligned}
\end{equation}

The value of the waveform normalization constant can be derived following ref.~\cite{phinney:GWBdefinition}. The integrated angle-average flux~$\left\langle S(t_{o}) \right\rangle$ from a distant GW source measured by an observer and we connect it with the GW strain as
\begin{equation}
\begin{aligned}
\int_{-\infty}^{+\infty} dt_o \left\langle S(t_o) \right\rangle &= \frac{c^3}{16\pi G} \int_{-\infty}^{+\infty} dt_o \left\langle \dot{h}^2_+(t_o) + \dot{h}^2_\times(t_o) \right\rangle \\
&= \frac{\pi c^3}{4 G} \int_{-\infty}^{+\infty} df_o f_o^2 \left\langle |h_+(f_o)|^2 + |h_\times(f_o)|^2 \right\rangle \\
&= \frac{\pi c^3}{2 G} \int_{0}^{+\infty} df_o f_o^2 \left\langle |h_+(f_o)|^2 + |h_\times(f_o)|^2 \right\rangle
\end{aligned}
\label{eq:integrated_flux_1}
\end{equation}
where the Fourier transform convention used here is~$h(t)=\int df h(f) e^{i2\pi ft}$ and we used the fact that~$h_{+,\times}$ are real functions. The angle average in this case refers only to the angle average over the inclination angle~$\iota$: using equation~\eqref{eq:inclination_angle_dependence} it reads as
\begin{equation}
\int_{-1}^{+1} \frac{d\cos\iota}{2}\left[ \left(\frac{1+\cos^2\iota}{2}\right)^2 + \cos^2\iota \right] = \frac{4}{5}.
\end{equation}
On the other hand the angle-averaged integrated flux is given by
\begin{equation}
\begin{aligned}
\int_{-\infty}^{+\infty} dt_o \left\langle S(t_o) \right\rangle &= \int_{-\infty}^{+\infty} dt_o \frac{L_e}{4\pi d^2_L} \\
&= \frac{1}{4\pi d^2_L} \int_{-\infty}^{+\infty} dt_e (1+z) \frac{dE_{\mathrm{GW},e}}{dt_e} \\
&= \frac{1+z}{4\pi d^2_L} \int_{-\infty}^{+\infty} df_e \frac{dE_{\mathrm{GW},e}}{df_e},
\end{aligned}
\label{eq:integrated_flux_2}
\end{equation}
where~$dE_{\mathrm{GW},e}/df_e = \pi^{2/3} G^{2/3} M^{5/3}_c f^{-1/3}_e /3$ for an inspiraling phase. By comparing equations~\eqref{eq:integrated_flux_1} and~\eqref{eq:integrated_flux_2} we find that
\begin{equation}
|h(f_o)|^2 = \frac{5}{24} \frac{(GM_c)^{5/3}}{\pi^{4/3} c^3} \frac{(1+z)^{5/3}}{d^2_L} f_o^{-7/3},
\end{equation}
thus the normalization constant is given by
\begin{equation}
C_o = \sqrt{\frac{5}{24}} \frac{(GM_c)^{5/6}}{\pi^{2/3} c^{3/2}} \frac{(1+z)^{5/6}}{d_L}.
\label{eq:observed_normalization_constant}
\end{equation}

Note that that the GW waveform template of equation~\eqref{eq:waveform_template} can be expressed also a function of non-redshifted variables as 
\begin{equation}
A(f_e) = C_e f^{-7/6}_1 \left\lbrace
\begin{aligned}
& \left(\frac{f_e}{f_1}\right)^{-7/6} \left(1 + \alpha_2 v^2 + \alpha_3 v^3 \right), \quad & f_e<f_1, \\
& \omega_m \left(\frac{f_e}{f_1}\right)^{-2/3} \left(1 + \epsilon_1 v + \epsilon_2 v^2 \right), \quad & f_1<f_e<f_2, \\
& \omega_r \frac{1}{2\pi} \frac{\sigma}{(f_e-f_2)^2+\sigma^2/4}, \quad & f_2<f_e<f_3,
\end{aligned} \right.
\end{equation}
where~$C_e = C_o (1+z)^{7/6}$.

Finally, by comparing equations~\eqref{eq:integrated_flux_1} and~\eqref{eq:integrated_flux_2} we derive the energy spectrum for the single source to be
\begin{equation}
\frac{dE_{\mathrm{GW},e}}{df_e d\Omega_e} = \frac{\pi d^2_L c^3 f_o^2}{2G(1+z)^2} \left(|h_+(f_o)|^2 + |h_\times(f_o)|^2 \right).
\label{eq:singleGW_energyspectrum}
\end{equation}


\section{Rotating coordinate systems}
\label{app:rotating_coordinate_systems}

Given that computing the antenna response function requires the use of rotation matrix theory to connect different reference frames, here we summarise the main ideas and results that might be helpful to the reader in charactering a detector. We refer the interested reader to classical books as ref.~\cite{goldstein:classicalmechanics} for all those technical aspects we do not discuss in detail here.  

Suppose to have two frames not aligned with each other with orthonormal right-handed bases~$\{\mathbf{e}_1, \mathbf{e}_2, \mathbf{e}_3 \}$ and~$\{\mathbf{e}'_1, \mathbf{e}'_2, \mathbf{e}'_3 \}$. The vectors of second base can be written in terms of the vectors of the first one as~$\mathbf{e}'_i = \sum_{j=1}^{3} a_{ij}\mathbf{e}_j$, where the~$a_{ij}$ are real coefficients, which can be conveniently organised in a~$3\times 3$ orthogonal matrix~$A$, i.e., $A^T = A^{-1}$ and~$\det A =1$. Suppose to have a vector~$\mathbf{n}$, which can be equivalently represented in the two basis as
\begin{equation}
\mathbf{n} = \tilde{n}_1 \mathbf{e}_1 + \tilde{n}_2 \mathbf{e}_2 + \tilde{n}_3 \mathbf{e}_3 = \tilde{n}'_1 \mathbf{e}'_1 + \tilde{n}'_2 \mathbf{e}'_2 + \tilde{n}'_3 \mathbf{e}'_3.
\end{equation}
In this case it can be proved that~$\tilde{\mathbf{n}}' = A \tilde{\mathbf{n}}$, where we have introduced the vectors~$\tilde{\mathbf{n}}^T=(\tilde{n}_1, \tilde{n}_2, \tilde{n}_3)$ and~$(\tilde{\mathbf{n}}')^T=(\tilde{n}'_1, \tilde{n}'_2, \tilde{n}'_3)$ containing the coordinates of the vector~$\mathbf{n}$ in the unprimed and primed basis, respectively. In the same way we can transform operators: suppose to have two vectors~$\mathbf{u}, \mathbf{v}$ related by the operator~$O$ through~$\mathbf{u} = O \mathbf{v}$. After the change of coordinates~$A$ the same relation reads as~$\mathbf{u}' = A\mathbf{u} = O' \mathbf{v}'$, where~$O' = AOA^{-1}$ is the form of the operator in the primed basis. 

As far as we are concerned, we are interested in a transformation that rotates one coordinate system into a second one. Every rotation in three dimensions can be decomposed into fundamental rotations around the axis, hence to describe any rotation we only need to know the explicit form of the three counter-clockwise rotations of angle~$\theta$ around the three axis of a generic orthonormal basis~$\{ \hat{x}, \hat{y}, \hat{z} \}$:
\begin{equation}
R^{\hat{x}}(\theta) = 
\begin{pmatrix}
1 & 0           & 0          \\
0 & \cos\theta  & \sin\theta \\
0 & -\sin\theta & \cos\theta \\
\end{pmatrix},\ 
R^{\hat{y}}(\theta) = 
\begin{pmatrix}
\cos\theta & 0 & -\sin\theta \\
0          & 1 & 0           \\ 
\sin\theta & 0 & \cos\theta  \\
\end{pmatrix},\ 
R^{\hat{z}}(\theta) = 
\begin{pmatrix}
\cos\theta  & \sin\theta & 0 \\
-\sin\theta & \cos\theta & 0 \\
0           & 0          & 1 \\
\end{pmatrix}.
\label{eq:fundamental_rotation_matrices}
\end{equation} 


\section{Halo properties}
\label{app:halo_properties}

In this appendix we summarise the main results concerning the different halo properties we use in this work. This appendix should be understood as independent from the rest of the paper: certain fitting parameter symbols overlap with other used for certain variables in the main text or in other appendices. In case of ambiguity, any symbol used here below refers exclusively to this appendix.

In this work we use the halo mass function, calibrated on numerical simulations, of ref.~\cite{rodriguezpuebla:halomassfunction}. The comoving halo number density distribution reads as
\begin{equation}
\frac{dn_h}{dM_h} = f(\sigma) \frac{\rho_{0c}\Omega_{m0}}{M_h} \frac{d\log\sigma^{-1}}{dM_h},
\end{equation}
where 
\begin{equation}
\begin{aligned}
f(\sigma) &= A \left[1 + \left(\frac{b}{\sigma}\right)^{a} \right] e^{-c/\sigma^2}, \\
A &= 0.144 - 0.011 z + 0.003 z^2, \\
a &= 1.351 + 0.068 z + 0.006 z^2, \\
b &= 3.113 - 0.077 z - 0.013 z^2, \\
c &= 1.187 + 0.009 z,
\end{aligned}
\end{equation}
and~$\sigma$ is the variance of the matter field smoother on the scale of the halo given by~\cite{rodriguezpuebla:halomassfunction}
\begin{equation}
\sigma = D(a) \frac{17.111 y^{0.405}}{1 + 1.306 y^{0.22} + 6.218 y^{0.317}},
\end{equation}
with~$y = 10^{12}\ h^{-1}M_\odot/M_h$ and~$D(a)=g(a)/g(1)$ being the linear growth factor~\cite{lahav:growthfactor, carroll:growthfactor} with
\begin{equation}
\begin{aligned}
g(a) &= \frac{5}{2}\frac{a\Omega_m(a)}{\Omega^{4/7}_m(a) - \Omega_\Lambda(a) + \left(1+\frac{\Omega_m(a)}{2}\right)\left(1+\frac{\Omega_\Lambda(a)}{70}\right)}, \\
\Omega_m(a) &= \frac{\Omega_{m0}a^{-3}}{\Omega_{\Lambda 0}+\Omega_{m0}a^{-3}} = 1 - \Omega_\Lambda(a), \\
\end{aligned}
\end{equation}

The functions characterizing the average star formation rate per halo of equation~\eqref{eq:total_sfr_per_halo} read as~\cite{behroozi:universemachine}
\begin{equation}
\begin{aligned}
f_Q &= Q_\mathrm{min} + (1-Q_\mathrm{min})\left[\frac{1}{2} + \frac{1}{2}\mathrm{erf}\left(\frac{\log_{10}\nu_Q}{\sqrt{2}\sigma_{V_Q}}\right)\right], \\
\mathrm{SFR}_{SF} &= \epsilon\left[ \frac{1}{\nu^\alpha + \nu^\beta} +\gamma\mathrm{exp}\left(-\frac{\log^2_{10}\nu}{2\delta^2}\right) \right], \\
\sigma_{SF} &= \mathrm{max}\left[0, \mathrm{min}\left[\sigma_{SF,0} + \frac{z}{1+z}\sigma_{SF,1}, 0.3 \right]\right] \ \mathrm{dex}, 
\end{aligned}
\label{eq:sfr_pdf_parameters_I}
\end{equation}
where
\begin{equation}
\begin{aligned}
Q_\mathrm{min} &= \mathrm{max}\left[0, Q_{\mathrm{min},0} + \frac{z}{1+z}Q_{\mathrm{min},1} \right], \\
\nu_Q &= \frac{\nu_{M_\mathrm{peak}}}{V_Q \ \mathrm{km\ s^{-1}}}, \\
\log_{10} V_Q &= V_{Q,0} + \frac{z}{1+z} V_{Q,1} + zV_{Q,2}, \\
\sigma_{V_Q} &= \sigma_{V_{Q},0} + \frac{z}{1+z} \sigma_{V_{Q},1} + \log(1+z)\sigma_{V_{Q},2}\ \mathrm{dex}, \\
\nu &= \frac{\nu_{M_\mathrm{peak}}}{V \ \mathrm{km\ s^{-1}}}, \\
\log_{10} V &= V_{0} + V_{1}\frac{z}{1+z} + V_2\log(1+z) + V_{3}z, \\
\log_{10} \epsilon &= \epsilon_{0} + \epsilon_{1}\frac{z}{1+z} + \epsilon_2\log(1+z) + \epsilon_{3}z, \\
\alpha &= \alpha_{0} + \alpha_{1}\frac{z}{1+z} + \alpha_2\log(1+z) + \alpha_{3}z, \\
\beta &= \beta_{0} + \beta_{1}\frac{z}{1+z} + \beta_{2}z, \\
\log_{10} \gamma &= \gamma_{0} + \gamma_{1}\frac{z}{1+z} + \gamma_{2}z, \\
\delta &= \delta_0, \\
\end{aligned}
\label{eq:sfr_pdf_parameters_II}
\end{equation}
and, most importantly,
\begin{equation}
\nu_{M_\mathrm{peak}} = 200 \left(\frac{M_h}{M_{200}}\right)^{1/3}\ \mathrm{km\ s^{-1}},\ \mathrm{with}\ M_{200} = \frac{1.64 \times 10^{12}\ M_\odot}{\left[0.378(1+z)\right]^{0.142}+\left[0.378(1+z)\right]^{1.79}}.
\end{equation}
The reference value for all the parameters appearing in equations~\eqref{eq:sfr_pdf_parameters_I} and~\eqref{eq:sfr_pdf_parameters_II} are reported in table~\ref{tab:sfr_pdf_parameters}.

\begin{table}[ht]
\centerline{
\begin{tabular}{|c|c||c|c||c|c|}
\hline
Parameter & Fiducial Value & Parameter & Fiducial Value & Parameter & Fiducial Value \\
\hline
\hline
$\sigma_{SF,0}$      & $-0.302$ & $V_{0}$        & $2.139$   & $\alpha_{2}$ & $8.601$  \\
$\sigma_{SF,1}$      & $4.961$  & $V_{1}$        & $-1.487$  & $\alpha_{3}$ & $-0.789$ \\
$Q_{\mathrm{min},0}$ & $-1.086$ & $V_{2}$        & $1.561$   & $\beta_{0}$  & $-1.813$ \\
$Q_{\mathrm{min},1}$ & $0.721$  & $V_{3}$        & $-0.216$  & $\beta_{1}$  & $-1.005$ \\
$V_{Q,0}$            & $2.232$  & $\epsilon_{0}$ & $0.219$   & $\beta_{2}$  & $1.772$  \\
$V_{Q,1}$            & $0.032$  & $\epsilon_{1}$ & $-3.844$  & $\gamma_{0}$ & $-1.914$ \\
$V_{Q,2}$            & $0.107$  & $\epsilon_{2}$ & $5.213$   & $\gamma_{1}$ & $5.053$  \\
$\sigma_{V_{Q},0}$   & $0.270$  & $\epsilon_{3}$ & $-0.768$  & $\gamma_{2}$ & $-0.921$ \\
$\sigma_{V_{Q},1}$   & $-0.258$ & $\alpha_{0}$   & $-6.135$  & $\delta_0$   & $0.066$  \\
$\sigma_{V_{Q},2}$   & $0.054$  & $\alpha_{1}$   & $-12.995$ & $ $          & $ $      \\
\hline
\end{tabular}}
\caption{Reference values for the star formation rate probability density distribution.}
\label{tab:sfr_pdf_parameters}
\end{table}

To account for the growth of dark matter halos between the time of binary formation and of the merger we evolve the halo mass using the fitting formulas for the median halo virial mass provided by ref.~\cite{behroozi:averagestarformationhistory}:
\begin{equation}
M_h(M_0,z) = M_{13}(z) 10^{f(M_0,z)},
\end{equation}
where~$M_0$ is the halo mass at redshift~$z=0$ and
\begin{equation}
\begin{aligned}
M_{13}(z) &= 10^{13.276} (1+z)^3 \left(1+\frac{z}{2}\right)^{-6.11} e^{-0.503 z} \ M_\odot, \\
f(M_0,z) &= \frac{g_h(M_0,1)}{g_h(M_0,a)} \log_{10}\frac{M_0}{M_{13}(0)}, \\
g_h(M_0,a) &= 1 + e^{-4.651\left[a-a_0(M_0)\right]}, \\
a_0(M_0) &= 0.205 - \log_{10}\left[1+\left(\frac{10^{9.649}\ M_\odot}{M_0}\right)^{0.18} \right].
\end{aligned}
\end{equation}

The halo bias is computed using the fitting formula (calibrated on numerical simulations) of ref.~\cite{tinker:halobias}. It reads as
\begin{equation}
\begin{aligned}
b_h(\nu) &= 1 - A\frac{\nu^a}{\nu^a + \delta^a_c} + B\nu^b + C\nu^c, \\
A &= 1+0.24ye^{-(4/y)^4}, \\
a &= 0.44y-0.88, \\
B &= 0.183, \\
b &= 1.5, \\
C &= 0.019+0.107y+0.19e^{-(4/y)^4}, \\
c &= 2.4, \\
y &= \log_{10}(200), \\
\end{aligned}
\end{equation}
where~$\nu = \delta_c/\sigma(M,z)$ is the ``peak height'' and~$\delta_c=1.686$ is the linearly extrapolated critical overdensity for spherical collapse in Einstein-de Sitter.


\section{Astrophysical external modules}
\label{app:external_modules}

As we have seen throughout the three parts of our worked example in sections~\ref{subsec:worked_example_Meff}, \ref{subsec:worked_example_bbeff} and~\ref{subsec:worked_example_signalnoise}, computing the anisotropies of the astrophysical GWB requires several steps that cannot be efficiently implemented into~\texttt{CLASS}, which receives as input only the three effective functions~$\mathcal{M}^\mathrm{eff},b^\mathrm{eff},b^\mathrm{eff}_\mathrm{evo}$. 

Here we just briefly present what these external modules do, also to shine light on the modularity of our approach. 
\begin{itemize}
    \item \texttt{merger\_rate}. This module computes the observed merger rate pdf of equation~\eqref{eq:observed_merger_rate_pdf} for different type of sources according to the desired astrophysical model. In particular in the worked example we implemented the~\textsc{UniverseMachine} approach to connect the halo mass to the star-formation rate of dark matter halos, hence to the cosmic star-formation rate.
    
    \item \texttt{catalog}. This module generate a catalog of a given number of GW merger events. Every parameter if the event is specified by a pdf given by the user.
    
    \item \texttt{detector}. This module receive as input a catalog of GW events and for each of them computes its SNR for a given detector network; then it creates two catalogs, containing only resolved or unresolved events.
    
    \item \texttt{effective\_functions}. This module takes the catalog of unresolved events and returns the three effective functions, that can later be passed to~\texttt{CLASS\_GWB}.
    
    \item \texttt{noise}. This last module computes the noise due to spatial and temporal clustering starting from the catalog of unresolved events.
\end{itemize}
Every module is sufficiently flexible to allow for different choice of models, for instance for different pdfs or a different detector network, even if in this paper we present a specific worked out example.


\bibliography{biblio}
\bibliographystyle{utcaps}

\end{document}